%% file: dynlift-model.tex
\documentclass[copyright]{eptcs}
\providecommand{\event}{QPL 2022}

\input{dynlift-model.sty}

\title{A Biset-Enriched Categorical Model for Proto-Quipper with Dynamic Lifting}
\author{Peng Fu
\institute{Dalhousie University}
\and
Kohei Kishida
\institute{University of Illinois at Urbana-Champaign}
\and
Neil J. Ross
\institute{Dalhousie University}
\and
Peter Selinger
\institute{Dalhousie University}
}
\def\titlerunning{A Biset-Enriched Categorical Model for Proto-Quipper with Dynamic Lifting}
\def\authorrunning{P. Fu, K. Kishida, N.J. Ross \& P. Selinger}
\begin{document}
\maketitle

\begin{abstract}
Quipper and Proto-Quipper are a family of quantum programming
languages that, by their nature as circuit description languages,
involve two runtimes: one at which the program generates a circuit and
one at which the circuit is executed, normally with probabilistic
results due to measurements. Accordingly, the language distinguishes
two kinds of data: parameters, which are known at 
circuit generation time, and states, which are known at
circuit execution time. Sometimes, it is desirable for the results of
measurements to control the generation of the next part of the
circuit. Therefore, the language needs to turn states, such as
measurement outcomes, into parameters, an operation we call \emph{dynamic
lifting}. The goal of this paper is to model this interaction between
the runtimes by providing a general categorical structure enriched in
what we call ``bisets''.  We demonstrate that the biset-enriched
structure achieves a proper semantics of the two runtimes and their
interaction, by showing that it models a variant of Proto-Quipper with
dynamic lifting. The present paper deals with the concrete categorical
semantics of this language, whereas a companion paper {\cite{FKRS-types-2022}}
deals with the syntax, type system, operational semantics, and
abstract categorical semantics.
\end{abstract}

\section{Introduction}

Quipper \cite{GLRSV2013-rc,GLRSV2013-pldi} is a functional programming
language for designing quantum circuits. It shares many properties
with hardware description languages. For example, Quipper
distinguishes two kinds of runtime: (i) Circuit generation time. This
is when a quantum circuit is generated on a classical computer. (ii)
Circuit execution time. This is when a quantum circuit is run on a
quantum computer or simulator.  As a result of these two runtimes,
Quipper makes a distinction between (i) \textit{parameters} and
(ii) \textit{states}.
A parameter is a value known at circuit generation time, such as a
boolean for an if-then-else expression.
A state is a value only known at circuit
execution time, such as the state of a qubit or a bit in a circuit.

The distinction between parameters and states reflects the assumption
that classical computers and quantum devices may reside in different
physical locations and that they cooperate to perform computations. This is
also an assumption shared by the quantum computing model QRAM
\cite{knill1996conventions}.  In practice, the computation in a
quantum device can interleave with the computation in a classical
computer. This means that there should be a mechanism to turn the
results of measurements, which are states, into parameters. Dynamic
lifting is a construct that makes this possible in the programming
language. It lifts the result of a measurement from a quantum computer
to a boolean in the programming language, where it can then be used as
a parameter in the construction of the rest of the circuit. This
enables more general post-processing for quantum computation than the
simpler model where all measurements are done at the end. Some quantum
algorithms, such as those involving magic state distillation, require
dynamic lifting, while many others do not.

Since Quipper is implemented as an embedded language in the host
language Haskell, it does not have a formal semantics. Proto-Quipper
\cite{FKRS2020-dpq-tutorial,FKS2020-lindep,RS2017-pqmodel,ross2015algebraic}
is a family of quantum programming languages that are intended to
provide Quipper with a formal foundation such as operational and
categorical semantics. Like Quipper, Proto-Quipper has the two runtimes
and distinguishes between parameters and states.

The semantics of the two runtimes depends on the meaning of
``circuit'' and ``quantum operation''. Rather than fixing one specific
kind of circuit or quantum operation, the programming language is
parametric on two small categories $\m$ and $\q$, which are assumed to be
given but otherwise arbitrary, subject to some conditions. The first
of these is a symmetric monoidal category $\m$, whose morphisms
represent quantum circuits. The second is a symmetric monoidal
category $\q$, whose morphisms represent quantum operations. We note
that there is an important conceptual difference between these
categories. The morphisms of $\m$ represent circuits as
\emph{syntactic} entities. For example, Quipper allows circuits to be
\emph{boxed}, which turns them into a data structure that can be
inspected and operated on. A boxed circuit may then be reversed,
printed, iterated over, etc. Thus, $\m$ is typically a free category
generated by some collection of (quantum and classical) gates.
Measurement can be supported in the category $\m$, but it will merely
be a gate in a circuit, turning a qubit into a classical bit of the
circuit. On the other hand, the category $\q$ represents quantum
operations, which are \emph{physical} entities. Typically, $\q$ is a
category of superoperators (which include unitary operations and
measurements). We assume that $\m$ and $\q$ have the same objects, and
that there is a symmetric monoidal \emph{interpretation functor} $J : \m \to \q$, which
interprets circuits by the quantum operations they embody.

We emphasize that measurement and dynamic lifting are two different
concepts that should not be confused. Measurement is merely a gate in
a quantum circuit, which turns a qubit (a state) into a classical bit
(also a state). On the other hand, dynamic lifting is an operation of
the programming language, which turns a classical bit (a state) into a
boolean of the programming language (a parameter). In the categorical
semantics, measurement is a morphism $\Qubit \to \Bit$ in the
categories $\m$ and $\q$.  On the other hand, dynamic lifting is not a
morphism in $\m$ or in $\q$; rather, it is a morphism in a certain
Kleisli category.

Specifically, in our recent work \cite{FKRS-types-2022}, we proposed a
type system, an operational semantics and an abstract categorical
semantics for a version of Proto-Quipper with dynamic lifting, which is called Proto-Quipper-Dyn. Dynamic
lifting is modeled as a map $\Bit \to T\Bool$, where $T$ is a
commutative strong monad, such that the following diagram commutes.
\[ \footnotesize
  \begin{tikzcd}
    & \Bit \arrow[d, "\Dyn"]\\
    \Bool \arrow[r, "\eta"] \arrow[ur, "\mathsf{init}"]  & T\Bool
  \end{tikzcd}
\]
We have shown in {\cite{FKRS-types-2022}} that our categorical model
is sound with respect to the type system and operational semantics of
the language. However, the categorical semantics in
{\cite{FKRS-types-2022}} is purely abstract, simply listing the
properties that such a categorical model must have, without showing
that such a category actually exists or giving an example of one.

In this paper, we construct a concrete model for the general
categorical semantics of \cite{FKRS-types-2022}. Constructing such a
model is challenging because it requires a novel combination of
quantum circuits (morphisms in $\m$) and quantum operations (morphisms
in $\q$): The categorical model must be able to account for both
quantum circuits and quantum operations, as well as operations such as
boxing, dynamic lifting, and of course higher-order functions.

Our technical innovation to make all of this work is \emph{biset
  enrichment}. A \emph{biset} is an object in the category
$\set^{\twoOp}$, or, more concretely, it is a triple $(X_{0}, X_{1}, f)$ of
sets $X_{0}, X_{1}$ and a function $f : X_{1} \to X_{0}$.
A morphism of bisets is an obvious commutative square. We will
consider categories enriched in bisets. Concretely, such a category
has one kind of objects, but two kinds of morphisms, which we use to
model quantum circuits and quantum operations,
respectively. Our
construction is based on a biset-enriched category $\C$ constructed
from $\m$ and $\q$. Its objects are the same as those of $\m$ and
$\q$, and its hom-bisets are $(\q(A, B), \m(A, B), J_{A, B})$, where
the function $J_{A,B} : \m(A, B) \to \q(A, B)$ is given by the
interpretation functor $J$.  A global element $f$ of $\C(A, B)$
consists of a pair of functions $f_{0}, f_{1}$ that makes the following
diagram commute.
\[\footnotesize
  \begin{tikzcd}
    1 \arrow[r, "f_{1}"]
    \arrow[dr, "f_{0}"]& \m(A, B) \arrow[d, "J_{A,B}"]\\
      & \q(A, B)
  \end{tikzcd}
\]
Thus, $f_{1}$ is a quantum circuit, which can be used as
a quantum operation $f_{0}$ by composing with $J_{A,B}$. The
biset-enriched category $\C$ therefore maintains a distinction between
$\m$ and $\q$ while taking the interpretation functor $J$ into
account. To model the higher-order features of the programming
language, we embed $\C$ in a monoidal closed biset-enriched category
$\ctilde$, which we construct as a certain subcategory of the
biset-enriched category of presheaves over $\C$.  We show that
$\ctilde$ satisfies the axiomatization specified in
\cite{FKRS-types-2022}. Therefore it is a concrete model for
Proto-Quipper with dynamic lifting.

Our approach to modeling dynamic lifting differs from recent work by
Lee et al.\@ \cite{LeePVX21}, where the category of \textit{quantum
  channels}, which generalize quantum circuits with a notion of
branching for measurement results, is used to model a single runtime.
Because our model accounts separately for circuit generation time
(category $\m$) and circuit execution time (category $\q$), we are
able to support a type system that distinguishes quantum circuits from
quantum computations that use dynamic lifting \cite{FKRS-types-2022}.
This prevents a class of runtime errors in Quipper caused by 
boxing a computation that uses dynamic lifting.

The rest of the paper is structured as follows. In Section
\ref{backgrounds}, we first review some basic concepts from enriched
category theory, and then recall from \cite{FKRS-types-2022} the axiomatization of an enriched
categorical semantics for dynamic lifting. In Section
\ref{biset-enrich}, we define the biset-enriched category $\C$. We
show its presheaf category $\cbar$ admits a commutative strong monad
and a linear-non-linear adjunction. In Section \ref{reflective}, we
construct a reflective subcategory $\ctilde$ of $\cbar$ and show that
it is an enriched categorical model for dynamic lifting.

\section{An enriched categorical semantics for dynamic lifting}
\label{backgrounds}
Enriched categories are a generalization of categories where, instead of
hom-sets, one works with hom-objects, which are objects in
a monoidal category.
\begin{definition}
  Let $\V$ be a monoidal category. 
  A $\V$-enriched category $\A$ is given by the following:
  \begin{itemize}
  \item A class of objects, also denoted $\A$. 
  \item For any $A, B \in \A$, an object $\A(A, B)$ in $\V$.
  \item For any $A \in \A$, a morphism $u_{A} : I \to
    \A(A, A)$ in $\V$, called the \emph{identity} on $A$.
  \item For any $A, B, C \in \A$, a morphism
    $c_{A,B,C} : \A(A, B) \otimes \A(B, C) \to
    \A(A, C)$ in $\V$, called \emph{composition}.
        
  \item The composition and identity morphisms must satisfy suitable diagrams in $\V$ (see \cite{borceux1994handbook2,kelly1982basic}).

  \end{itemize}
\end{definition}

\noindent \textbf{Remarks}
\begin{itemize}
\item Many concepts from the theory of non-enriched categories can be generalized  to the enriched setting. For example,
  $\V$-functors, $\V$-natural transformations, $\V$-adjunctions, and the $\V$-Yoneda embedding are all
  straightforward generalizations of their non-enriched counterparts. We refer the reader to \cite{borceux1994handbook2,kelly1982basic}
  for comprehensive introductions. Symmetric monoidal categories can also be generalized to
the enriched setting (see Appendix \ref{app:sym} for a definition). 
\item In the rest of this paper, when we speak of a map $f : A \to B$ in a $\V$-enriched category $\A$, we mean a morphism of the form $f : I \to \A(A, B)$ in $\V$. Furthermore, when $g : B\to C$ is also a map in $\A$, we write $g \circ f : A \to C$ as a shorthand for
  \[ I \stackrel{f\otimes g}{\to} \A(A, B)\otimes \A(B, C) \stackrel{c}{\to} \A(A, C).\]

\item A $\V$-enriched category $\A$ gives rise to an ordinary (non-enriched) category $V(\A)$, called the \textit{underlying category} of $\A$.\footnote{We use $V(\A)$ to denote the underlying category, rather than the usual $U(\A)$, because the letter $U$ will serve another purpose in this paper.}
The objects of $V(\A)$
  are the objects of $\A$ and the hom-sets of $V(\A)$ are defined as $V(\A)(A, B) = \V(I, \A(A, B))$, for any $A, B\in V(\A)$. Similarly,
  a $\V$-functor $F : \A \to \B$ gives rise to a functor $VF : V(\A) \to V(\B)$ and a $\V$-natural transformation $\alpha : F \to G$
  gives rise to a natural transformation $V\alpha : VF \to VG$. 
\end{itemize}

The construction in this paper is parameterized by two small symmetric
monoidal categories, denoted by $\m$ and $\q$. We fix $\m$ and $\q$
once and for all and require the following:

\begin{itemize}

\item[(1)] $\m$ and $\q$ have the same objects, including a distinguished object called $\Bit$. The category
    $\m$ has distinguished morphisms $\mathrm{zero}, \mathrm{one} : I \to \Bit$.
  
\item[(2)] $\q$ has a coproduct $\Bit = I + I$, and the tensor product in $\q$ distributes over this coproduct.

\item[(3)] There is a strict symmetric monoidal functor $J : \m \to \q$ that is the identity on objects
  and $J(\mathrm{zero}) = \mathrm{inj}_{1} : I \to I+I, J(\mathrm{one}) = \mathrm{inj}_{2} : I \to I+I$. We call
  $J$ the \textit{interpretation functor}.

\item[(4)] The category $\q$ is enriched in \textit{convex spaces}. That is, for any real numbers $p_{1}, p_{2}\in [0,1]$ such that $p_{1}+p_{2} = 1$,
  and any maps $f, g \in \q(A, B)$, there is a \textit{convex sum} $p_{1}f + p_{2}g \in \q(A, B)$, and the
  convex sum satisfies certain standard conditions which are detailed in Appendix \ref{def-convex}. Moreover, composition is \textit{bilinear} with respect
  to convex sum, i.e., $(p_{1}f_{1} + p_{2}f_{2}) \circ g = p_{1}(f_{1}\circ g) + p_{2}(f_{2}\circ g)$
  and $h \circ (p_{1}f_{1} + p_{2}f_{2}) = p_{1}(h \circ f_{1}) + p_{2}(h \circ f_{2})$.

\item[(5)] For any $A\in \q$, and $f : I \to \Bit \otimes A \in \q$, we have 
      $f = p_{1} (\mathrm{inj}_{1} \otimes f_{1}) + p_{2}( \mathrm{inj}_{2} \otimes f_{2})$, where $\mathrm{inj}_{1}, \mathrm{inj}_{2} : I \to I + I$ and $p_1,p_2\in [0,1]$ are uniquely determined real numbers such that $p_1 + p_2 = 1$. When $p_{i} \not = 0$, the map $f_{i} : I \to A$ is also unique.  
\end{itemize}

Perhaps it is useful to explain more specifically what we mean when we
say that $\m$ and $\q$ are fixed ``once and for all''. The point is
that these categories are not only used in the categorical semantics,
but also in the operational semantics of Proto-Quipper-Dyn (i.e., to run
the program, we must know what a circuit is and what a quantum
operation is). Therefore, these categories should be regarded as given
as part of the language specification, rather than as a degree of
freedom in the semantics. On the other hand, nothing in the
operational or denotational semantics depends on particular properties
of $\m$ and $\q$ other than properties (1)--(5) above. Therefore,
Proto-Quipper-Dyn can handle a wide variety of possible circuit models and
physical execution models.

In practice, the category $\m$ will be a category of quantum circuits
and the category $\q$ will be a category of quantum operations. These
categories will typically have additional objects, such as $\Qubit$
and perhaps $\Qutrit$, and additional morphisms, such as $H : \Qubit
\to \Qubit$ and $\meas : \Qubit \to \Bit$. Requirement (5) is only
needed in the operational semantics of Proto-Quipper-Dyn; it is not needed for the
denotational semantics.

We now recall the enriched categorical semantics for dynamic lifting specified in \cite{FKRS-types-2022}.
\begin{definition}
  \label{abst-model}
  Let $\V$ be a cartesian closed category with
  coproducts. 
  A $\V$-category $\A$ is a \emph{model for Proto-Quipper with dynamic lifting} if it satisfies the following properties.
  \begin{enumerate}[label=\textbf{\alph*}]
  \item \label{closed} $\A$ is symmetric monoidal closed,
    i.e., it is symmetric monoidal and there is a $\V$-adjunction $- \otimes A \dashv A \multimap -$
    for any $A \in \A$.
    
  \item \label{coproducts} $\A$ has coproducts. 
    Note that the tensor products distribute over coproducts, because $-\otimes A$ is a left adjoint functor for any $A \in \A$,
    which preserves coproducts. 
  \item \label{lin-nonlin} $\A$ is equipped with a $\V$-adjunction
    $p : \V \to \A \dashv \flat : \A \to \V$ such that $p$ is a strong monoidal $\V$-functor. This implies that $p(1) \cong I$
    and $p(X\times Y) \cong pX \otimes pY$.

  \item \label{monad-t} $\A$ is equipped with a commutative strong $\V$-monad $T$.
    For any $A, B\in \A$, we write $t_{A,B} : A\otimes TB \to T(A\otimes B)$ for the
    strength and $s_{A, B} : TA \otimes B \to T(A\otimes B)$ for the costrength.
    
  \item \label{convexity}
    Let $V(\A)$ be the underlying category of $\A$, 
    $VT$ be the underlying monad of $T$, 
    and $\Kl_{VT}(V(\A))$ be the Kleisli category of $VT$.  The Kleisli category $\Kl_{VT}(V(\A))$ is
    enriched in convex spaces. In other words, for any $A, B, C \in \A$,
    if $f, g : A \to TB$ and $p, q\in \left[0,1\right], p+q = 1$, then there exists a convex sum $pf + qg : A \to TB$. Moreover,
    for any $h : C \to T A, e : B \to TC$, we have the following:
    \[
    \mu \circ T(pf + qg)\circ h = p (\mu \circ T f\circ h) + q (\mu \circ T g\circ h),
    \]
    \[
    \mu \circ T e\circ (pf + qg) = p (\mu \circ T e\circ f) + q (\mu \circ T e\circ g).
    \]    

  \item \label{simple-types}

    There are  
    fully faithful embeddings $\m \stackrel{\psi}{\hookrightarrow} V(\A)$
    and 
    $\q \stackrel{\phi}{\hookrightarrow} \Kl_{VT}(V(\A))$.
    These embedding functors are strong monoidal, and $\phi$ preserves
    the convex sum. 
    Moreover, the following
    diagram commutes for any $S, U \in \m$. 
    \[ \footnotesize
      \begin{tikzcd}
        \m(S, U)
        \arrow[r, "\psi_{S, U}"]
        \arrow[d, "J_{S, U}"]
        & V(\A)(S, U) \arrow[d, "E_{S, U}"]\\
        \q(S, U) \arrow[r, "\phi_{S, U}"]& \Kl_{VT}(V(\A))(S, U)
      \end{tikzcd}
    \]
    Here, $E : V(\A) \to \Kl_{VT}(V(\A))$ is the
    functor such that
    $E(A) = A$ and $E(f) = \eta \circ f$. 

  \item \label{box-unbox} Let $\mathcal{S}$ denote the set of objects in the image of $\psi$.
    For any $S, U\in \mathcal{S}$, there is an isomorphism
    \[ \flat(S \multimap U) \stackrel{e}{\cong}\A(S, U).\]

  \item \label{dynlift}
    There are maps $\Dyn : \Bit \to T\Bool$ and $\mathrm{init} : \Bool \to \Bit$ in $\A$
    such that the following diagram commutes.
    \[\footnotesize
      \begin{tikzcd}
         & \Bit \arrow[d, "\Dyn"]\\
        \Bool \arrow[r, "\eta"] \arrow[ur, "\mathrm{init}"]  & T\Bool
    \end{tikzcd}
  \]
  \end{enumerate}
\end{definition}

\subsection*{Remarks}
\begin{itemize}

\item Condition \ref{lin-nonlin} gives rise to a comonoid structure
  $\mathsf{dup}_{X} : pX \to pX\otimes pX$ and $\mathsf{discard}_{X} : pX \to I$ for any $X \in \V$. Moreover, for any map $f : X \to Y$
  in $\V$, we have the following in $\A$. 
  \[ \mathsf{dup}_{Y}\circ pf = (pf \otimes pf) \circ \mathsf{dup}_{X}.\]
\item Objects in the image of the functor $p$ are called
  \emph{parameter objects} in $\A$. Such objects are equipped with
  maps $\mathsf{dup} : A \to A\otimes A$ and $\mathsf{discard} : A \to
  I$. In particular, 
  $\Bool := I+I = p(1) + p(1) = p(2)$ is a parameter object.
\item Using condition \ref{box-unbox}, we define $\boxt = p(e)$ and $\unboxt = p(e^{-1})$,
    and we have 
    \[p\flat(S \multimap U) \stackrel{\boxt/ \unboxt}{\cong} p\A(S, U).\]    

\item Note that \[ \Kl_{VT}(V(\A))(A, B) = V(\A)(A, VTB) = \V(1, \A(A, TB)) = \V(1, \Kl_{T}(\A)(A, B)) = V(\Kl_{T}(\A))(A, B).\]  
\item The Kleisli category $\Kl_{VT}(V(\A))$ is monoidal because $VT$ is a commutative strong monad
  and $V(\A)$ is monoidal. For any $f : A_{1} \to VTB_{1}$ and $g : A_{2} \to VTB_{2}$ in $\Kl_{VT}(V(\A))$,
  we define $f\otimes g \in \Kl_{VT}(V(\A))(A_{1}\otimes A_{2}, B_{1}\otimes B_{2})$ by
  \[ A_{1}\otimes A_{2} \stackrel{f\otimes g}{\to} VTB_{1}\otimes VTB_{2}
    \stackrel{s}{\to} VT(B_{1}\otimes VTB_{2})
    \stackrel{Tt}{\to} VTVT(B_{1}\otimes B_{2})
    \stackrel{\mu}{\to} VT(B_{1}\otimes B_{2}).
  \]
\item Since $\psi(S) = \phi(S)$ for any $S \in \m, \q$, 
    we define $\Bit = \psi(\Bit) = \phi(\Bit) \in \A$.
  
\item Condition \ref{simple-types} expresses the requirement that the enriched category $\A$
  must combine
  both categories $\m$ and $\q$, i.e., they are subcategories
  of $V(\A)$ and its Kleisli category, respectively. Thus $\A$ has both quantum circuits and quantum operations. The commutative diagram implies that
  a circuit in $\A$ can be used as a quantum operation. 
\item In {\cite{FKRS-types-2022}}, we have shown that conditions \ref{closed}-\ref{dynlift} are sufficient to give a model 
  of Proto-Quipper-Dyn that is sound with respect to its type system and an operational semantics.
\end{itemize}

\section{A biset-enriched category $\C$ and its category of presheaves \texorpdfstring{$\cbar$}{C-bar}}
\label{biset-enrich}

\subsection{Biset enrichment}

We now begin our construction of a concrete model satisfying Definition \ref{abst-model}. 
Let $\mathbf{2}$ be the category with two objects $0,1$ and one nontrivial arrow $0 \to 1$.
Let $\V=\set^{\twoOp}$ be the category of functors from $\twoOp$ to $\set$. Concretely, the objects of $\V$ are  triples $(A_{0}, A_{1}, f)$, where $A_{0}, A_{1}$ are sets and
$f$ is a function $A_{1} \to A_{0}$. We call such a triple a \emph{biset}. 
A morphism in $\V$ from $(A_{0}, A_{1}, f)$ to $(B_{0}, B_{1}, g)$ is a pair $(h_{0}, h_{1})$, where
$h_{0} : A_{0} \to B_{0}$ and $h_{1} : A_{1} \to B_{1}$ are functions such that the following diagram commutes.
\[\footnotesize
  \begin{tikzcd}
    A_{1} \arrow[r, "h_{1}"] \arrow[d, "f"]  & B_{1} \arrow[d, "g"]\\
    A_{0} \arrow[r, "h_{0}"] & B_{0}
  \end{tikzcd}
\] 
Because it is a presheaf category, the category of bisets $\V=\set^{\twoOp}$  is complete, cocomplete, and cartesian closed. We write $A \Rightarrow B$ to denote an exponential object in $\V$.

The category $\V$ is itself a $\V$-category where the hom-object $\V(A, B)$ is given by the exponential object $A \Rightarrow B$. We write
$\mathrm{Hom}_{\V}(A, B)$ to denote a hom-set when viewing $\V$ as an ordinary category. 
Any set $X$ can be viewed as a trivial biset $(X,X,\id)$. Therefore,
any ordinary category can be viewed as a trivial biset-enriched category. For example,  
 $\set$ can be viewed as a $\V$-category, where
   the hom-objects are given by $\set(A,B) = (\set(A,B), \set(A,B), \id)$ for any $A, B\in \set$.
\begin{definition}
  We define $\V$-functors $U_{0}(A_{0}, A_{1}, a) = A_{0} : \V \to \set$, and $\Delta(X) = (X, X, \id) :  \set \to \V$. 
\end{definition}

The $\V$-functor $\Delta$ is fully faithful and $U_{0}$ is strong monoidal. Note that there is also another functor $U_{1}(A_{0}, A_{1}, a) = A_{1} : \V \to \set$, but it is only an ordinary functor, not a $\V$-functor. This
is because for $A, B \in \V$, there does not in general exist a morphism $A\Rightarrow B \to \set(A_{1}, B_{1})$ in $\V$.
The functor $U_{1}$ will play no role in this paper, but the two $\V$-functors $U_{0}$ and $\Delta$ will be important.

\begin{proposition}
  There is a $\V$-adjunction $U_{0} : \V \to \set \dashv \Delta : \set \to \V$. We write $T$ for the $\V$-monad $\Delta \circ U_{0}$, it
  is a  commutative strong $\V$-monad.
\end{proposition}

\subsection{The $\V$-category $\C$}

In the following we define a non-trivial $\V$-category $\C$. 
\begin{definition}
 We define the $\V$-category $\mathbf{C}$ as following.
  \begin{itemize}
  \item The objects of $\C$ are the same as those of $\m$ and $\q$. 
  \item For objects $A, B \in \mathbf{C}$, we define $\C(A,B)$ as the following object of $\V$,
    \[\mathbf{C}(A, B) = (\q(A, B), \m(A, B), J_{AB} : \m(A, B)\to \q(A, B)),\] 
    where $J : \m \to \q$ is the interpretation functor.
  
  \item For every object $A \in \mathbf{C}$, we have a morphism $u_{A} = (\tilde{\id}_{0}, \tilde{\id}_{1}):  1 \to \mathbf{C}(A, A)$ in $\V$, where
   $\tilde{\id}_{1}(*) = \id_{A} : A \to A$ in $\m$ and $\tilde{\id}_{0}(*) = \id_{A} : A \to A$ in $\q$. 
 \item For any $A, B, C \in \mathbf{C}$, we have a morphism $c_{A,B,C} = (c_{0}, c_{1}) :  \mathbf{C}(A, B) \times \mathbf{C}(B, C)  \to \mathbf{C}(A, C)$ in $\V$, where $c_{0} :  \q(A, B) \times \q(B, C) \to \q(A, C)$ and $c_{1} :  \m(A, B) \times \m(B, C) \to \m(A, C)$ are the compositions in $\q$ and $\m$, respectively. 
  \end{itemize}
\end{definition}

\subsection{The $\V$-category \texorpdfstring{$\cbar$}{C-bar}}

The biset-enriched category $\C$ is symmetric monoidal. However, it is not closed. For that,
we will need to work in the $\V$-enriched presheaf category $\cbar$. 

\begin{definition}
  We define the $\V$-category $\overline{\mathbf{C}} = \V^{\mathbf{C}^{\mathrm{op}}}$. 
  Concretely, an object $F \in \overline{\mathbf{C}}$ is a $\V$-functor $\mathbf{C}^{\mathrm{op}} \to \V$. Because $\V$ is complete, for any $F, G \in \overline{\mathbf{C}}$,
  we have a hom-object $\overline{\mathbf{C}}(F, G) \in \V$ 
 that represents $\V$-natural transformations $F \to G$. 
\end{definition}

An object in $\overline{\mathbf{C}}$ is a $\V$-functor
    $F : \mathbf{C}^{\mathrm{op}} \to \V$.  This means that for each
    $A \in \mathbf{C}^{\mathrm{op}}$, there is an object $F A \in \V$.
    And for any $A, B\in \mathbf{C}^{\mathrm{op}}$ there is a
    morphism
    $F_{AB} : \mathbf{C}^{\mathrm{op}}(A, B) \to F A\Rightarrow F B$ in
    $\V$, which is the following commutative diagram.  
  \[ \footnotesize
    \begin{array}{lll}
      \begin{tikzcd}
        \m(B, A) \arrow[d, "J_{B,A}"] \arrow[r, "F_{AB}^{1}"] & (FA \Rightarrow FB)_{1} = \mathrm{Hom}_{\V}(FA, FB) \arrow[d, "p_{0}"]\\
        \q(B, A) \arrow[r, "F_{AB}^{0}"] & (FA \Rightarrow FB)_{0} = \set((FA)_{0}, (FB)_{0}).
      \end{tikzcd}
    \end{array}
  \]
  Note that an element $h \in \mathrm{Hom}_{\V}(FA, FB)$ is a pair of function $(h_{0}, h_{1})$ such that
  the following commutes.
  \[\footnotesize
  \begin{tikzcd}
    (FA)_{1} \arrow[r, "h_{1}"] \arrow[d, "f"]  & (FB)_{1} \arrow[d, "g"]\\
    (FA)_{0} \arrow[r, "h_{0}"] & (FB)_{0}
  \end{tikzcd}
\] 
  Thus we define $p_{0}(h_{0}, h_{1}) = h_{0}$. 
  So a $\V$-functor $F : \mathbf{C}^{\mathrm{op}} \to \V$ induces an ordinary functor $F^{0} : \q^{\mathrm{op}} \to \set$, where $F^{0}(A) = (FA)_{0}$ and
 the function $\q(B, A) \to \set(F^{0}A, F^{0}B)$ is given by $F^{0}_{AB}$ for any $A, B\in \q$.

  \begin{proposition}
    The $\V$-category $\cbar$ is a $\V$-monoidal closed category, where the tensor product $\otimes_{\mathrm{Day}}$ and
    linear exponential $\multimap_{\mathrm{Day}}$ are given by Day's convolution \cite{day1970closed}. The tensor unit
    is defined by $I := yI = \C(-, I)$, where $y$ is the $\V$-enriched Yoneda embedding functor. 
  \end{proposition}
  The $\V$-category $\cbar$ has coproducts. Day's construction implies that the Day tensor product distributes over the coproducts,
  and that the $\V$-enriched Yoneda embedding $y : \C \hookrightarrow \cbar$ is strong monoidal.

The $\V$-adjunction $U_{0} \dashv \Delta$ and the $\V$-monad $T$ can be lifted to $\cbar$. 
\begin{definition}
  We define $\V$-functors $\overline{U_{0}}(F) := U_{0} \circ F : \V^{\mathbf{C}^{\mathrm{op}}} \to \set^{\mathbf{C}^{\mathrm{op}}}$, $\overline{\Delta}(F) := \Delta \circ F : \set^{\mathbf{C}^{\mathrm{op}}} \to \V^{\mathbf{C}^{\mathrm{op}}}$, and $\overline{T} := \overline{\Delta} \circ \overline{U_{0}} : \V^{\mathbf{C}^{\mathrm{op}}} \to \V^{\mathbf{C}^{\mathrm{op}}}$.
\end{definition}

Note that $\overline{\Delta}$ is fully faithful and that $\overline{U_{0}}$ is strong monoidal.
\begin{proposition}
   There is a $\V$-adjunction $\overline{U_{0}} :  \V^{\mathbf{C}^{\mathrm{op}}} \to \set^{\mathbf{C}^{\mathrm{op}}} \dashv \overline{\Delta} : \set^{\mathbf{C}^{\mathrm{op}}} \to \V^{\mathbf{C}^{\mathrm{op}}}$.
 \end{proposition}
 \begin{proof}
   For any $F \in \set^{\mathbf{C}^{\mathrm{op}}}, G\in \cbar$, we need to show that $\set^{\C^{\mathrm{op}}}(\overline{U_0}F, G) \cong \V^{\C^{\mathrm{op}}}(F, \overline{\Delta}G)$ that is $\V$-natural in $F$ and $G$. This is true since  
  the following isomorphisms follow from properties of \textit{end}.  
  \begin{align*}\V^{\C^{\mathrm{op}}}(F, \overline{\Delta}G) &\cong \int_{A\in \C} \V(FA, \Delta GA) \cong \int_{A\in \C}\set(U_0 FA, GA)  \cong \int_{A\in \C}\V(\Delta U_0 F A, \Delta GA)  \\
  & \cong \V^{\C^{\mathrm{op}}}(\overline{\Delta}\overline{U_0}F, \overline{\Delta} G) \cong  \set^{\C^{\mathrm{op}}}(\overline{U_0}F, G). \qedhere
  \end{align*}
\end{proof}

\begin{proposition}
  \label{prop:commutative-strong}
  The monad $\overline{T}$ is a commutative strong monad.
\end{proposition}

Proposition \ref{prop:commutative-strong} is a consequence of the following more general theorem, whose proof can be found in Appendix~\ref{pf:strong-monad-presheaves}. 
\begin{theorem}
  \label{thm:strong-monad-presheaves}
  Let $\V$ be a complete, cocomplete, symmetric monoidal closed category. Let $\A$ be a $\V$-category. If $T$ is a commutative strong $\V$-monad on $\V$, then $\overline{T}(F) = T\circ F$ is a
  commutative strong $\V$-monad on $\V^{\A^{\mathrm{op}}}$.
\end{theorem}

Consider a $\V$-functor $F : \C^{\mathrm{op}} \to \set$. For any $A, B\in \C$, $FA \in \set$, and
the map $F_{AB} : \C(B, A) \to \set(FA, FB)$ is uniquely determined by the function $F^{0}_{AB} : \q(B, A) \to \set(FA, FB)$.
So $F$ is uniquely determined by $F^{0} : \q^{\mathrm{op}} \to \set$. In fact, the following theorem holds (the proof is in Appendix \ref{app:biset-functors}).

\begin{theorem}
  \label{thm:iso}
  We have $\set^{\mathbf{C}^{\mathrm{op}}} \cong \set^{\q^{\mathrm{op}}}$. 
\end{theorem}

The following proposition shows the maps in the Kleisli category of $\tbar$ are
essentially maps in $\set^{\q^{\mathrm{op}}}$.
\begin{proposition}
  \label{prop:t-bar}
  For any $F, G \in \cbar$, we have \[\overline{\mathbf{C}}(F, \overline{T}G) = \overline{\mathbf{C}}(F, \overline{\Delta}\overline{U_{0}}G) \cong \set^{\mathbf{C}^{\mathrm{op}}}(\overline{U_{0}} F,  \overline{U_{0}} G) \cong \set^{\q^{\mathrm{op}}}(F^{0},  G^{0}).\]
\end{proposition}

\subsection{A linear-non-linear adjunction in \texorpdfstring{$\cbar$}{C-bar}}
\label{sec:linear-non-linear}
Suppose $F\in \cbar$ and $V\in \V$. By definition, the \textit{copower} $V \odot F$, if it exists,
is an object $V \odot F \in \cbar$ such that the isomorphism $\cbar(V \odot F, G) \cong V \Rightarrow \cbar(F, G)$ is $\V$-natural in $G \in \cbar$.

\begin{definition}
  Let $V \in \V, F \in \overline{\mathbf{C}}$. We define the copower $V \odot F$ in $\overline{\mathbf{C}}$ as follows:
    \[ (V \odot F)(A) = V\times FA : \C^{\mathrm{op}}\to \V .\]
  \end{definition}
  The fact that the above is indeed a copower can
  be verified using the calculus of \textit{ends}. For any $F, G \in \cbar$, we have 
  \begin{align*}
  \cbar(V\odot F, G) &\cong \int_{A\in \C} V\times FA \Rightarrow GA \cong \int_{A\in \C} V \Rightarrow (FA \Rightarrow GA)\\
  &\cong V \Rightarrow \int_{A\in \C}(FA \Rightarrow GA) \cong V \Rightarrow \cbar(F, G).
  \end{align*}
  
  \begin{definition}
    We define $\V$-functors $p(X) = X \odot I : \V \to \overline{\mathbf{C}}$
    and
    $\flat(F) = \overline{\mathbf{C}}(I, F) : \overline{\mathbf{C}} \to \V$.
  \end{definition}

  The $\V$-functors $p$ and $\flat$ form a linear-non-linear adjunction
  in the sense of Benton \cite{benton1994mixed}.
  \begin{theorem}
    We have a $\V$-adjunction $p \dashv \flat$. Moreover, $p$ is strong monoidal.  
  \end{theorem}
  \begin{proof}
    We have $\overline{\mathbf{C}}(pX, G) \cong \overline{\mathbf{C}}(X \odot I, G) \cong X \Rightarrow \overline{\mathbf{C}}(I, G) \cong X \Rightarrow \flat(G)$. Moreover, $p$ is a strong monoidal $\V$-functor. We have $p(1) = 1 \odot yI \cong 1 \times \C(-, I) \cong yI$ and 
    \begin{align*} p(X) \otimes_{\mathrm{Day}} p(Y) &= \int^{A,B}\mathbf{C}(-, A\otimes B) \times X \times yI(A) \times Y \times yI(B) \\
    &\cong X \times Y \times \int^{A,B}\mathbf{C}(-, A\otimes B) \times yI(A) \times yI(B) \cong  X \times Y \times yI = p(X \times Y).\qedhere
    \end{align*} 
  \end{proof}

  \begin{theorem}
    For any $S, U\in \C$, there is an isomorphism $\flat (yS \multimap_{\mathrm{Day}} yU) \cong \C(S, U)$.
  \end{theorem}
  \begin{proof}
    We have
    $\flat (yS \multimap_{\mathrm{Day}} yU) = \cbar(I, yS \multimap_{\mathrm{Day}} yU) \cong \cbar(yS, yU) \cong \C(S, U)$.
  \end{proof}
  Applying $p$ to the above isomorphism yields $p\flat (yS \multimap_{\mathrm{Day}} yU) \cong p\C(S, U)$. 
  This isomorphism is called the \textit{box/unbox} isomorphism in \cite{RS2017-pqmodel}. 
  
  \section{A reflective subcategory \texorpdfstring{$\ctilde$}{C-tilde} of \texorpdfstring{$\cbar$}{C-bar}}
  \label{reflective}

  The $\V$-category $\cbar$ itself is not a model for Proto-Quipper
  with dynamic lifting. For example, it does not have a map $\Bit \to
  \overline{T}\Bool$ for dynamic lifting.  Namely, we define $\Bool :=
  yI + yI = \mathbf{C}(-, I) + \mathbf{C}(-, I)$ and $\Bit := y
  \mathrm{Bit} = \mathbf{C}(-, \mathrm{Bit}) \in \cbar$, where
  $\mathrm{Bit} \in \mathbf{C}$. Note that $\mathrm{Bit} = I + I$ in
  $\q$.  Consider the following
  \begin{align*} \overline{\mathbf{C}}(\Bit, \overline{T}\Bool) &
    \cong \set^{\mathbf{C}^{\mathrm{op}}}(\overline{U_{0}}\Bit,
    \overline{U_{0}}\Bool)
    \cong \set^{\q^{\mathrm{op}}}(\Bit^{0}, \Bool^{0})\\ &\cong
    \set^{\q^{\mathrm{op}}}(\q(-, \mathrm{Bit}) , \q(-, I) + \q(-, I))
    = \set^{\q^{\mathrm{op}}}(\q(-, I+I), \q(-, I) + \q(-,
    I)).
  \end{align*}
  So a map in $\overline{\mathbf{C}}(\Bit, \overline{T}\Bool)$ is the
  same as a natural transformation from $\q(-, I+I)$ to $\q(-, I) +
  \q(-, I)$ in $\set^{\q^{\mathrm{op}}}$. Moreover, for condition
  \ref{dynlift} to be satisfied, this natural transformation should be
  a left inverse of the canonical natural transformation $\q(-, I) +
  \q(-, I)\to \q(-, I+I)$. On the other hand, by the Yoneda lemma,
  every natural transformation from $\q(-, I+I)$ to $\q(-, I) + \q(-,
  I)$ either takes all of its values in the left component or in the
  right component of the disjoint union. Therefore, it can't be a left
  inverse to $\q(-, I) + \q(-, I)\to \q(-, I+I)$. It follows that
  dynamic lifting cannot be interpreted in $\cbar$.  To fix this, we
  now consider a reflective subcategory of $\cbar$, in the style of
  Lambek {\cite{lambek2006completions}}.

  \begin{definition}
    A $\V$-functor $F : \mathbf{C}^{\mathrm{op}} \to \V$ is called \textit{smooth} if
     $F^{0} : \q^{\mathrm{op}} \to \set$ is a product-preserving functor, i.e., $F^{0}(A+B) \cong F^{0}A \times F^{0}B$ for any $A, B \in \q$.   \end{definition}
  
   Observe that for any $A\in \mathbf{C}$,
   the $\V$-enriched Yoneda embedding $y$ of $A$, which is $\mathbf{C}(-, A)$,
   is smooth.
   Because $\mathbf{C}(-, A)^{0} = \q(-, A)$, and
   for any $B_{1}, B_{2} \in \q$, we have $\q(B_{1}+B_{2}, A) \cong \q(B_{1}, A) \times \q(B_{2}, A)$. 
   Thus, the codomain of $y$ consists of smooth $\V$-functors.  
   
\begin{definition}
  We define $\widetilde{\mathbf{C}}$ to be the full $\V$-subcategory of smooth functors. 
\end{definition}

\begin{definition}
  We define
  the \emph{Lambek embedding} $\overline{y} : \mathbf{C} \to \widetilde{\mathbf{C}}$ to be the corestriction of the Yoneda embedding $y$,
  i.e., it is the unique $\V$-functor such that the following diagram commutes.
  \[\footnotesize
  \begin{array}{lll}
    \begin{tikzcd}
       \mathbf{C} \arrow[dr, "y"] \arrow[d, "\overline{y}"] & \\
       \widetilde{\mathbf{C}} \arrow[r, hook, "i"] & \overline{\mathbf{C}}
    \end{tikzcd}
  \end{array}
\]  
\end{definition}

The details of the proof of the following theorem are in Appendix \ref{app:reflective-sub}.
\begin{theorem} 
  \label{thm:reflective-sub}
 The $\V$-category $\ctilde$ is a reflective $\V$-subcategory of $\cbar$, i.e., the inclusion $\V$-functor $i : \widetilde{\mathbf{C}} \hookrightarrow \cbar$ has
    a left adjoint $\tilde{L}$. 
\end{theorem}

Using results of  Day \cite{day1972reflection,day1973note} (see also \cite{malherbe2013categorical} for a more recent exposition), we can
furthermore show that that $\ctilde$ is symmetric monoidal 
and $\ltilde$ is strong monoidal. See Appendix \ref{app:day-reflection} for further details. We now give an explicit definition of the monoidal closed structure
in $\ctilde$.
\begin{definition}
  For any $F, G \in \widetilde{\mathbf{C}}$, we define the tensor product, internal hom and tensor unit in $\widetilde{\mathbf{C}}$ as $F \otimes_{\mathrm{Lam}} G := \tilde{L}(iF \otimes_{\mathrm{Day}} iG)$, $F \multimap_{\mathrm{Lam}} G  :=  iF \multimap_{\mathrm{Day}} iG$, and $I := \overline{y}I = \C(-, I)$, respectively.
\end{definition}

In the above definition, the linear exponential $F \multimap_{\mathrm{Lam}} G$ is well-defined because $iF\multimap_{\mathrm{Day}} iG$ 
is an object in $\ctilde$ (Theorem \ref{thm:hom-preserves-smooth}).

  \begin{theorem}
    \label{thm:closed}
    The $\V$-category $\ctilde$ is symmetric monoidal closed. 
    For any $F, G, H \in \ctilde$, there is a $\V$-natural isomorphism
    $\widetilde{\mathbf{C}}(F \otimes_{\mathrm{Lam}} G, H) \cong  \widetilde{\mathbf{C}}(F , G \multimap_{\mathrm{Lam}} H)$.
  \end{theorem}
  \begin{proof}
  We have $
    \widetilde{\mathbf{C}}(F \otimes_{\mathrm{Lam}} G, H) \cong \widetilde{\mathbf{C}}(\tilde{L}(iF \otimes_{\mathrm{Day}}i G), H) \cong  \overline{\mathbf{C}}(iF \otimes_{\mathrm{Day}}i G, iH) \cong \overline{\mathbf{C}}(iF,i G \multimap_{\mathrm{Day}} iH) \cong \widetilde{\mathbf{C}}(F, G \multimap_{\mathrm{Lam}} H)$.
  \end{proof}

  \subsection{A linear-non-linear adjunction in \texorpdfstring{$\ctilde$}{C-tilde}}
  
  The $\V$-category $\ctilde$ also admits a linear-non-linear adjunction and, as in $\cbar$, there is a box/unbox isomorphism in $\ctilde$. 
  \begin{definition}
    We define the $\V$-functors
    $\tilde{p}(X) = \tilde{L}(p(X)) : \V \to \widetilde{\mathbf{C}}$ and
    $\tilde{\flat}(F) = \flat(iF) : \widetilde{\mathbf{C}} \to \V$.
  \end{definition}

  \begin{theorem}
    \label{thm:linear-nonlinear}
    We have a $\V$-adjunction $\tilde{p} \dashv \tilde{\flat}$.
    Moreover, $\tilde{p}$ is strong monoidal.
  \end{theorem}
  \begin{proof}
    We have 
    $\widetilde{\mathbf{C}}(\tilde{p}X, F) = \widetilde{\mathbf{C}}(\tilde{L}(p(X)), F) \cong \overline{\mathbf{C}}(pX, iF) \cong
      X \Rightarrow \flat(iF) \cong X \Rightarrow \tilde{\flat}(F)$.
    Moreover, $\tilde{p}$ is strong monoidal because both $\ltilde$ and
    $p$ are strong monoidal.
  \end{proof}

  \begin{theorem}
    \label{thm:box-unbox}
    For any $S, U \in \mathbf{C}$, we have
    $\mathbf{C}(S, U) \cong \tilde{\flat}(\overline{y}S \multimap_{\mathrm{Lam}} \overline{y}U)$. 
  \end{theorem}
  \begin{proof}
    We have $    \tilde{\flat}(\overline{y}S \multimap_{\mathrm{Lam}}\overline{y}U) = \flat(i(\overline{y}S \multimap_{\mathrm{Lam}}\overline{y}U)) = \overline{\mathbf{C}}(I, i(\overline{y}S \multimap_{\mathrm{Lam}} \overline{y}U) ) 
    \cong\widetilde{\mathbf{C}}(I, \overline{y}S \multimap_{\mathrm{Lam}} \overline{y}U ) \cong \widetilde{\mathbf{C}}(\overline{y}S, \overline{y}U) \cong \mathbf{C}(S, U)$.
  \end{proof}

  \subsection{A commutative strong monad on \texorpdfstring{$\ctilde$}{C-tilde}}
  The $\V$-category $\ctilde$ has a commutative strong monad. 
  In the following we write $[\V^{\mathbf{C}^{\mathrm{op}}}]_{\mathrm{prod}}$ for $\widetilde{\mathbf{C}}$
  and $\V^{\mathbf{C}^{\mathrm{op}}}$
  for $\cbar$. We write $[\set^{\q^{\mathrm{op}}}]_{\mathrm{prod}}$ for the full subcategory
  of product-preserving functors of $\set^{\q^{\mathrm{op}}}$.
  Consider the following diagram. 
  \[ \footnotesize
    \begin{tikzcd}
      \set^{\mathbf{C}^{\mathrm{op}}} \arrow[r, shift right=.75ex, hook, "\overline{\Delta}", swap]
      \arrow[d, shift right=.75ex, "L", swap]& \V^{\mathbf{C}^{\mathrm{op}}} \arrow[l, shift right=.75ex, "\overline{U_{0}}", swap]\arrow[d, shift right=.75ex, "\tilde{L}", swap] \\
      \left[\set^{\mathbf{C}^{\mathrm{op}}}\right]_{\mathrm{prod}} \arrow[u, shift right=.75ex, hook, "j", swap]\arrow[r, shift right=.75ex, hook, "\overline{\Delta}'", swap] & \left[\V^{\mathbf{C}^{\mathrm{op}}}\right]_{\mathrm{prod}}. \arrow[l, shift right=.75ex, "\overline{U_{0}}'", swap] \arrow[u, shift right=.75ex, hook, "i", swap]
    \end{tikzcd}
  \]
  We define the $\V$-functor $\overline{U_{0}}' : [\V^{\mathbf{C}^{\mathrm{op}}}]_{\mathrm{prod}} \to [\set^{\mathbf{C}^{\mathrm{op}}}]_{\mathrm{prod}}$
  by restricting the domain of $\overline{U_{0}}$ to $[\V^{\mathbf{C}^{\mathrm{op}}}]_{\mathrm{prod}}$.
  Here $[\set^{\mathbf{C}^{\mathrm{op}}}]_{\mathrm{prod}}$ is the full $\V$-subcategory of smooth $\V$-functors. 
  Similarly, $\overline{\Delta}' : [\set^{\mathbf{C}^{\mathrm{op}}}]_{\mathrm{prod}} \to [\V^{\mathbf{C}^{\mathrm{op}}}]_{\mathrm{prod}}$ is
a restriction of $\overline{\Delta}$. 
  We have a monoidal adjunction $L \dashv j$, since $\qtilde \cong [\set^{\mathbf{C}^{\mathrm{op}}}]_{\mathrm{prod}}$, the full subcategory of product-preserving functors, is
  reflective in $\qbar \cong \set^{\mathbf{C}^{\mathrm{op}}}$. 
  We write $\widetilde{T} = \overline{\Delta}' \circ \overline{U_{0}}'$. Observe that $\widetilde{T}$ is $\overline{T}$ with
  a restricted domain.

  \begin{proposition}
    By definition, we have $i \circ \overline{\Delta}' \cong \overline{\Delta} \circ j$ and $j \circ \overline{U_{0}}' \cong \overline{U_{0}} \circ i$,
    therefore $i \circ \ttilde \cong \tbar \circ i$. Moreover, $\overline{U_{0}}' \circ \ltilde \cong L \circ \overline{U_{0}}$. 
  \end{proposition}

    \begin{theorem}
    We have a $\V$-adjunction $\overline{U_{0}}' \dashv \overline{\Delta'} : [\set^{\mathbf{C}^{\mathrm{op}}}]_{\mathrm{prod}} \to [\V^{\mathbf{C}^{\mathrm{op}}}]_{\mathrm{prod}}$. And $\overline{U_{0}}'$ is strong monoidal.
  \end{theorem}
  \begin{proof}
    For any $X \in [\V^{\mathbf{C}^{\mathrm{op}}}]_{\mathrm{prod}}, Y \in [\set^{\mathbf{C}^{\mathrm{op}}}]_{\mathrm{prod}}$, we have
    \begin{align*} \left[\set^{\mathbf{C}^{\mathrm{op}}}\right]_{\mathrm{prod}} (\overline{U_{0}}'X, Y) &\cong \set^{\mathbf{C}^{\mathrm{op}}} (j\overline{U_{0}}'X, jY) \cong \set^{\mathbf{C}^{\mathrm{op}}} (\overline{U_{0}}iX, jY) \\
    &\cong \V^{\mathbf{C}^{\mathrm{op}}} (iX, \overline{\Delta}jY) \cong \V^{\mathbf{C}^{\mathrm{op}}} (iX, i \overline{\Delta'}Y) \cong \left[\V^{\mathbf{C}^{\mathrm{op}}}\right]_{\mathrm{prod}} (X, \overline{\Delta'}Y).\end{align*}
    The $\V$-functor $\overline{U_{0}}'$ is strong monoidal. 
    For any $F, G \in [\V^{\mathbf{C}^{\mathrm{op}}}]_{\mathrm{prod}}$, we have
    $\overline{U_{0}}' I \cong \overline{U_{0}} I \cong I$ and
    \begin{align*}
    \overline{U_{0}}'(F\otimes_{\mathrm{Lam}}G) &= \overline{U_{0}}'\ltilde(iF\otimes_{\mathrm{Day}} iG)\cong L\overline{U_{0}}(iF\otimes_{\mathrm{Day}} iG) \cong L (\overline{U_{0}}i F \otimes_{\mathrm{Day}}  \overline{U_{0}}i G)\\
    &\cong L (j\overline{U_{0}}' F \otimes_{\mathrm{Day}}  j\overline{U_{0}}' G) = \overline{U_{0}}' F\otimes_{\mathrm{Lam}}\overline{U_{0}}' G. \qedhere
    \end{align*} 
  \end{proof}

  \begin{theorem}
     There is a $\V$-natural transformation
      $\rho : \tilde{L}\circ \overline{T} \to \widetilde{T} \circ \tilde{L}$.
  \end{theorem}
  \begin{proof}
    For any $F \in \cbar$, let $\eta_{F} : F \to i\ltilde F$ be the unit and $\epsilon_{F} : \ltilde i F \to F$ be the counit (which is an isomorphism).
      We define $\rho_{F}$ to be the composition
      $\ltilde \tbar F \stackrel{\ltilde \tbar \eta_{F}}{\to} \ltilde \tbar i \ltilde F \stackrel{\cong}{\to}
      \ltilde i \ttilde \ltilde F \stackrel{\epsilon_{\ttilde \ltilde F}}{\to}\ttilde \ltilde F$.
  \end{proof}
  The natural transformation $\rho$ is one of the components for defining the strength for $\ttilde$.

  \begin{theorem}
    \label{thm:tmonad}
    The $\V$-functor $\widetilde{T}$ is a commutative strong monad. 
  \end{theorem}
  \begin{proof}
    For any $F, G \in \ctilde$, the strength of $\ttilde$ is given by 
    \[
    F\otimes_{\mathrm{Lam}} \ttilde G = \ltilde(iF\otimes_{\mathrm{Day}} i\ttilde G) \stackrel{\cong}{\to} \ltilde(iF\otimes_{\mathrm{Day}} \tbar i G) 
    \stackrel{\ltilde \overline{t}}{\to} \ltilde \tbar (iF\otimes_{\mathrm{Day}} i G) \stackrel{\rho}{\to} \ttilde \ltilde (iF\otimes_{\mathrm{Day}} i G) = \ttilde (F\otimes_{\mathrm{Lam}} G).
    \]
    Note that $\overline{t}$ is the strength for $\overline{T}$. 
    The verification of the strength diagrams is in Appendix \ref{app:strength-ttilde}. 
  \end{proof}

  Similarly to Proposition \ref{prop:t-bar}, we have the following theorem for $\ttilde$.
  \begin{theorem}
    \label{thm:monad-ttilde}
    For any $F, G \in \ctilde$, we have the following $\V$-natural isomorphisms.
    \[\widetilde{\mathbf{C}}(F, \tilde{T}G)\cong [\set^{\mathbf{C}^{\mathrm{op}}}]_{\mathrm{prod}}(\overline{U_{0}}'F, \overline{U_{0}}'G) \cong [\set^{\q^{\mathrm{op}}}]_{\mathrm{prod}}(F^{0}, G^{0}).\]
  \end{theorem}
  \begin{proof}
    We have $\widetilde{\mathbf{C}}(F, \tilde{T}G) = \widetilde{\mathbf{C}}(F, \overline{\Delta}'\overline{U_{0}}' G)\cong [\set^{\mathbf{C}^{\mathrm{op}}}]_{\mathrm{prod}}(\overline{U_{0}}'F, \overline{U_{0}}'G) \cong [\set^{\q^{\mathrm{op}}}]_{\mathrm{prod}}(F^{0}, G^{0})$. Note that by Theorem \ref{thm:iso}, $[\set^{\mathbf{C}^{\mathrm{op}}}]_{\mathrm{prod}} \cong [\set^{\q^{\mathrm{op}}}]_{\mathrm{prod}}$. 
    \end{proof}

    \subsection{Dynamic lifting in \texorpdfstring{$\ctilde$}{C-tilde}}

    Since $\cbar$ has coproducts and $\ctilde$ is a reflective
    subcategory, the coproduct of $A, B\in \ctilde$ is defined
    as $A +' B = \tilde{L}(iA + iB)$.
    In $\widetilde{\mathbf{C}}$, we define $\Bool :=
    \overline{y}I +' \overline{y} I = \tilde{L}(\overline{y}I + \overline{y}I)$ and $\Bit := \overline{y}(\mathrm{Bit})$, where $I, \mathrm{Bit} \in \mathbf{C}$. There exists maps $\mathsf{zero}, \mathsf{one} : \overline{y}I \to \Bit$ in $\ctilde$. We are now ready to define a map for dynamic lifting.

    \begin{theorem}
      \label{thm:dynlift}
      There are $\V$-natural transformations $\mathsf{init} : \Bool \to \Bit$ and $\Dyn : \Bit \to \widetilde{T}\Bool$ in $\widetilde{\mathbf{C}}$ such that the following diagram commutes.
      \[\footnotesize
        \begin{tikzcd}
          & \Bit \arrow[d, "\Dyn"]\\
          \Bool \arrow[r, "\eta"] \arrow[ur, "\mathsf{init}"]  & \ttilde \Bool
        \end{tikzcd}
      \]      
  \end{theorem}

  \begin{proof}
    We define $\mathsf{init} = [\mathsf{zero}, \mathsf{one}] : \Bool \to \Bit$. Firstly, we want to show that $\widetilde{T}\mathsf{init} : \widetilde{T}\Bool \to \widetilde{T}\Bit$
    is an isomorphism.
    Using Yoneda's principle, we just need to show $\ctilde(F, \widetilde{T}\mathsf{init}) : \ctilde(F, \widetilde{T}\Bool) \to \ctilde(F, \widetilde{T}\Bit)$ is
    an isomorphism
    for any $F\in \ctilde$. By Theorem \ref{thm:monad-ttilde}, this is equivalent to showing that 
    \[[\set^{\q^{\mathrm{op}}}]_{\mathrm{prod}}(F^{0}, \mathsf{init}^{0}) : [\set^{\q^{\mathrm{op}}}]_{\mathrm{prod}}(F^{0}, \Bool^{0}) \to [\set^{\q^{\mathrm{op}}}]_{\mathrm{prod}}(F^{0}, \Bit^{0})\]
    is an isomorphism. This is the case because the Lambek embedding $\kappa: \q \hookrightarrow \qtilde$ preserves coproducts, $\mathrm{Bit}= I + I \in \q$, and the map $\mathsf{init}^{0} :  \kappa I + \kappa I \to \kappa(I+I)$ is
    an isomorphism in $[\set^{\q^{\mathrm{op}}}]_{\mathrm{prod}}$. We therefore define $\Dyn$ as the composition $(\widetilde{T}\mathsf{init})^{-1} \circ \eta: \Bit \to \widetilde{T} \Bit \to \widetilde{T} \Bool$.
    As a result, we have the following commutative diagram.
    \[\footnotesize
      \begin{tikzcd}
        \Bool \arrow[r, "\eta"]
        \arrow[d, "\mathsf{init}"]& \widetilde{T} \Bool \arrow[d, "{\widetilde{T}\mathsf{init}}"] \arrow[dr, "\id"]& \\
        \Bit \arrow[r, "\eta"] & \widetilde{T} \Bit \arrow[r, "{(\widetilde{T}\mathsf{init})^{-1}}", swap]& \widetilde{T} \Bool
      \end{tikzcd} \vspace{-2.5em} 
    \]
  \end{proof}

  \subsection{\texorpdfstring{$\ctilde$}{C-tilde} is a model for Proto-Quipper with dynamic lifting}
  Recall that the category $\q$ is enriched in convex spaces, i.e.,
  the hom-sets of $\q$ are convex spaces and
  the composition is bilinear with respect to the convex sum.
  We have the following theorem, whose proof is in Appendix~\ref{def-convex}.
  
  \begin{theorem}
    \label{prop:convex}
    The category $\qtilde$ is enriched in convex spaces. Moreover, the Lambek embedding $\kappa : \q \hookrightarrow \qtilde$ preserves
    the convex sum in $\q$. 
  \end{theorem}

  The above theorem implies that for any $A, B\in \ctilde$, the Kleisli-hom $\ctilde(A, \ttilde B)$ is convex because
    of the isomorphism $\ctilde(A, \ttilde B) \cong \qtilde(A^{0}, B^{0})$ from Theorem~\ref{thm:monad-ttilde}. We are now ready to state our main theorem (see Appendix~\ref{app:main} for the proof).
    \begin{theorem}
      \label{thm:model}
      The $\V$-category $\ctilde$ is a model for Proto-Quipper with
      dynamic lifting, i.e., it satisfies conditions \ref{closed}--\ref{dynlift} in Definition \ref{abst-model}. 
    \end{theorem}

\section{Conclusion}
\label{conclude}

We constructed a categorical model for dynamic lifting using biset enrichment. We defined a biset-enriched category $\C$, which combines the categories $\m$ and $\q$. 
We then considered the full subcategory $\ctilde$ of smooth functors and showed that $\ctilde$ is a reflective subcategory in the enriched presheaf category of $\C$. Finally, we proved that $\ctilde$ is categorical model for dynamic lifting in the sense of \cite{FKRS-types-2022}.

\section*{Acknowledgements}

This work was supported by the Natural Sciences and Engineering
Research Council of Canada (NSERC) and by the Air Force Office of
Scientific Research under Award No.\@ FA9550-21-1-0041.

\bibliographystyle{eptcs}
\bibliography{dynlift-model}

\newpage

\appendix
\section{Enriched symmetric monoidal categories}
\label{app:sym}
\begin{definition}
  Let $\V$ be a symmetric monoidal category. A $\V$-category $\A$ is
  symmetric monoidal if it is equipped with the following:
  \begin{itemize}
  \item There is an object $I$, called the \emph{tensor unit}.
    For all $A, B\in \A$, there is an object $A \otimes B\in \A$. Moreover,
    for all $A_{1}, A_{2}, B_{1}, B_{2} \in \A$, there is a map
    \[\Tensor : \A(A_{1}, B_{1})\otimes \A(A_{2}, B_{2}) \to \A(A_{1}\otimes A_{2}, B_{1}\otimes B_{2})\]
in $\V$.
    The tensor product is a bifunctor in the sense that $\Tensor \circ (u_{A}\otimes u_{B}) = u_{A\otimes B}$ for the identity maps $u_{A}, u_{B}, u_{A\otimes B}$, and the following diagram commutes for any $A_{1}, A_{2}, B_{1}, B_{2}, C_{1}, C_{2}\in \A$. 
    \[
      \begin{tikzcd}
        \A(A_{1}, B_{1})\otimes \A(A_{2}, B_{2}) \otimes \A(B_{1}, C_{1})\otimes \A(B_{2}, C_{2})
        \arrow[r, "c\otimes c"]
        \arrow[d, "\Tensor\otimes \Tensor"]& \A(A_{1}, C_{1})\otimes \A(A_{2}, C_{2}) \arrow[d, "\Tensor"]\\
        \A(A_{1}\otimes A_{2}, B_{1}\otimes B_{2}) \otimes \A(B_{1}\otimes B_{2}, C_{1}\otimes C_{2})\arrow[r, "c"]& \A(A_{1}\otimes A_{2}, C_{1}\otimes C_{2})
      \end{tikzcd}
    \]
  \item There are the following $\V$-natural isomorphisms in $\A$ and they satisfy
    the same coherence diagrams as for symmetric
    monoidal categories, and analogous naturality conditions.
    \[l_{A} : I\otimes A \to A\]
    \[r_{A} : A\otimes I \to A\]
    \[\gamma_{A,B} : A\otimes B \to B\otimes A\]
    \[\alpha_{A,B,C} : (A\otimes B)\otimes C \to  A \otimes (B\otimes C)\]
    
  \end{itemize}
  
\end{definition}

If the $\V$-category $\A$ is symmetric monoidal, for all maps $f : A_{1} \to B_{1}, g : A_{2} \to B_{2}$ in $\A$, we write $f \otimes g : A_{1}\otimes A_{2} \to B_{1}\otimes B_{2}$ as a shorthand for the
    following composition.
    \[ I \stackrel{f\otimes g}{\to} \A(A_{1}, B_{1})\otimes \A(A_{2}, B_{2}) \stackrel{\Tensor}{\to} \A(A_{1}\otimes A_{2}, B_{1}\otimes B_{2})\]

\section{Biset-enriched functor categories}
\label{app:biset-functors}
\textbf{Notations}. Let $\A, \B$ be $\V$-categories. For all $A, B \in \A$,
we have
\[\A(A, B) = (\A(A, B)_{0}, \A(A, B)_{1}, \varphi^{\A} : \A(A, B)_{1} \to \A(A, B)_{0}).\] So we write $A \to_{1} B := \A(A, B)_{1}$ and $A\to_{0}B := \A(A, B)_{0}$. Moreover, for all $f : A \to_{1} B$, we have $\varphi^{\A}(f) : A \to_{0} B$. 
A $\V$-functor $F : \A \to \B$ gives rise to the following commutative diagram for all $A, B\in \A$.

\[
  \begin{tikzcd}
    \A(A, B)_{1} \arrow[r, "F_{A,B}^{1}"]\arrow[d, "\varphi^{\A}"] &  \B(FA, FB)_{1} \arrow[d, "\varphi^{\B}"]\\
    \A(A, B)_{0} \arrow[r, "F_{A,B}^{0}"] &  \B(FA, FB)_{0}
  \end{tikzcd}
\]
For all $f : A \to_{1} B$, we
have $F_{A, B}^{1}f : FA \to_{1} FB$. Similarly, for all $g : A \to_{0} B$,
we have $F_{A, B}^{0}g : FA \to_{0} FB$.

For any $\V$-functors $F, G : \A \to \B$,
we define a biset
$(F \Rightarrow_{0} G, F \Rightarrow_{1} G, p : F \Rightarrow_{1} G \to F \Rightarrow_{0} G)$ as follows.
\[F \Rightarrow_{0} G := \{ (\beta_{A} : FA \to_{0} GA)_{A\in \A}\ |\ \forall A, B \in \A, \forall g : A \to_{0} B, \beta_{B} \circ F_{AB}^{0}g = G_{AB}^{0}g \circ \beta_{A}\}\]

\[F \Rightarrow_{1} G := \{ (\alpha_{A} : FA \to_{1} GA)_{A\in \A}\ |\ \forall A, B \in \A, \forall f : A \to_{1} B, \alpha_{B} \circ F_{AB}^{1}f = G_{AB}^{1}f \circ \alpha_{A}, \]
\[\forall A, B \in \A, \forall g : A \to_{0} B, \varphi^{\B}(\alpha_{B}) \circ F_{AB}^{0}g = G_{AB}^{0}g \circ \varphi^{\B}(\alpha_{A})\}\]

\[p((\alpha_{A} : FA \to_{1} GA)_{A\in \A}) := (\varphi^{\B}(\alpha_{A}) : FA \to_{0} GA)_{A\in \A} : F \Rightarrow_{1} G \to F \Rightarrow_{0} G\]

\begin{proposition}
  Suppose $\A, \B$ are $\V$-categories.
  Since the category of bisets $\V$ is complete, the functor category $\B^{\A}$ is $\V$-enriched. For all $\V$-functors $F, G : \A\to \B$, 
  we have
  \[\B^{\A}(F, G) := \int_{A\in \A}\B(FA, GA) \cong (F \Rightarrow_{0} G, F \Rightarrow_{1} G, p : F \Rightarrow_{1} G \to F \Rightarrow_{0} G)\]
\end{proposition}
\begin{proof}
  By definition of end, we have the following equalizer diagram in $\V$.
  \[
  \begin{array}{lll}
    \begin{tikzcd}
     \int_{A\in \A}\B(FA, GA) := eq(u, v)  \arrow[r, "k", tail]
      & \prod_{A\in \A} \B(F A, G A)
      \arrow[r,shift left=.75ex,"u"]
      \arrow[r,shift right=.75ex, "v", swap] 
      & \prod_{A, B \in \A} \A(A, B) \Rightarrow \B(F A, G B)
    \end{tikzcd}
  \end{array}
\] 
Note that $u = \langle \mathrm{curry}(c \circ (\pi_{A}\times G_{AB}))\rangle_{A,B\in \A}$, where $c \circ (\pi_{A}\times G_{AB})$ is the following. 
\[
  \begin{array}{lll}
    \begin{tikzcd}[column sep=large]
      (\prod_{A} \B(F A, G A)) \times \A(A, B)
      \arrow[r, "\pi_{A} \times G_{AB}"]
      & 
      \B(FA, GA)\times \B(GA, GB)
      \arrow[r, "c"]
      & \B(FA, GB)
    \end{tikzcd}
  \end{array}
\]  
We have $v = \langle \mathrm{curry}(c \circ (\pi_{B}\times F_{AB}))\rangle_{A,B\in \A}$, where $c \circ (\pi_{B}\times F_{AB})$ is the following. 
\[
  \begin{array}{lll}
    \begin{tikzcd}[column sep=large]
      (\prod_{A} \B(F A, G A)) \times \A(A, B)
      \arrow[r, "\pi_{B}\times F_{AB}"]
      & 
      \B(FB, GB)\times \B(FA, FB)
      \arrow[r, "c"]
      & \B(FA, GB)
    \end{tikzcd}
  \end{array}
\]
We can show $(\int_{A\in \A}\B(FA, GA))_{1} = eq(u_{1}, v_{1}) \cong F \Rightarrow_{1} G$
and $(\int_{A\in \A}\B(FA, GA))_{0} = eq(u_{0}, v_{0}) \cong F \Rightarrow_{0} G$.
\end{proof}

\begin{theorem}
  The biset-enriched categories $\set^{\mathbf{C}^{\mathrm{op}}}$ and $\set^{\q^{\mathrm{op}}}$ are isomorphic. 
\end{theorem}
\begin{proof}
  Let us define a $\V$-enriched functor $\Omega : \set^{\mathbf{C}^{\mathrm{op}}} \to \set^{\q^{\mathrm{op}}}$.
  On objects, $\Omega(F) = F^{0}$ for any $F\in \set^{\mathbf{C}^{\mathrm{op}}}$. Since $F : \C^{\mathrm{op}} \to \set$
  is uniquely determined by $F^{0}$, the function $\Omega$ is bijective on objects.

  Suppose $F, G : \C^{\mathrm{op}} \to \set$. We claim that $\set^{\mathbf{C}^{\mathrm{op}}}(F, G) \cong \set^{\q^{\mathrm{op}}}(F^{0}, G^{0})$.
  This will allow us to define $\Omega_{F,G}$ to be this isomorphism. To show $\set^{\mathbf{C}^{\mathrm{op}}}(F, G) \cong \set^{\q^{\mathrm{op}}}(F^{0}, G^{0})$, first of all, we have
  \[\set^{\q^{\mathrm{op}}}(F^{0}, G^{0}) = (X, X, \id),\]
  where
  \[X = \{ (\alpha_{A} : F^{0}A \to G^{0}A)_{A\in \q} \ | \ \forall A, B\in \q, \forall f : A \to B \in \q, \alpha_{B} \circ F^{0}_{AB}f = G^{0}_{AB}f\circ \alpha_{A}\}.\]
  Next,
  \[\set^{\mathbf{C}^{\mathrm{op}}}(F, G) = (F \Rightarrow_{1}G, F \Rightarrow_{0}G, p),\]
  where
  \[F \Rightarrow_{0}G = \{ (\alpha_{A} : FA \to_{0} GA)_{A\in \C} \ | \ \forall A, B\in \C, \forall f : A \to_{0} B \in \C, \alpha_{B} \circ F^{0}_{AB}f = G^{0}_{AB}f\circ \alpha_{A}\}\cong X\]
  and
\[F \Rightarrow_{1} G := \{ (\alpha_{A} : FA \to_{1} GA)_{A\in \A}\ |\ \forall A, B \in \C, \forall f : A \to_{1} B, \alpha_{B} \circ F_{AB}^{1}f = G_{AB}^{1}f \circ \alpha_{A}, \]
\[\forall A, B \in \C, \forall g : A \to_{0} B, \varphi^{\set}(\alpha_{B}) \circ F_{AB}^{0}g = G_{AB}^{0}g \circ \varphi^{\set}(\alpha_{A})\}.\]

Since $F^{1}_{A,B} = F^{0}_{A,B}\circ \varphi^{\C^{\mathrm{op}}}$, and $\varphi^{\set} = \id$, and $\varphi^{\C^{\mathrm{op}}}(f) : A\to_{0}B$
for any $f : A\to_{1}B$ with $A, B\in \C$, therefore $\forall A, B \in \C, \forall g : A \to_{0} B, \varphi^{\set}(\alpha_{B}) \circ F_{AB}^{0}g = G_{AB}^{0}g \circ \varphi^{\set}(\alpha_{A})$ implies $\forall A, B \in \C, \forall f : A \to_{1} B, \alpha_{B} \circ F_{AB}^{1}f = G_{AB}^{1}f \circ \alpha_{A}$. So $F \Rightarrow_{1} G \cong F \Rightarrow_{0} G \cong X$ and $p = \id$.  
\end{proof}

\section{Convexity}
\label{def-convex}
Let $\runit$ denote the real unit interval. 
\begin{definition}
  A \emph{convexity structure} on a set $X$ is an operation that assigns to all $p,q\in\runit$ with $p+q=1$ and all $x,y\in X$ an element $px+qy\in X$, subject to the following properties. Throughout, we assume $p+q=1$.
  \begin{itemize}
  \item[(a)] $px + qx = x$ for all $x\in X$. 
    
  \item[(b)]  
    $p x + q y = q y + p x$ for all $x, y \in X$.
    
  \item[(c)] $0x + 1y = y$ for all $x, y \in X$.
    
  \item[(d)] $(a+b)(\frac{a}{a+b} x + \frac{b}{a+b} y) +
    (c+d)(\frac{c}{c+d} z + \frac{d}{c+d} w) =
    (a+c)(\frac{a}{a+c} x + \frac{c}{a+c} z) +
    (b+d)(\frac{b}{b+d} y + \frac{d}{b+d} w)$, where
    $a,b,c,d\in\runit$ with $a+b+c+d=1$ and
    all denominators are non-zero.
  \end{itemize}
\end{definition}

\noindent \textbf{Remark.} Property $(d)$ can best be understood by
realizing that both sides of the equation are equal to $ax+by+cz+dw$,
decomposed in two different ways into convex sums of two elements at a
time. In the literature, we sometimes find a different, but equivalent
condition of the form $s(px + qy) + r z = spx +
(qs+r)(\frac{qs}{qs+r}y + \frac{r}{qs+r}z)$. The latter axiom is arguably
shorter, but harder to read.

We often expand the binary $+$ operation to a multi-arity operation,
i.e., $\sum_{i} p_{i}x_{i}$, where $\sum_{i}p_{i} = 1$ and $x_{i} \in X$ for all $i$.

We say that a category $\A$ is \emph{enriched in convex spaces} if for
all $A,B\in\A$, the hom-set $\A(A, B)$ is convex, and composition is
bilinear, i.e., for all $f, g\in \A(A, B), e \in \A(C, A), h\in
\A(B, C)$ and $p, q\in [0,1]$ with $p+q=1$, we have
\[(p f + q g) \circ e = p f\circ e + q g \circ e\] and
\[h \circ (p f + q g) = p h \circ f + q h \circ g. \]

\begin{theorem}
  \label{thm:convex2}
  Let $\A$ be a symmetric monoidal category with a coproduct $I+I$,
  such that tensor distributes over this coproduct. The following are
  equivalent.
  \begin{enumerate}
  \item The category $\A$ is enriched in convex spaces.
  \item  There exists a family of maps $\llangle p, q \rrangle : I \to
    I + I$, where $p,q\in \runit$ with $p+q =1$, such that the
    following diagrams commute:
    \[
    \begin{tikzcd}
      I
      \arrow[rr, bend right = 20, "\id", swap]
      \arrow[r, "{\llangle p, q\rrangle}"] & I + I\arrow[r, "{\left[\id, \id\right]}"] & I\\
    \end{tikzcd}
    \qquad
    \begin{tikzcd}
      I\arrow[r, "{\llangle p, q\rrangle}"]
      \arrow[dr, "{\llangle q, p\rrangle}", swap]& I + I \arrow[d, "{\left[\mathrm{inj}_{2}, \mathrm{inj}_{1}\right]}"]\\
      & I + I
    \end{tikzcd}
    \qquad
    \begin{tikzcd}
      I \arrow[r, bend right = 20, "{\llangle 0, 1 \rrangle}", swap]
      \arrow[r, bend left = 20, "\mathrm{inj}_{2}"]& I + I
    \end{tikzcd}
    \]
    \[
    \begin{tikzcd}
      & I \arrow[dl, "{\llangle a+b, c+d\rrangle}",swap] \arrow[dr, "{\llangle a+c, b+d\rrangle}"]& \\
      I + I \arrow[d, "{\llangle \frac{a}{a+b}, \frac{b}{a+b}\rrangle}+{\llangle \frac{c}{c+d}, \frac{d}{c+d}\rrangle}", swap] & & I + I \arrow[d, "{\llangle \frac{a}{a+c}, \frac{c}{a+c} \rrangle}+{\llangle\frac{b}{b+d}, \frac{d}{b+d}\rrangle}"]\\
      (I + I) + (I+I) \arrow[rr, "\mathrm{iso}"]&& (I+I) + (I + I)
    \end{tikzcd}
    \]
    Here, in the last diagram, we have $a,b,c,d\in\runit$ with
    $a+b+c+d=1$, and we assume the denominators are non-zero. The map
    ``iso'' is the canonical isomorphism $(A+B)+(C+D) \cong (A+C)+(B+D)$.
  \end{enumerate}
\end{theorem}

\begin{proof}
  For the left-to-right implication, suppose $\A$ is enriched in
  convex spaces. We can define \[\llangle p, q \rrangle :=
  p\,\mathrm{inj}_{1} + q\,\mathrm{inj}_{2} : I \to I + I.\] It is
  easy to verify that this definition of $\llangle p, q \rrangle$
  satisfies the four diagrams above.

  We now focus on the right-to-left implication.
      \begin{itemize}
      \item First we need to show that $\A(A, B)$ is convex for all $A, B \in \A$.
      Given $f, g \in \A(A, B)$, we define $pf + qg$ as follows.
      \[ A \xrightarrow{\lambda^{-1}} A \otimes I \xrightarrow{A\otimes \llangle p, q \rrangle} A \otimes(I+I)
        \xrightarrow{d} A\otimes I + A\otimes I  \xrightarrow{\lambda + \lambda} A + A \xrightarrow{\left[f, g\right]} B.
      \]
      \item $pf + qf = f$. This holds because the following diagram commutes.
        \[
          \begin{tikzcd}
            A \arrow[r, "\lambda^{-1}"]
            \arrow[d, "f"]
            & A\otimes I \arrow[r, "A\otimes {\llangle p, q\rrangle}"]
            \arrow[d, "f\otimes I"]
            & A\otimes (I+I) \arrow[r, "d"]
            \arrow[d, "f\otimes (I+I)"]
            & A\otimes I + A\otimes I \arrow[r, "\lambda + \lambda"]
            \arrow[d, "f\otimes I+ f\otimes I"]
            & A + A \arrow[r, "{\left[f, f\right]}"]
            \arrow[d, "f+f", swap]
            & B \\
            B \arrow[r, "\lambda^{-1}"] & B\otimes I \arrow[r, "B\otimes {\llangle p, q\rrangle}"]
            \arrow[dr, "\id"]
            & B\otimes (I+I) \arrow[r, "d"]\arrow[d, "B\otimes {\left[\id, \id\right]}"]
            & B\otimes I + B\otimes I \arrow[r, "\lambda + \lambda"]
            \arrow[dl, "{\left[\id, \id\right]}", bend left = 10]
            & B + B \arrow[ur, "{\left[\id, \id\right]}"]&  \\
            &&B\otimes I \arrow[uurrr, bend right = 40, "\lambda"]&&&
          \end{tikzcd}
        \]
        
      \item $pf + qg = qg + pf $. This holds because the following diagram commutes.
        \[
          \begin{tikzcd}
            A \arrow[r, "\lambda^{-1}"]
            & A\otimes I \arrow[r, "A\otimes {\llangle p, q\rrangle}"]
            \arrow[dr, "A\otimes {\llangle q, p\rrangle}",swap]
            & A\otimes (I+I) \arrow[r, "d"]
            \arrow[d, "A\otimes {\left[\mathrm{inj}_{2}, \mathrm{inj}_{1}\right]}"]
            & A\otimes I + A\otimes I \arrow[r, "\lambda + \lambda"]
            & A + A \arrow[r, "{\left[f, g\right]}"]
            \arrow[d, "{\left[\mathrm{inj}_{2}, \mathrm{inj}_{1}\right]}", swap]
            & B \\
            & 
            & A\otimes (I+I) \arrow[r, "d"]
            & A\otimes I + A\otimes I \arrow[r, "\lambda + \lambda"]
            & A + A \arrow[ur, "{\left[g, f\right]}", swap]&  \\
          \end{tikzcd}
        \]
        
    \item $0f + 1g = g$. We have the following commutative diagram.

      
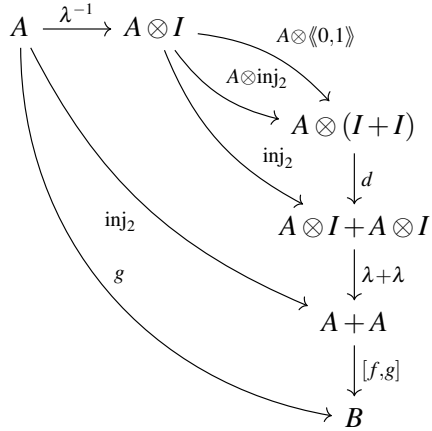
\[
        \begin{tikzcd}
          A\arrow[r, "\lambda^{-1}"]
          \arrow[ddddrr, bend right = 40, "g"]
          \arrow[dddrr, bend right = 20, "\mathrm{inj}_{2}", swap]
          & A\otimes I \arrow[dr, "A\otimes {\llangle 0, 1\rrangle}", bend left = 20]
          \arrow[ddr, "\mathrm{inj}_{2}", bend right = 20, near end]
          \arrow[dr, bend right = 20, "A\otimes \mathrm{inj}_{2}"]& \\
          &&A\otimes(I+I) \arrow[d, "d"]\\
          &&A\otimes I + A \otimes I \arrow[d, "\lambda + \lambda"]\\
          &&A + A  \arrow[d, "{\left[f, g\right]}"]\\
          &&B
        \end{tikzcd}
      \]
      
    \item $(a+b)(\frac{a}{a+b} f + \frac{b}{a+b} g) +
      (c+d)(\frac{c}{c+d} h + \frac{d}{c+d} w) = (a+c)(\frac{a}{a+c} f
      + \frac{c}{a+c} h) + (b+d)(\frac{b}{b+d} g + \frac{d}{b+d} w)$.

      Let us write $\alpha=\frac{a}{a+b} f + \frac{b}{a+b} g$ and
      $\beta=\frac{c}{c+d} h + \frac{d}{c+d} w$. We have the following
      commutative diagram.
  \[ \small
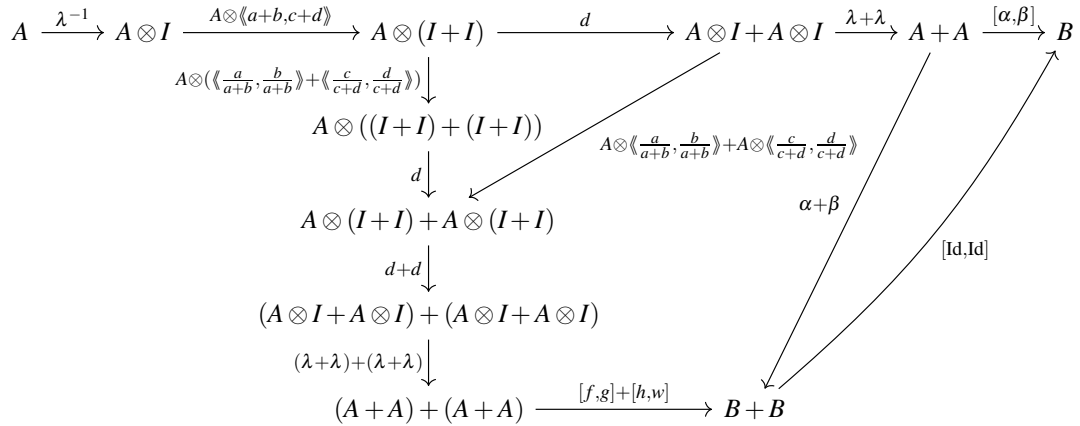
    \begin{tikzcd}
      A \arrow[r, "\lambda^{-1}"] & A\otimes I \arrow[r, "A\otimes{\llangle a+b , c+d\rrangle}"] & A\otimes (I+I) \arrow[r, "d"] \arrow[d, "A \otimes{(\llangle \frac{a}{a+b},\frac{b}{a+b}\rrangle+ \llangle \frac{c}{c+d},\frac{d}{c+d}\rrangle)}" , swap] & A\otimes I + A\otimes I \arrow[ddl, "A\otimes{\llangle \frac{a}{a+b},\frac{b}{a+b}\rrangle}+A\otimes{\llangle \frac{c}{c+d},\frac{d}{c+d}\rrangle}"]
      \arrow[r, "\lambda + \lambda"]  & A+A \arrow[r, "{[\alpha, \beta]}"] \arrow[ddddl, "\alpha + \beta",swap]& B \\
      && A\otimes((I+I)+(I+I)) \arrow[d, "d",swap]&&& \\
      &&A\otimes (I+I) + A\otimes (I+I)\arrow[d, "d+d",swap] &  && \\
      && (A\otimes I + A\otimes I) + (A\otimes I + A\otimes I)\arrow[d, "(\lambda + \lambda)+(\lambda+\lambda)",swap]&&& \\
      && (A+A) + (A+A) \arrow[r, "{[f,g]+[h,w]}"]& B+B \arrow[uuuurr, "{[\id, \id]}", swap, bend right = 10]&&
    \end{tikzcd}
  \]

  Thus
  \[
    \def\arraystretch{1.4}
    \begin{array}{l}
      (a+b)(\frac{a}{a+b} f + \frac{b}{a+b} g) +                           
      (c+d)(\frac{c}{c+d} h + \frac{d}{c+d} w) \\
      \quad {} = {[\id, \id]}\circ ([f, g]+[h,w]) \circ((\lambda + \lambda)+(\lambda
      + \lambda)) \circ (d+d) \circ d \\
      \qquad {}\circ (A \otimes{(\llangle
      \frac{a}{a+b},\frac{b}{a+b}\rrangle+ \llangle
      \frac{c}{c+d},\frac{d}{c+d}\rrangle)}) \circ (A\otimes \llangle a+b, c+d\rrangle) \circ \lambda^{-1}.
    \end{array}
  \]
  Similarly, we can show that
  \[
    \def\arraystretch{1.4}
    \begin{array}{l}
      (a+c)(\frac{a}{a+c} f + \frac{c}{a+c} h) +                           
      (b+d)(\frac{b}{b+d} g + \frac{d}{b+d} w) \\
      \quad {} = {[\id, \id]}\circ ([f, h]+[g,w]) \circ((\lambda + \lambda)+(\lambda
      + \lambda)) \circ (d+d) \circ d \\
      \qquad {} \circ (A \otimes{(\llangle
      \frac{a}{a+c},\frac{c}{a+c}\rrangle+ \llangle
      \frac{b}{b+d},\frac{d}{b+d}\rrangle)}) \circ (A\otimes \llangle a+c, b+d\rrangle) \circ \lambda^{-1}.
    \end{array}
  \]
  Thus we can show
  \[\textstyle
    (a+b)(\frac{a}{a+b} f + \frac{b}{a+b} g) +
    (c+d)(\frac{c}{c+d} h + \frac{d}{c+d} w) = (a+c)(\frac{a}{a+c} f
    + \frac{c}{a+c} h) + (b+d)(\frac{b}{b+d} g + \frac{d}{b+d} w)
  \]
  by the following commutative diagram.
  \[
    \begin{tikzcd}
      & A \otimes I \arrow[dl, "A\otimes{\llangle a+b,c+d\rrangle}", swap] \arrow[dr, "A\otimes {\llangle a+c,b+d\rrangle}"]& \\
      A\otimes (I+I) \arrow[d, "A\otimes({\llangle \frac{a}{a+b},\frac{b}{a+b}\rrangle}+{\llangle \frac{c}{c+d},\frac{d}{c+d}\rrangle})"]&& A\otimes (I+I) \arrow[d, "A\otimes({\llangle \frac{a}{a+c},\frac{c}{a+c}\rrangle+\llangle \frac{b}{b+d},\frac{d}{b+d}\rrangle})"]\\
      A\otimes ((I+I)+(I+I)) \arrow[d, "d"] \arrow[rr, "A\otimes \mathrm{iso}"]&& A\otimes ((I+I)+(I+I)) \arrow[d, "d"]\\
      A\otimes (I+I) + A\otimes (I+I) \arrow[d, "d+d"]&& A\otimes (I+I) + A\otimes (I+I)\arrow[d, "d+d"]\\
      (A\otimes I+A\otimes I) + (A\otimes I+A\otimes I) \arrow[rr, "\mathsf{iso}"]\arrow[d, "(\lambda +\lambda)+(\lambda +\lambda)"] && (A\otimes I+A\otimes I) + (A\otimes I+A\otimes I) \arrow[d, "(\lambda +\lambda)+(\lambda +\lambda)"]\\
      (A+A)+ (A+A) \arrow[rr, "\mathsf{iso}"] \arrow[d, "{[f,g]+[h, w]}"] && (A+A)+ (A+A)\arrow[d, "{[f,w]+[g,h]}"] \\
      B+B\arrow[r, "{[\id,\id]}"] & B & B+B\arrow[l, "{[\id,\id]}", swap]
    \end{tikzcd}
  \]

      \item $(pf+qg) \circ e = p (f \circ e) + q (g \circ e)$. This is by the
        following commutative diagram.
        \[
          \begin{tikzcd}
            C \arrow[r, "e"] \arrow[drr, "\lambda^{-1}"]& A \arrow[r, "\lambda^{-1}"] & A \otimes I \arrow[r, "A\otimes {\llangle p, q\rrangle}"]
            & A\otimes (I + I) \arrow[r, "d"] & A\otimes I + A\otimes I \arrow[r, "\lambda + \lambda"] & A + A \arrow[r, "{\left[f, g\right]}"] & B\\
            && C\otimes I \arrow[u, "e\otimes I"]
            \arrow[r, "C\otimes {\llangle p, q\rrangle}"]& C\otimes (I+I) \arrow[u, "e\otimes (I+I)"]\arrow[r, "d"]
            & C\otimes I + C\otimes I \arrow[r, "\lambda + \lambda"] \arrow[u, "e\otimes I + e\otimes I"]& C+C
            \arrow[u, "e+e"]\arrow[ur, "{\left[f \circ e, g \circ e\right]}",swap] &
          \end{tikzcd}
        \]
        
      \item $h \circ (pf+q g) = p (h\circ f) + q (h \circ g)$. This is by the following.
        \[
          \begin{tikzcd}
            A \arrow[r, "\lambda^{-1}"] & A \otimes I \arrow[r, "A\otimes {\llangle p, q\rrangle}"]
            & A\otimes (I + I) \arrow[r, "d"] & A\otimes I + A\otimes I \arrow[r, "\lambda + \lambda"]
            & A + A \arrow[r, "{\left[f, g\right]}"] \arrow[rr, bend right = 20, "{\left[h\circ f, h\circ g\right]}",swap]& B \arrow[r, "h"]& C
          \end{tikzcd}\qedhere
        \]
  \end{itemize}
  
\end{proof}

\begin{theorem}
    The category $\qtilde$ is enriched in convex spaces. Moreover, the Lambek embedding $\kappa : \q \hookrightarrow \qtilde$ preserves
    the convex sum in $\q$. 
\end{theorem}
\begin{proof}
    By Theorem~\ref{thm:convex2} (2), there exists a map $\llangle p, q\rrangle : I \to I+I$ in $\q$ for any $p, q\in [0,1]$, $p+q = 1$, and it satisfies the four diagrams. Since $\kappa$ preserves coproducts in $\q$, the map $\kappa \llangle p, q\rrangle : \kappa I \to \kappa I +' \kappa I$ in $\qtilde$ also satisfies the four diagrams
    in Theorem~\ref{thm:convex2} (2). Therefore $\qtilde$ is enriched in convex spaces.

    For all $f, g \in \q(A, B)$, the convex sum $pf + qg \in \q(A, B)$ is defined to be the following.
      \[ A \xrightarrow{\lambda^{-1}} A \otimes I \xrightarrow{A\otimes \llangle p, q \rrangle} A \otimes(I+I)
        \xrightarrow{d} A\otimes I + A\otimes I  \xrightarrow{\lambda + \lambda} A + A \xrightarrow{\left[f, g\right]} B\]
      Since $\kappa$ preserves coproducts in $\q$ and it is strong monoidal, we have $\kappa(pf + qg) = p \kappa(f) + q \kappa(g) \in \qtilde(\kappa A, \kappa B)$.
      
\end{proof}

\section{Proof of Theorem~\ref{thm:strong-monad-presheaves}}
\label{pf:strong-monad-presheaves}
In this section, we assume $\V$ to be a complete, cocomplete, symmetric monoidal closed category. 
The following proposition is due to Kock \cite{kock1972strong}.

\begin{proposition}
  \label{v-to-strength}
  Let $T : \V \to \V$ be a $\V$-monad. Then $T$ is a strong monad with
  strength $t : A\otimes TB\to T(A\otimes B)$ given by the following commutative diagram. Note that $\eta$
  is the unit of the adjunction $-\otimes A \dashv A\multimap -$. 
  \[
    \begin{tikzcd}
      A\arrow[d, "\eta"] \arrow[r, "\mathrm{curry}(t)"] & TB \multimap T(A \otimes B) \\
      B \multimap A\otimes B \arrow[ur, "T_{B, A\otimes B}", swap] &
    \end{tikzcd}
  \]
\end{proposition}

\begin{theorem}
  \label{enriched-triangle}
  Let $T$ be a strong monad on $\V$ and $F : \A^{\mathrm{op}} \to \V$ be a $\V$-functor.
  For all $A, B\in \A$, we have maps 

  \[F_{A B} : \A(B, A) \to FB \multimap FA\] and 
  \[(TF)_{A B} : \A(B, A) \to TFB \multimap TFA.\]
  We have the following commutative diagram.

  \[
    \begin{tikzcd}
      \A(B, A)\otimes TFA \arrow[drr, "{\mathrm{uncurry}((TF)_{A B})}"] \arrow[d, "t"]& &\\
      T(\A(B, A)\otimes FA) \arrow[rr, "{T\mathrm{uncurry}(F_{A B})}", swap] & & TFB
    \end{tikzcd}
  \]
\end{theorem}
\begin{proof}
  By currying the diagram above, we just need to show the right triangle commutes
  in the following diagram. 

    \[
      \begin{tikzcd}
        & \A(B, A)
        \arrow[ddl, bend right= 10, "F_{AB}", swap]
        \arrow[d, "\eta"]
        \arrow[ddr, bend left = 10, "(TF)_{AB}"]
        \arrow[drr, "\mathrm{curry}(t)"]
        & & \\
        & FA\multimap \A(B, A)\otimes FA
        \arrow[dl, "FA\multimap \mathrm{uncurry}(F_{AB})", near start]
        \arrow[rr, "T_{FA, \A(B,A)\otimes FA}", swap] & & TFA \multimap T(\A(B, A)\otimes FA)
        \arrow[dl, "TFA\multimap T\mathrm{uncurry}(F_{A B})"]\\
        FA\multimap FB \arrow[rr, "T_{FA, FB}", swap] & & TFA\multimap TFB & 
      \end{tikzcd}
    \]
  Note that the bottom square commutes because of the $\V$-naturality of $T$. The left triangle commutes by
  the property of monoidal closedness. The front triangle commutes by definition of $(TF)_{AB}$. The
  back triangle commutes by Proposition~\ref{v-to-strength}.
\end{proof}

\begin{theorem}\label{xi}
  Let $F : \A^{\mathrm{op}}\otimes \A \to \V$ be a $\V$-functor and let $T$ be a strong monad on $\V$. 
  Then there exists a natural map \[\xi_{F} : \int^{A\in \A}TF(A, A) \to T \int^{A\in \A}F(A, A).\]

\end{theorem}
\begin{proof}
  Recall that by definition of coend, we have the following coequalizers.

  \[
    \begin{array}{lll}
      \begin{tikzcd}
        \sum_{A, B \in \A} \A(B, A) \otimes F(A, B)
        \arrow[r,shift left=.75ex,"\rho_{1}"]
        \arrow[r,shift right=.75ex, "\rho_{2}", swap]
        &
        \sum_{A\in \A} F(A, A)  \arrow[r, "e"]     
        & \int^{A\in \A}F(A, A) 
      \end{tikzcd}
    \end{array}
  \]
  \[
    \begin{array}{lll}
      \begin{tikzcd}
        \sum_{A, B \in \A} \A(B, A) \otimes TF(A, B)
        \arrow[r,shift left=.75ex,"\rho_{1}'"]
        \arrow[r,shift right=.75ex, "\rho_{2}'", swap]
        &
        \sum_{A\in \A} TF(A, A)  \arrow[r, "e'"]     
        & \int^{A\in \A}TF(A, A) 
      \end{tikzcd}
    \end{array}
  \]
  For any $A\in \A$, the functor $F(A, -) : \A \to \V$ gives rise
  to a map
  \[ F(A, -)_{BA} : \A(B, A) \to F(A, B) \multimap F(A, A)\]
  for each $B\in \A$. The map $\rho_{1}$ is defined as the coproduct pairing $[\mathrm{inj}_{A} \circ \mathrm{uncurry}(F(A, -)_{BA})]_{A,B \in \A}$.
  For any $B\in \A$, the functor $F(-, B) : \A^{\mathrm{op}} \to \V$ gives rise
  to a map
  \[ F(-, B)_{AB} : \A(B, A) \to F(A, B) \multimap F(B, B)\]
  for each $A\in \A$. The map $\rho_{2}$ is defined as the coproduct pairing $[\mathrm{inj}_{B} \circ \mathrm{uncurry}(F(-, B)_{AB})]_{A,B \in \A}$.
  The maps $\rho_{1}', \rho_{2}'$ are induced similarly.
  
  Consider the following diagram. 
  
  \[
    \begin{array}{lll}
      \begin{tikzcd}
        \int^{A}TF(A, A) \arrow[rr, "\xi", dashed] & & T \int^{A}F(A, A) \\
        \sum_{A}TF(A, A)\arrow[rr, "{\left[T\mathrm{inj}_{A}\right]_{A}} "]
        \arrow[u, "e'"]
        & & T\sum_{A}F(A, A) \arrow[u, "Te"]\\
        \sum_{A,B}\A(B, A)\otimes TF(A, B)
        \arrow[r, "\sum_{A,B} t"]
        \arrow[u,shift left=.75ex,"\rho_{1}'"]
        \arrow[u,shift right=.75ex, "\rho_{2}'", swap]
        & \sum_{A, B}T(\A(B, A)\otimes F(A, B))
        \arrow[r, "{\left[T\mathrm{inj}_{A, B}\right]_{A, B}}", shift left=.75ex]
        & T\sum_{A,B}\A(B, A)\otimes F(A, B)
        \arrow[u,shift left=.75ex,"T\rho_{1}"]
        \arrow[u,shift right=.75ex, "T\rho_{2}", swap]
      \end{tikzcd}
    \end{array}
  \]
  Note that $[T\mathrm{inj}_{A}]_{A}$ and $[T\mathrm{inj}_{A,B}]_{A,B}$ are coproduct pairings. The morphism
  $t : \A(B, A)\otimes TF(A, B) \to T(\A(B, A)\otimes F(A, B))$ is the strength map for $T$. 
  
  To show the existence of $\xi$, we just need to show $Te \circ \left[T\mathrm{inj}_{A} \right]_{A} \circ \rho_{1}' = Te \circ \left[T\mathrm{inj}_{A}\right]_{A} \circ \rho_{2}'$, which is to show the bottom square commutes for $\rho_{1}'$ and $T\rho_{1}$ ($\rho_{2}'$ and $T\rho_{2}$). This is the case
  because of the following commutative diagram.
  Note that the left triangle commutes by Theorem~\ref{enriched-triangle}.
  \[
    \begin{tikzcd}
      \A(B, A) \otimes TF(A, B)
      \arrow[r, "{\rho_{1}'(A,B)}"]
      \arrow[d, "t"]
      & TF(A, A) \arrow[r, "T\mathrm{inj}_{A}"]
      &  T \sum_{A}F(A, A) \\
      T(\A(B, A) \otimes F(A, B))
      \arrow[rr, "T\mathrm{inj}_{A,B}", swap]
      \arrow[ur, "{T \rho_{1}(A,B)}"]
      &
      &
      T \sum_{A,B}\A(B, A) \otimes F(A, B)
      \arrow[u, "{T \rho_{1}}", swap]
    \end{tikzcd}
  \]
  Note that $\rho_{1}'(A,B)$ is a component of $\rho_{1}'$ and $\rho_{1}(A,B)$ is a component of $\rho_{1}$. 
  By the universal property of the coequalizer $e'$,
  there exists a unique arrow
  \[\xi : \int^{A\in \A}TF(A, A) \to T \int^{A\in \A}F(A, A).
  \qedhere
  \]
\end{proof}

\begin{proposition}
  \label{yoneda-as-diagram}
  Suppose $F : \A^{\mathrm{op}} \to \V$. For all $B, C\in \A$, the following diagram commutes.

  \[
    \begin{tikzcd}
      \A(C, B) \otimes F(B)\arrow[d, "{\mathsf{uncurry}(F_{BC})}",swap] \arrow[r, "e_{B}"] & \int^{B} \A(C, B) \otimes F(B) \arrow[dl, "y"]\\
      F(C)
    \end{tikzcd}
  \]
  Note that $y$ is an isomorphism expressing the Yoneda lemma in the language of coends, and $F_{BC} : \A(C,B) \to F(B) \multimap F(C)$,
  and $e_{B}$ is the unit of the coend. 
\end{proposition}

\begin{proof}[Proof sketch]
  Note that the map $\mathsf{uncurry}(F_{BC}): \A(C, B) \otimes F(B) \to F(C)$ is $\V$-natural
  in $B$. By the universal property of coends, there exists a map $y : \int^{B} \A(C, B) \otimes F(B) \to F(C)$
  such that the diagram above commutes. Moreover, $y$ is an isomorphism \cite[Chapter 2.4]{kelly1982basic}.
\end{proof}

\begin{theorem}
  \label{thm:xi}
  Let $T$ be a strong monad on $\V$. For all $F : \A^{\mathrm{op}}\otimes \A \to \V$, the map $\xi_{F} : \int^{A\in \A}TF(A, A) \to T \int^{A\in \A}F(A, A)$
 makes the following diagrams commute.
  \begin{enumerate}
  \item
    \[
      \begin{tikzcd}
        \int^{A}F(A, A) \arrow[d, "\eta"]
        \arrow[dr, "\int \eta"]
        & \\
        T\int^{A}F(A, A)
        & \int^{A}TF(A, A)
        \arrow[l, "\xi"]
      \end{tikzcd}
    \]

  \item
    \[
      \begin{tikzcd}
        \int^{A}TTF(A, A) 
        \arrow[r, "\xi"]
        \arrow[d, "\int \mu"]
        & T\int^{A}TF(A, A)
        \arrow[r, "T\xi"]
        &
        TT\int^{A}F(A, A) \arrow[d, "\mu"]\\
        \int^{A}TF(A, A)
        \arrow[rr, "\xi"]
        &
        & T\int^{A}F(A, A)
      \end{tikzcd}
    \]

  \item Suppose $G : \A^{\mathrm{op}} \to \V$ and $A \in \A$.

    \[
      \begin{tikzcd}
        \int^{B}\A(A, B) \otimes TG B \arrow[d, "\int t"]
        \arrow[r, "y'"]
        & TG A \\
        \int^{B}T(\A(A, B) \otimes GB)
        \arrow[r, "\xi"]
        & T\int^{B}\A(A, B) \otimes GB \arrow[u, "Ty"]
      \end{tikzcd}
    \]
    Note that $y', y$ are isomorphisms induced by the Yoneda lemma.
    
  \item Suppose $F : \A^{\mathrm{op}}\otimes \A  \to \V$
    and $X \in \V$. 
    \[
      \begin{tikzcd}
        (\int^{A}F(A,A)) \otimes T X
        \arrow[r, "t"]
        \arrow[d, "\cong"]
        & T((\int^{A}F(A, A)) \otimes X)
        \arrow[dd, "\cong"]\\
        \int^{A}(F(A,A) \otimes T X )
        \arrow[d, "\int t"]
         & \\
        \int^{A}T(F(A,A) \otimes X) \arrow[r, "\xi"]
        &T\int^{A}(F(A,A) \otimes X) 
      \end{tikzcd}
    \]

  \item Suppose $F : \A^{\mathrm{op}}\otimes \A \to \V$ and $X \in \V$. 
    
    \[
      \begin{tikzcd}
        \int^{A}(X \otimes TF(A,A))
        \arrow[r, "\int t"]
        \arrow[d, "\cong"]
        & \int^{A}T(X \otimes F(A, A))
        \arrow[d, "\xi"]\\
        X \otimes \int^{A}TF(A,A)
        \arrow[d, "X \otimes \xi"]
        & T \int^{A}(X \otimes F(A,A)) \arrow[d, "\cong"]\\
        X \otimes T\int^{A}F(A,A)  \arrow[r, "t"]
        &T(X \otimes \int^{A}F(A,A)) 
      \end{tikzcd}
    \]

  \item Suppose $F : \A^{\mathrm{op}}\otimes \A^{\mathrm{op}} \otimes \A \otimes \A \to \V$. 
    \[
      \begin{tikzcd}
        \int^{A}\int^{B}TF(A,B,A,B) \arrow[r, "\int^{A}\xi"] \arrow[d, "\cong"]& \int^{A} T \int^{B}F(A,B,A,B) \arrow[r, "\xi"] 
        & T \int^{A}\int^{B}F(A,B,A,B)  \arrow[d, "\cong"]\\
        \int^{A, B}TF(A,B,A,B) \arrow[rr, "\xi"]& & T\int^{A, B}F(A,B,A,B) 
      \end{tikzcd}
    \]
  \end{enumerate}  
\end{theorem}
\begin{proof}

  \begin{enumerate}
  \item We need to show that the following commutes. 
    \[
      \begin{tikzcd}
        \int^{A}F(A, A) \arrow[d, "\eta"]
        \arrow[dr, "\int \eta"]
        & \\
        T\int^{A}F(A, A)
        & \int^{A}TF(A, A)
        \arrow[l, "\xi"]
      \end{tikzcd}
    \]
    Consider the following diagram. We write $\eta_{1}$ for the map $F(A, A) \to TF(A, A)$
    and $\eta_{2}$ for the map $\int^{A}F(A, A) \to T\int^{A}F(A, A)$. 
    \[
      \begin{tikzcd}
        \int^{A}F(A, A) \arrow[r, "\int \eta_{1}"]
        \arrow[drr, "\eta_{2}", near start, swap]
        &
        \int^{A}TF(A, A) \arrow[dr, "\xi"] & \\
        & & T\int^{A}F(A, A) \\
        \sum_{A}F(A, A)
        \arrow[uu, "e", two heads]
        \arrow[r,"\sum \eta_1"]
        \arrow[drr, "\eta"]
        &
        \sum_{A}TF(A, A)
        \arrow[uu, "e'"]
        \arrow[dr, "{[T\mathrm{inj}_{A}]_{A}}"]
        & \\
        & & T \sum_{A}F(A, A)
        \arrow[uu, "Te"]
      \end{tikzcd}
    \]
    We need to show that the top triangle commutes.
    Since $e$ is an epimorphism, we just need to show
    $ \xi \circ \int \eta_{1}\circ e = \eta_{2}\circ e$.
    This
    is the case because the bottom triangle commutes and
    all three square faces commute. The bottom triangle
    commutes by the universal property of coproducts. The square with $\xi$
    commutes by definition of $\xi$. 
    Also note that $e' \circ \sum \eta_{1} = \int \eta_{1} \circ e$ is
    a property of coends (see \cite[4.2]{kelly1982basic}). 

  \item The proof is similar to (1).    
  \item Next we need to show that the following commutes (where $y, y'$ are isomorphisms induced by the Yoneda lemma).
    \[
      \begin{tikzcd}
        \int^{B}\A(A, B) \otimes TG B \arrow[d, "y'"]
        \arrow[r, "\int t"]
        & \int^{B}T(\A(A, B) \otimes GB)
        \arrow[d, "\xi"]
        \\
        TGA
        & T\int^{B}\A(A, B) \otimes GB \arrow[l, "Ty"]
      \end{tikzcd}
    \]
    The above diagram commutes because the following diagram commutes for all $A, B\in \A$.
    \[\footnotesize
      \begin{tikzcd}
        &
        \A(A, B) \otimes TGB
        \arrow[rr, two heads, "e_{1}"]
        \arrow[dl, "{\mathrm{uncurry}((TG)_{BA})}", swap]
        \arrow[d, "t"]
        &
        &
        \int^{B}\A(A, B) \otimes TGB
        \arrow[dl, "y'"]
        \arrow[d, "\int t"]
        \\
        TG A\arrow[rr, bend right=10, dashed, "\id", swap]
        &
        T(\A(A, B) \otimes GB)
        \arrow[dl, "\id"]
        \arrow[rr, "e_{2}", near start]
        &
        TGA
        &
        \int^{B} T(\A(A, B) \otimes GB)
        \arrow[dl, "\xi"]
        \\
        T(\A(A, B) \otimes G(B))
        \arrow[u, "{T\mathrm{uncurry}(G_{BA})}"]
        \arrow[rr, "Te_{3}"]
        &
        &
        \arrow[u, "Ty"]
        T\int^{B}\A(A, B) \otimes GB
        &
      \end{tikzcd}
    \]
    Since $G$ and $TG$ are contravariant $\V$-functors, there are the following maps in $\V$.
    \[G_{BA} : \A(A, B) \to GB \Rightarrow GA\]
    \[(TG)_{BA} : \A(A, B) \to TGB \Rightarrow TGA\]
    The bottom square commutes by the definition of $\xi$, and the back square (with $e_{1}, e_{2}$) commutes by
    naturality of coends. The top and the front squares commutes because of Proposition~\ref{yoneda-as-diagram}.
    Thus we just need to show that the left square commutes, i.e.,
    \[
      \begin{tikzcd}
        \mathbf{C}(C, B)\otimes TFB\arrow[r, "t"] \arrow[d, "u'"] & T(\mathbf{C}(C, B)\otimes FB) \arrow[dl, "Tu"]\\
        TFC
      \end{tikzcd}.
    \]
    This commutes by Proposition~\ref{enriched-triangle}. 
    
  \item Next we need to prove that the following commutes.
    \[
      \begin{tikzcd}
        (\int^{A}F(A,A)) \otimes TX
        \arrow[r, "t"]
        \arrow[d, "\cong"]
        & T((\int^{A}F(A, A)) \otimes X)
        \arrow[dd, "\cong"]\\
        \int^{A}(F(A,A) \otimes TX)
        \arrow[d, "\int t"]
         & \\
        \int^{A}T(F(A,A) \otimes X) \arrow[r, "\xi"]
        &T\int^{A}(F(A,A) \otimes X)
      \end{tikzcd}
    \]
    First observe that the following commutes (each arrow is canonical). 
    \[
      \begin{tikzcd}
        (\sum_{A}F(A,A)) \otimes TX
        \arrow[r]
        \arrow[d, "\cong"]
        & T((\sum_{A}F(A, A)) \otimes X)
        \arrow[dd, "\cong"]\\
        \sum_{A}(F(A,A) \otimes TX)
        \arrow[d]
         & \\
        \sum_{A}T(F(A,A) \otimes X) \arrow[r]
        &T\sum_{A}(F(A,A) \otimes X)
      \end{tikzcd}
    \]
    Now let us consider the following cube.
    \[ \footnotesize
      \begin{tikzcd}
        & (\int^{A}F(A,A)) \otimes TX
        \arrow[dd, "\cong", near start]
        \arrow[rr, "t"]& & T((\int^{A}F(A,A))\otimes X)\arrow[dddd, "\cong"] \\
        (\sum_{A}F(A,A)) \otimes TX
        \arrow[rr, "t", near start]
        \arrow[ur, two heads]
        \arrow[dd, "\cong"]& & T((\sum_{A}F(A,A)) \otimes X) \arrow[ur, "T(e \otimes X)"] \arrow[dddd, "\cong"]&  \\
        & \int^{A}(F(A,A)\otimes TX) \arrow[dd, "\int t"] & & \\
        \sum_{A}(F(A, A)\otimes TX)\arrow[ur, two heads]
        \arrow[dd, "\sum_{A}t"]
        & & & \\
        & \int^{A}T(F(A,A)\otimes X) \arrow[rr, "\xi", near start]& & T\int^{A}(F(A,A)\otimes X)\\
        \sum_{A}T(F(A, A)\otimes X)\arrow[ur, two heads] \arrow[rr, "{[T\mathrm{inj}_{A}]_{A}}"]& & T\sum_{A}(F(A, A)\otimes X)
        \arrow[ur, "T e'"]&
      \end{tikzcd}
    \]
    Note that the top square commutes, by naturality of $t$. The bottom square commutes, by
    definition of $\xi$. The left square involving $\sum_{A}t, \int t$ commutes by naturality of coend.
    The right square and the left top square commute for the same reason. For simplicity, consider the following
    diagram.
    \[
      \begin{tikzcd}
        (\sum_{A,B} \A(B, A)\otimes F(A, B))\otimes X
        \arrow[r,shift left=.75ex]
        \arrow[r,shift right=.75ex]
        \arrow[d, "\cong"]
        & (\sum_{A}F(A, A))\otimes X \arrow[r, two heads] \arrow[d, "\cong"]& (\int^{A}F(A, A))\otimes X \arrow[d, "\cong"]\\
        \sum_{A,B} (\A(B, A)\otimes F(A, B)\otimes X)
        \arrow[r,shift left=.75ex]
        \arrow[r,shift right=.75ex]
        & \sum_{A}(F(A, A)\otimes X) \arrow[r, two heads] & \int^{A}(F(A, A)\otimes X)
      \end{tikzcd}
    \]
    Note that the right square is the same square as the right square in the cube.
    And $-\otimes X$ preserves coequalizers. 
    The left square commutes by naturality. This implies that the right square commutes, by the universal
    property of coequalizers. Therefore the cube above commutes.

  \item Next, we need to show that the following diagram commutes.
    \[
      \begin{tikzcd}
        \int^{A}(X \otimes TF(A,A))
        \arrow[r, "\int t"]
        \arrow[d, "\cong"]
        & \int^{A}T(X \otimes F(A, A))
        \arrow[d, "\xi"]\\
        X \otimes \int^{A}TF(A,A)
        \arrow[d, "\id \otimes \xi"]
        & T \int^{A}(X \otimes F(A,A)) \arrow[d, "\cong"]\\
        X \otimes T\int^{A}F(A,A)  \arrow[r, "t"]
        &T(X \otimes \int^{A}F(A,A)) 
      \end{tikzcd}
    \]
    First, observe that the following diagram commutes.
    \[
      \begin{tikzcd}
        \sum_{A}(X \otimes TF(A,A))
        \arrow[r]
        \arrow[d, "\cong"]
        & \sum_{A}T(X \otimes F(A, A))
        \arrow[d]\\
        X \otimes \sum_{A}TF(A,A)
        \arrow[d]
        & T \sum_{A}(X \otimes F(A,A)) \arrow[d, "\cong"]\\
        X \otimes T\sum_{A}F(A,A)  \arrow[r]
        &T(X \otimes \sum_{A}F(A,A)) 
      \end{tikzcd}
    \]
    Now let us consider the following cube.
    \[ \footnotesize
      \begin{tikzcd}
        & \int^{A}(X  \otimes T F(A,A))
        \arrow[dd, "\cong", near start]
        \arrow[rr, "\int t"]& & \int^{A}T(X \otimes F(A,A))\arrow[dd, "\xi"] \\
        \sum_{A}(X \otimes TF(A,A))
        \arrow[rr, "\sum t", near start]
        \arrow[ur, two heads]
        \arrow[dd, "\cong"]& & \sum_{A}T(X \otimes F(A,A) ) \arrow[ur] \arrow[dd]&  \\
        & X \otimes \int^{A}TF(A,A) \arrow[dd, "X \otimes \xi"]& & T\int^{A}(X\otimes F(A,A))\arrow[dd, "\cong"] \\
        X \otimes \sum_{A}TF(A, A)\arrow[ur]
        \arrow[dd]
        & & T\sum_{A}(X\otimes F(A,A))\arrow[ur] \arrow[dd, "\cong", near start] & \\
        & X \otimes T\int^{A}F(A,A) \arrow[rr, "t", near start]& & T (X \otimes \int^{A}(F(A,A) \\
        X \otimes T \sum_{A}F(A, A)\arrow[ur, two heads] \arrow[rr, "t"]& & T(X \otimes \sum_{A}F(A, A))
        \arrow[ur, "T e'"] &
      \end{tikzcd}
    \]
    Note that the top square commutes by naturality of coends. The bottom square commutes by naturality of $t$.
    The left bottom and the right top square involving $\xi$ commute by definition. The left top and the right bottom
    square commute for the same reason. Consider the following diagram. 
    \[
      \begin{tikzcd}
        \sum_{A,B} (\A(B, A) \otimes X\otimes F(A, B))
        \arrow[r,shift left=.75ex]
        \arrow[r,shift right=.75ex]
        \arrow[d, "\cong"]
        & \sum_{A}(X \otimes F(A, A)) \arrow[r, two heads] \arrow[d, "\cong"]& \int^{A}(X \otimes F(A, A)) \arrow[d, "\cong"]\\
        X \otimes \sum_{A,B} (\A(B, A)\otimes F(A, B))
        \arrow[r,shift left=.75ex]
        \arrow[r,shift right=.75ex]
        & X \otimes \sum_{A}F(A, A)\arrow[r, two heads] & X \otimes \int^{A}F(A, A) 
      \end{tikzcd}
    \]
    Note that the right square is the same square as the right bottom square in the cube under the functor $T$.
    And $X\otimes -$ preserves coequalizers. 
    The left square commutes by naturality. This implies that the right square commutes, by the universal
    property of coequalizers. Therefore the cube above commutes.
    
  \item Let $F : \A^{\mathrm{op}}\otimes \A^{\mathrm{op}} \otimes \A \otimes \A \to \V$.
    We now need to show that the following diagram commutes. 
    \[
      \begin{tikzcd}
        \int^{A}\int^{B}TF(A,B,A,B) \arrow[r, "\int^{A}\xi"] \arrow[d, "f"]& \int^{A} T \int^{B}F(A,B,A,B) \arrow[r, "\xi"] 
        & T \int^{A}\int^{B}F(A,B,A,B)  \arrow[d, "Tf"]\\
        \int^{A, B}TF(A,B,A,B) \arrow[rr, "\xi"]& & T\int^{A, B}F(A,B,A,B) 
      \end{tikzcd}
    \]
    First, the isomorphisms $f$, $Tf$ above are instances of so-called Fubini theorem for coends, which
    also gives rise to the following commutative diagram for any $F : \A^{\mathrm{op}}\otimes \A^{\mathrm{op}} \otimes \A \otimes \A \to \V$.
    \[
      \begin{tikzcd}
        F(A,B,A,B)\arrow[r, two heads, "e_{A,B}"] \arrow[d, two heads, "e_{B}"]& \int^{A,B}F(A,B,A,B)\\
        \int^{B}F(A,B,A,B)\arrow[r, two heads, "e_{A}"] & \int^{A}\int^{B}F(A,B,A,B) \arrow[u, "f"]
      \end{tikzcd}
    \]
    We have the following commutative diagram.
    \[ \footnotesize
      \begin{tikzcd}
        &\int^{A}\int^{B}TF(A,B,A,B) \arrow[dl, "f",swap]\arrow[r, "\int^{A}\xi"]&\int^{A}T\int^{B}F(A,B,A,B) \arrow[r, "\xi"] & T \int^{A}\int^{B}F(A,B,A,B)\arrow[d, "Tf"] \\
        \int^{A,B}TF(A,B,A,B) \arrow[rrr, bend right = 25, "\xi"]
        & \int^{B}TF(A,B, A,B) \arrow[r, "\xi"]
        \arrow[u, "e'_{A}"]
        & T \int^{B}F(A,B, A,B) \arrow[u, "e''_{A}"] \arrow[ur, "T e_{A}"] & T\int^{A,B}F(A,B, A,B)\\
        TF(A,B,A,B) \arrow[u, "e'_{A,B}"] \arrow[ur, "e'_{B}"] \arrow[urr, "Te_{B}"] \arrow[urrr, "Te_{A,B}", swap] & & &
      \end{tikzcd}
    \]
    Note that above diagram commutes, by properties of the
    Fubini theorem, definition of $\xi$, $\V$-naturality of
    coends, and because $e'_{A, B}$ is an epimorphism.
  \end{enumerate}  
\end{proof}

\begin{theorem}
  Let $\A$ be a $\V$-category. If $T$ is a commutative strong monad on $\V$ (the strength is given by the map $t_{A,B} : A \otimes TB\to T(A\otimes B)$ for any $A, B\in \A$), then $\overline{T}(F) = T\circ F$ is a
  commutative strong $\V$-monad on $\V^{\A^{\mathrm{op}}}$.
\end{theorem}

\begin{proof}
  It is straightforward to verify that $\overline{T}$ is a monad. 
  We define the strength $\overline{t}$ to be the following composition.
  \[ (F \otimes_{\mathrm{Day}} \overline{T}G)(C) = \int^{(A, B) \in \A\otimes \A } \A(C, A\otimes B)\otimes F A \otimes T G B \]
  \[\stackrel{\int^{(A, B)} t }{\to} \int^{(A, B) \in \A\otimes \A } T(\A(C, A\otimes B) \otimes F A \otimes G B)\]
  \[\stackrel{\xi}{\to} T\int^{(A, B) \in \A\otimes \A }(\A(C, A\otimes B) \otimes F A \otimes G B)\]
  \[= \overline{T}(F \otimes_{\mathrm{Day}} G)(C)\]
  Now to show that $\overline{T}$ is a commutative strong monad, we need to show the following diagrams commute.
  \begin{itemize}
  \item
  \[
    \begin{tikzcd}
      F \otimes_{\mathrm{Day}} G \arrow[r, "F \otimes_{\mathrm{Day}} \eta" ]
      \arrow[d, "\eta"]
      & F \otimes_{\mathrm{Day}} \overline{T} G \arrow[dl, "\overline{t}"] \\
      \overline{T}(F\otimes_{\mathrm{Day}} G)
      \end{tikzcd}
  \]    
  To show this, we just need to show that the following diagram commutes for any $C \in \A$.
    \[
    \begin{tikzcd}
      \int^{A, B}\A(C, A \otimes B)\otimes F A  \otimes G B
      \arrow[r, "\int^{A, B} \eta'" ]
      \arrow[d, "\eta"]
      \arrow[dr, "\int^{A, B} \eta"]
      &\int^{A, B}\A(C, A \otimes B)\otimes F A  \otimes T G B \arrow[d, "\int^{A, B} t"]\\
      T(\int^{A, B}\A(C, A \otimes B)\otimes F A  \otimes G B)
      & \int^{A, B}T(\A(C, A \otimes B)\otimes F A  \otimes G B) \arrow[l, "\xi"]
    \end{tikzcd}
    \]    
    Note that $\int^{A, B} \eta'$ is a shorthand for $\int^{A, B}{\A(C, A \otimes B)\otimes F A} \otimes \eta$. Similarly,
    $\int^{A, B} t$ is a shorthand for
    $\int^{A, B} t_{\A(C, A \otimes B)\otimes FA ,  GB}$ in the above diagram. The bottom triangle commutes because of Theorem~\ref{thm:xi}(1). The top triangle commutes by properties of $t$.

  \item We need to show that the following diagram commutes.
  \[
    \begin{tikzcd}
      I \otimes_{\mathrm{Day}} \overline{T}F
      \arrow[r, "\overline{t}_{I, F}"]
      \arrow[dr, "\lambda_{\overline{T}F}", swap]
      & \overline{T}(I \otimes_{\mathrm{Day}} F)
      \arrow[d, "\overline{T}\lambda_{F}"]
      \\
      & \overline{T}F
    \end{tikzcd}
  \]
  If we unfold the definition of the Day tensor, we
  have the following diagram for any $C \in \A$.
  {\footnotesize
  \[
    \begin{tikzcd}
      & \int^{B} \A(C, I\otimes B) \otimes TFB
      \arrow[r, "\int t"]
      \arrow[dl, "\int \lambda"]
      & \int^{B} T(\A(C, I\otimes B) \otimes FB)
      \arrow[dl, "\int T\lambda"]
      \arrow[dr, "\xi"]
      & \\
      \int^{B} \A(C, B) \otimes TFB
      \arrow[r, "\int t' "]
      \arrow[dr, "\cong"]
      & \int^{B} T(\A(C, B) \otimes FB)
      \arrow[dr, "\xi"] &  &
      T \int^{B} \A(C, I\otimes B) \otimes FB
      \arrow[dl, "T\int \lambda"]\\
      & TFC \arrow[r, "\cong"] & T\int^{B}\A(C, B) \otimes FB & 
    \end{tikzcd}
  \]}
  Note that for any $C\in \A$, the following
  commutes by naturality of $t$. 
  \[
    \begin{tikzcd}
      \int^{B}\A(C,I\otimes B) \otimes {T}FB
      \arrow[r, "\int t"]
      \arrow[d, "\int \lambda_{B}", swap]
      & \int^{B}T(\A(C, I\otimes B) \otimes FB)
      \arrow[d, "\int T\lambda_{B}"]
      \\
      \int^{B}\A(C, B) \otimes TFB
      \arrow[r, "\int t'"]& \int^{B}T(\A(C, B) \otimes FB)
    \end{tikzcd}
  \]
  Note that $I = \A(-, I) \in \V^{\A^{\mathrm{op}}}$,
  and $\int t$ is a shorthand for $\int^{B} t_{\A(C,I\otimes B), FB}$,
  and $\int \lambda_{B}$ is a shorthand for
  $\int^{B}\A(C, \lambda_{B}) \otimes \id_{TFB}$,
  and $\int T \lambda_{B}$ is a shorthand for
  $\int^{B}T\A(C, \lambda_{B}) \otimes \id_{FB}$,
  and $\int t'$ is a shorthand for $\int^{B} t_{\A(C, B), FB}$. 

  The bottom square commutes because of Theorem~\ref{thm:xi}(3).
  The right square commutes by the naturality of $\xi$.

  \item Next we need to show that the following diagram commutes.
    \[
      \begin{tikzcd}
        F \otimes_{\mathrm{Day}} \overline{T} \overline{T} G
        \arrow[r, "t"]
        \arrow[d, "\id_{F}\otimes_{\mathrm{Day}} \mu"]
        & \overline{T}(F \otimes_{\mathrm{Day}} \overline{T} G)
        \arrow[r, "\overline{T}t"]
        &
        \overline{T} \overline{T}(F \otimes_{\mathrm{Day}} G)
        \arrow[d, "\mu"]
        \\
        F \otimes_{\mathrm{Day}} \overline{T} G
        \arrow[rr, "t"]
        & & \overline{T}(F \otimes_{\mathrm{Day}} G) 
      \end{tikzcd}
    \]    

    The above diagram commutes because for any $C \in \A$, we have the following commutative diagram. 

    {\footnotesize
    \[
      \begin{tikzcd}
        \int^{A, B}\A(C, A \otimes B)\otimes FA \otimes TTGB
        \arrow[r, "\int t"]
        \arrow[d, "\int \id \otimes \mu"]
        & \int^{A, B}T(\A(C, A \otimes B)\otimes FA \otimes TGB)
        \arrow[r, "\xi"]
        \arrow[d, "\int T t"]
        &
        T\int^{A, B}(\A(C, A \otimes B)\otimes FA \otimes TGB)
        \arrow[d, "T\int t"]
        \\
        \int^{A, B}\A(C, A \otimes B)\otimes FA \otimes TGB
        \arrow[d, "\int t"]
        & \int^{A, B}TT(\A(C, A \otimes B)\otimes FA \otimes GB)
        \arrow[r, "\xi"]
        \arrow[dl, "\int \mu", swap]
        & T\int^{A, B}T(\A(C, A \otimes B)\otimes FA \otimes GB) \arrow[d, "T \xi"]\\
        \int^{A, B}T(\A(C, A \otimes B)\otimes FA \otimes GB) \arrow[r, "\xi"]
        &
        T\int^{A, B}(\A(C, A \otimes B)\otimes FA \otimes GB) 
        &
        TT\int^{A, B}(\A(C, A \otimes B)\otimes FA \otimes GB) \arrow[l, "\mu"]
      \end{tikzcd}
    \]    
  }
 
  Note that the top right square commutes by naturality of $\xi$,
  the bottom diagram commutes by Theorem~\ref{thm:xi} (2), and
  the left diagram commutes by properties of $t$.
\item Next we need to show that the following diagram commutes.
    \[
      \begin{tikzcd}
        (F\otimes_{\mathrm{Day}} G) \otimes_{\mathrm{Day}} \overline{T}H
        \arrow[rr,"t"]
        \arrow[d, "\alpha"]
        & & \overline{T}((F\otimes_{\mathrm{Day}} G) \otimes_{\mathrm{Day}} H) \arrow[d, "\overline{T}\alpha"]\\
        F\otimes_{\mathrm{Day}} (G\otimes_{\mathrm{Day}} \overline{T}H)
        \arrow[r, "\id_{F}\otimes_{\mathrm{Day}} t"]
        &
        F\otimes_{\mathrm{Day}} \overline{T}(G\otimes_{\mathrm{Day}} H)
        \arrow[r, "t"]
        &
        \overline{T}(F\otimes_{\mathrm{Day}} (G\otimes_{\mathrm{Day}} H))
      \end{tikzcd}
    \]    
    For any $C\in \A$, we have 
    \[ ((F\otimes_{\mathrm{Day}} G)\otimes_{\mathrm{Day}}\overline{T}H) (C) 
    {\cong} \int^{ B \in \A }\int^{(X, Y) \in \A\otimes \A }  \A(C, (X\otimes Y) \otimes B) \otimes F X \otimes G Y \otimes T H B\]

    and 
    \[ (F \otimes_{\mathrm{Day}} (G \otimes_{\mathrm{Day}} \overline{T}H))(C) 
    {\cong} \int^{X \in \A } \int^{(Y, B) \in \A\otimes \A } \A(C, X\otimes (Y \otimes B)) \otimes FX \otimes G Y \otimes THB\]
  \[\cong \int^{X, Y, B}  FX \otimes G Y \otimes THB \otimes \int^{A} \A(A, Y\otimes B) \otimes \A(C, X\otimes A).\]

  Consider the following diagram. We need to show that the outermost diagram commutes. Note that $\int y, T\int y, a, a'$ are
  all isomorphisms.
  \smallskip
   
  \includestandalone[width=0.9\textwidth]{assoc0}

  Note that the top square and the top right square commute by naturality of $t$ and $\xi$.  
  We just need to show that the bottom diagram commutes.
  The expanded bottom diagram is the following. 
  \smallskip
  
  \includestandalone[width=0.9\textwidth]{assoc}

  Our goal is to show that the outermost diagram commutes.
  Note that all the inner diagrams commute,
  by Theorem~\ref{xi}(4)--(6) and naturality.
  Therefore the whole diagram commutes.
  
\item Lastly, since $\V^{\A^{\mathrm{op}}}$ is a symmetric monoidal $\V$-category with $\gamma_{F, G} : F\otimes_{\mathrm{Day}}G \to G\otimes_{\mathrm{Day}}F$,
  we can define the costrength as the following for any $F, B\in \V^{\A^{\mathrm{op}}}$. 
  \[\sigma_{F, G} := \overline{T} \gamma_{G, F} \circ t_{G, F}\circ \gamma_{\overline{T}F, G} : \overline{T}F\otimes_{\mathrm{Day}} G \to \overline{T}(F\otimes_{\mathrm{Day}}G)\]
  We need to show that the following diagram commutes.
  \[
    \begin{tikzcd}
      & \overline{T} F \otimes_{\mathrm{Day}} \overline{T} G
      \arrow[dl, "\sigma"]
      \arrow[dr, "t"]
      & \\
      \overline{T} (F \otimes_{\mathrm{Day}} \overline{T} G)
      \arrow[d, "\overline{T}t"]
      & & \overline{T} (\overline{T}F \otimes_{\mathrm{Day}}  G)
      \arrow[d, "\overline{T}\sigma"]
      \\
      \overline{T}\overline{T} (F \otimes_{\mathrm{Day}} G)
      \arrow[dr, "\mu"]
      & & \overline{T} \overline{T}(F \otimes_{\mathrm{Day}} G)
      \arrow[dl, "\mu"]\\
      & \overline{T}(F \otimes_{\mathrm{Day}} G)&
    \end{tikzcd}
  \]
  For any $C\in \A$, the above diagram can
  be expanded to the following diagram. 
  We need to show the outermost diagram commutes.

  \includestandalone[width=0.9\textwidth]{commute}

 It commutes because every inner diagram commutes (by naturality and Theorem~\ref{xi}(2)). \qedhere
  \end{itemize}
\end{proof}

\section{Proof of Theorem~\ref{thm:reflective-sub}}
\label{app:reflective-sub}
  We write $\qtilde$ for the full subcategory of $\set^{\q^{\mathrm{op}}}$ consisting of product-preserving functors.  
  We write $L : \set^{\q^{\mathrm{op}}} \to \qtilde $ for the left
  adjoint of the inclusion functor $i : \qtilde \hookrightarrow \set^{\q^{\mathrm{op}}}$. We write $\eta : \id \to iL$ for the unit of the adjunction. 
\begin{definition}
  We define a function $\tilde{L} : \overline{\mathbf{C}} \to \widetilde{\mathbf{C}}$ as follows.
  \begin{itemize}
  \item For any $F \in \overline{\mathbf{C}}$, we define $\tilde{L}(F) = G$ such that for all $A \in \mathbf{C}$,
    \[G(A) = ((iLF^{0})A, (FA)_{1}, (\eta_{F^{0}})_{A} \circ h_{A} : (FA)_{1} \to (iLF^{0})A) \in \V,\] where $h_{A} : (FA)_{1}\to (FA)_{0}$ and $(\eta_{F^{0}})_{A} : F^{0}A \to (iLF^{0})A$. 

    For any $A, B\in \C$, we define
    $G^{0}_{AB}$ and $G^{1}_{AB}$ by the following.
    \[
      \begin{tikzcd}
        f \in \m^{\mathrm{op}}(A, B) \arrow[r, mapsto, "G^{1}_{AB}"] \arrow[mapsto, d, "J_{A B}"] &  (F^{1}_{AB}f, (iLF^{0})(J_{AB} f)) \in \V(GA, GB) \arrow[d, mapsto, "p_{0}"]\\
        J_{AB} f \in \q^{\mathrm{op}}(A, B) \arrow[r, mapsto, "G^{0}_{AB}"] & (iLF^{0})(J_{AB} f)\in \set((iLF^{0})A, (iLF^{0})B)
      \end{tikzcd}
    \]

    Note that $G$ is smooth, hence $G \in \widetilde{\mathbf{C}}$.
  \end{itemize}
\end{definition}

\begin{proposition}
$\tilde{L} : \overline{\mathbf{C}} \to \widetilde{\mathbf{C}}$ is a $\V$-functor.
\end{proposition}
\begin{proof}
  We just need to show that for all $F, G\in \cbar$, there is a morphism
  \[\tilde{L}_{FG} : \overline{\mathbf{C}}(F, G) \to  \widetilde{\mathbf{C}}(\tilde{L}F, \tilde{L}G)\]
  in $\V$. This is provided by the following commuting square in $\set$.
  {\footnotesize
  \[
  \begin{tikzcd}
    \{(\alpha_A : FA \to GA)_{A\in \mathbf{C}}\ | \ \alpha \in \operatorname{\V-\mathrm{Nat}}(F, G)\} \arrow[r, "\tilde{L}_{FG}^{1}"] \arrow[d] &  \{(\beta_A : (i\tilde{L}F)A \to (i\tilde{L}G)A)_{A\in \mathbf{C}}\ | \ \beta \in \operatorname{\V-\mathrm{Nat}}(i\tilde{L}F, i\tilde{L}G)\} \arrow[d]\\
    \set^{\q^{\mathrm{op}}}(F^0, G^0) \arrow[r, "\tilde{L}_{FG}^{0}"] & \left[\set^{\q^{\mathrm{op}}}\right]_{\mathrm{prod}}(L F^0, L G^0)
  \end{tikzcd}
  \]}%
We write $\operatorname{\V-\mathrm{Nat}}(F, G)$ for the set of $\V$-natural transformations from $F$ to $G$.
The arrow $\tilde{L}_{FG}^{0}$ is given by the functor $L : \set^{\q^{\mathrm{op}}} \to [\set^{\q^{\mathrm{op}}}]_{\mathrm{prod}}$.
And the arrow $\tilde{L}_{FG}^{1}$ is given by extending the commuting square $\alpha_A : FA \to GA$ with $\eta : \id \to iL$, as in
the following diagram. Note that for each $\V$-natural transformation $\alpha \in \V\mathrm{Nat}(F, G)$, we have $\alpha^{0} : F^{0} \to G^{0}$. 

  \[
  \begin{tikzcd}
    (FA)_{1} \arrow[r, "\alpha_{A}^{1}"] \arrow[d] & (GA)_{1} \arrow[d]\\
    (FA)_{0} \arrow[r, "\alpha_{A}^{0}"] \arrow[d, "(\eta_{F^{0}})_{A}"] & (GA)_{0} \arrow[d, "(\eta_{G^{0}})_{A}"]\\
    (iLF^{0})A  \arrow[r, "(iL)\alpha^{0}_{A}"] & (iLG^{0})A
  \end{tikzcd}
\]
\end{proof}

\begin{theorem}
 The $\V$-category $\ctilde$ is a reflective $\V$-subcategory of $\cbar$, i.e., the inclusion $\V$-functor $i : \widetilde{\mathbf{C}} \hookrightarrow \cbar$ has
    a left adjoint $\tilde{L}$. 
\end{theorem}
\begin{proof}[Proof sketch]
  We need to show $\cbar(F, iG) \cong \ctilde(\ltilde F, G)$ for any $F\in \cbar, G\in \ctilde$ and it is
  $\V$-natural in $F$ and $G$. We just need to show the following diagram commutes. 

  \[
    \begin{tikzcd}
      \operatorname{\V-\mathrm{Nat}}(F, iG)\arrow[d] \arrow[r, "\cong"]  & \operatorname{\V-\mathrm{Nat}}(i\ltilde F, iG) \arrow[d]\\
      \set^{\q^{\mathrm{op}}}(F^{0}, iG^{0}) \arrow[r, "\cong"] & \qtilde(LF^{0}, G^{0})
    \end{tikzcd}
  \]
  The bottom arrow is an isomorphism because $L \dashv i$. The top arrow is an isomorphism because for any $A \in \C$ and
  $\V$-natural transformation $\gamma : F \to iG$,  we have the following commutative diagram.
  \[
    \begin{tikzcd}
      (FA)_{1} \arrow[r, "\gamma_{A}^{1}"] \arrow[d] & ((iG)A)_{1} \arrow[d]\\
      (FA)_{0} \arrow[r, "\gamma_{A}^{0}"] \arrow[d, "(\eta_{F^{0}})_{A}", swap] & ((iG)A)_{0}\\
      (iLF^{0})A \arrow[ur, "i(\widehat{\gamma^{0}})_{A}", swap]&
    \end{tikzcd}
  \]

\end{proof}  

\section{Day's reflection theorem for \texorpdfstring{$\ctilde$}{C-tilde}}
\label{app:day-reflection}

\begin{theorem}
    \label{thm:hom-preserves-smooth}
  If $H\in \ctilde$,
  then $G \multimap_{\mathrm{Day}} iH$ is also a smooth functor
  for any $G\in \cbar$.
\end{theorem}
\begin{proof}
  Suppose $H \in \ctilde$ and $G\in \cbar$. 
  For any $C \in \mathbf{C}$,
  we have
  \[
    (G \multimap_{\mathrm{Day}} iH)(C) = \int_{A \in \mathbf{C}} G A \Rightarrow iH(C \otimes A) \cong \overline{\mathbf{C}}(G, iH(C\otimes -)).
  \]
  Thus
  \[
    \begin{array}{lll}
      \displaystyle
      (G \multimap_{\mathrm{Day}} H)(C)_{0}
      &\cong&
      \displaystyle \overline{\mathbf{C}}(G, iH(C\otimes -))_{0} 
      ~~\cong~~ \set^{\q^{\mathrm{op}}}(G^{0}, (iH)^{0}(C\otimes -))
      \\\\[-2ex]
      &\cong& \displaystyle \int_{A\in \q} \set(G^{0}A, (iH)^{0}(C\otimes A))
      ~~\cong~~ (G^{0} \multimap_{\mathrm{Day}} (iH)^{0})(C),
    \end{array}
  \]
  where $G^{0} \multimap_{\mathrm{Day}} (iH)^{0}$ is an exponential in $\set^{\q^{\mathrm{op}}}$. Since $H^{0}$ preserves products, so does $G^{0} \multimap_{\mathrm{Day}} (iH)^{0}$, thus $G \multimap_{\mathrm{Day}} iH$ is smooth and $G \multimap_{\mathrm{Day}} iH\in \widetilde{\mathbf{C}}$. The
  functor $G^{0} \multimap_{\mathrm{Day}} (iH)^{0}$
  preserves products in $\q^{\mathrm{op}}$ because for any $C_{1}, C_{2} \in \C$, we have
  \[
    \begin{array}[b]{lll}
    \displaystyle(G^0 \multimap_{\mathrm{Day}} iH^0)(C_{1}+C_{2}) 
      &=& \displaystyle\int_{A} \set(G^0A, iH^0((C_{1}+C_{2})\otimes A))
      \\\\[-2ex]
      &\cong& \displaystyle \int_{A} \set(G^0A, iH^0(C_{1}\otimes A +C_{2} \otimes A))
      \\\\[-2ex]
    &\cong& \displaystyle\int_{A} \set(G^0A, iH^0(C_{1}\otimes A) \times iH^0(C_{2}\otimes A))
      \\\\[-2ex] 
    &\cong& \displaystyle \int_{A} \set(G^0A, iH^0(C_{1}\otimes A)) \times  \set(G^0A, iH^0(C_{2}\otimes A))
      \\\\[-2ex] 
    &\cong& \displaystyle\int_{A} \set(G^0A, iH^0(C_{1}\otimes A)) \times  \int_{A} \set(G^0A, iH^0(C_{2}\otimes A))
      \\\\[-2ex] 
    &=& \displaystyle (G^0 \multimap_{\mathrm{Day}} iH^0)(C_{1}) \times (G^0 \multimap_{\mathrm{Day}} iH^0)(C_{2}).
    \end{array}\qedhere
  \]
\end{proof}

The above theorem implies that
for any $G\in \ctilde, F\in \cbar$, the unit
$\eta_{F \multimap_{\mathrm{Day}} iG} : F \multimap_{\mathrm{Day}} iG \to i\ltilde(F \multimap_{\mathrm{Day}} iG)$
is an isomorphism, which gives rise to the following theorem. 

\begin{theorem}
  \label{thm:day-reflection}
  For any $F, H \in \cbar$, we have
  \[
    \tilde{L}(F\otimes_{\mathrm{Day}}H) \stackrel{\tilde{L}(\eta_{F}\otimes_{\mathrm{Day}} H)}{\cong} \tilde{L}(i\tilde{L} F\otimes_{\mathrm{Day}}H).
  \]
\end{theorem}
\begin{proof}
  For any $G \in \ctilde$, we have the following commutative diagram.
  \[
    \begin{tikzcd}
    \ctilde(\ltilde(i\ltilde F\otimes_{\mathrm{Day}}H), G)   \arrow[rr, "{\ctilde(\ltilde(\eta_{F}\otimes_{\mathrm{Day}} H), G)}"]
    \arrow[d, "\cong"]& & \ctilde(\ltilde(F\otimes_{\mathrm{Day}}H), G) \arrow[d, "\cong"] \\
    \cbar(i\ltilde F\otimes_{\mathrm{Day}}H, iG)
    \arrow[rr, "{\cbar(\eta_{F}\otimes_{\mathrm{Day}} H, iG)}"]
    \arrow[d, "\cong"]
       & & \cbar(F\otimes_{\mathrm{Day}}H, iG) \arrow[d, "\cong"]\\
       \cbar(i\ltilde F, H \multimap_{\mathrm{Day}} iG)
      \arrow[rr, "{\cbar(\eta_{F}, H \multimap_{\mathrm{Day}}iG)}"]
      \arrow[d, "{\cbar(i\ltilde F, \eta_{H\multimap_{\mathrm{Day}} iG})}", swap]&& \cbar(F, H \multimap_{\mathrm{Day}} iG) \arrow[d, "{\cbar(F, \eta_{H\multimap_{\mathrm{Day}} iG})}"] \\
      \cbar(i\ltilde F, i\ltilde(H \multimap_{\mathrm{Day}} iG))
      \arrow[rr, "{\cbar(\eta_{F}, i\ltilde(H \multimap_{\mathrm{Day}}iG))}"]
       \arrow[d, "\cong"]
      & &
      \cbar(F, i\ltilde(H \multimap_{\mathrm{Day}} iG))
      \arrow[ddll, "\id"]
      \\
      \ctilde(\ltilde F, \ltilde(H \multimap_{\mathrm{Day}} iG)) \arrow[d, "\cong"] & & \\
      \ctilde( F, i\ltilde(H \multimap_{\mathrm{Day}} iG)) & & 
    \end{tikzcd}
  \]
  The top two squares commute by naturality of the adjunctions, the third square commutes by the bi-functoriality of $\cbar(-,-)$ and the bottom triangle commutes by
  properties of the adjunction $\ltilde \dashv i$. 
\end{proof}

  With the help of Theorem~\ref{thm:day-reflection}, one can verify that $\ltilde$ is strong monoidal, e.g.,
  for any $F, G\in \cbar$,
  \[\ltilde F\otimes_{\mathrm{Lam}}\ltilde G = \ltilde (i\ltilde F\otimes_{\mathrm{Day}}i\ltilde G) \cong \ltilde (F\otimes_{\mathrm{Day}}G).\]

  \section{Proof of Theorem~\ref{thm:tmonad}}
  \label{app:strength-ttilde}
  We write $\beta : i\circ \ttilde \to \tbar\circ i$ to denote
  the isomorphism $i\circ \ttilde \cong \tbar\circ i$.
  \begin{theorem}
    \label{thm:rho}
    The following diagrams commute.

    \begin{enumerate}
    \item Suppose $F \in \ctilde$. 
      \[
        \begin{tikzcd}
          iF\arrow[r, "i\eta^{\widetilde{T}}"] \arrow[dr, "\eta^{\overline{T}}"] & i \tilde{T}F\arrow[d, "\beta"] \\
          & \overline{T}iF
        \end{tikzcd}
      \]
      
    \item Suppose $F \in \cbar$. 
      \[
        \begin{tikzcd}
          \tilde{L}F \arrow[r, "\tilde{L}\eta^{\overline{T}}"] \arrow[dr, "\eta^{\widetilde{T}}"] &  \tilde{L}\overline{T}F \arrow[d, "\rho"]\\
          & \widetilde{T}\tilde{L}F
        \end{tikzcd}
      \]

    \item Suppose $F \in \ctilde$. 
      \[
        \begin{tikzcd}
          i\widetilde{T}\widetilde{T}F
          \arrow[r, "\beta"]
          \arrow[d, "i\mu^{\ttilde}"]
          & \overline{T}i\widetilde{T}F
          \arrow[r, "\overline{T}\beta"]
          & \overline{T}\overline{T}iF
          \arrow[d, "\mu^{\tbar}"]
          \\
          i\widetilde{T}F\arrow[rr, "\beta"] && \overline{T} iF
        \end{tikzcd}
      \]
      
    \item Suppose $F \in \cbar$. 
      \[
        \begin{tikzcd}
          \tilde{L}\overline{T}\overline{T}F
          \arrow[r, "\rho"]
          \arrow[d, "\tilde{L}\mu^{\tbar}"]
          & \widetilde{T}\tilde{L}\overline{T}F
          \arrow[r, "\widetilde{T}\rho"]
          & \widetilde{T}\widetilde{T}\tilde{L}F
          \arrow[d, "\mu^{\ttilde}"]
          \\
          \tilde{L}\overline{T}F\arrow[rr, "\rho"] && \widetilde{T} \tilde{L}F
        \end{tikzcd}
      \]

    \item Suppose $F \in \ctilde$. 
      \[
        \begin{tikzcd}
          \ltilde i \ttilde F
          \arrow[r, "\ltilde \beta"]
          \arrow[d, "\epsilon"]
          & \ltilde \tbar iF\arrow[d, "\rho"]
          \\
          \ttilde F
          &
          \ttilde \ltilde i F
          \arrow[l, "\ttilde \epsilon"]
        \end{tikzcd}
      \]
    \item Suppose $F \in \cbar, G\in \ctilde$. 
      \[
        \begin{tikzcd}
          \ltilde \tbar (i\ttilde G\otimes_{D}F)
          \arrow[r, "\ltilde \tbar(\beta \otimes_{D}F)"]
          \arrow[d, "\rho"]
          & \ltilde \tbar (\tbar i G\otimes_{D}F)
          \arrow[r, "\ltilde \tbar \gamma"]
          & \ltilde \tbar (F \otimes_{D}\tbar i G)
          \arrow[d, "\rho"]
          \\
          \ttilde \ltilde (i\ttilde G\otimes_{D}F)
          \arrow[r, "\ttilde\ltilde \gamma"]
          &
          \ttilde \ltilde (F\otimes_{D}i\ttilde G)
          \arrow[r, "\ttilde \ltilde(F \otimes_{D}\beta)"]
          &
          \ttilde \ltilde (F\otimes_{D}\tbar i G)
        \end{tikzcd}
      \]
      
    \item Suppose $F, G \in \cbar$. 
      \[
        \begin{tikzcd}
          \ltilde(F \otimes_{D}\tbar G)\arrow[r, "\ltilde \overline{t}"]
          \arrow[d, "\ltilde(F\otimes_{D} \eta^{i\ltilde})"]
          & \ltilde\tbar (F \otimes_{D}G)
          \arrow[r, "\rho"]
          & \ttilde\ltilde (F \otimes_{D}G)
          \arrow[r, "\ttilde \ltilde(F\otimes_{d} \eta^{i\ltilde})"]& \ttilde\ltilde (F \otimes_{D}i\ltilde G)
          \\
          \ltilde(F \otimes_{D}i \ltilde \tbar G)
          \arrow[r, "\ltilde(F \otimes_{D} i\rho)"]
          &
          \ltilde(F \otimes_{D}i\ttilde \ltilde G)
          \arrow[r, "\ltilde(F \otimes_{D} \beta)"]
          &
          \ltilde(F \otimes_{D} \tbar i \ltilde G)
          \arrow[r, "\ltilde \overline{t}"]
          &
          \ltilde\tbar(F \otimes_{D}  i \ltilde G)
          \arrow[u, "\rho"]
        \end{tikzcd}
      \]

    \end{enumerate}
  \end{theorem}
  \begin{proof}
    \begin{enumerate}
    \item We have 
    \[\cbar(iF, \tbar i F) =  \cbar(iF, \overline{\Delta}\overline{U_{0}}iF) \cong
      \cbar(\overline{U_{0}}iF, \overline{U_{0}}iF) \stackrel{}{\cong} \cbar(j\overline{U_{0}}' F, j\overline{U_{0}}' F) \cong \ctilde(\overline{U_{0}}' F, \overline{U_{0}}' F) \cong \ctilde(F, \ttilde F)\].
      
    \item If we unfold the definition of $\rho$, we have
      the following diagram.

      \[
        \begin{tikzcd}
          \tilde{L}F \arrow[r, "\tilde{L}\eta^{\overline{T}}"]
          \arrow[dddr, "\eta^{\widetilde{T}}", bend right = 10, swap] &  \tilde{L}\overline{T}F \arrow[d, "\tilde{L}\overline{T}\eta^{i\tilde{L}}"]\\
          & \tilde{L}\overline{T}i\tilde{L}F
          \arrow[d, "\tilde{L}\beta^{-1}"]
          \\
          &\tilde{L}i\widetilde{T}\tilde{L}F
          \arrow[d, "\epsilon"]\\
          & \widetilde{T}\tilde{L}F\\ 
        \end{tikzcd}
      \]

      Note that we have the following commutative diagram,
      by naturality and (1). 
      \[
        \begin{tikzcd}
          F \arrow[r, "\eta^{\overline{T}}"]
          \arrow[d, "\eta^{\tilde{L}}"]
          & \overline{T}F \arrow[r, "\overline{T}\eta^{i\tilde{L}}"]
          &  \overline{T}i \tilde{L} F
          \arrow[r, "\beta^{-1}"]
          &  i \widetilde{T} \tilde{L} F
          \\
          i\tilde{L} F \arrow[urr, "\eta^{\overline{T}}"]
          \arrow[urrr, "i \eta^{\widetilde{T}}", swap ]
          & & & 
        \end{tikzcd}
      \]
      Therefore we just need to show the following diagram
      commutes (and indeed it does).
      \[
        \begin{tikzcd}
          \tilde{L}F \arrow[r, "\tilde{L}\eta^{i\tilde{L}}"]
          \arrow[dr, "\id"]&
          \tilde{L}i\tilde{L}F \arrow[d, "\epsilon"]
          \arrow[r, "\tilde{L}i\eta^{\widetilde{T}}"]
          & \tilde{L}i\widetilde{T}\tilde{L}F
          \arrow[d, "\epsilon"]\\
          &\tilde{L}F \arrow[r, "\eta^{\widetilde{T}}"]&
          \widetilde{T} \tilde{L}F
        \end{tikzcd}
      \]
    \item Since each component of $\mu$ and $\beta$ is an identity, the diagram commutes. 
    \item
      If we unfold the definition of $\rho$, we have the following diagram.
      \[
        \begin{tikzcd}
          \tilde{L} \overline{T} \overline{T}F \arrow[r, "\tilde{L}\overline{T} \eta^{i\tilde{L}}"]
          \arrow[d, "\ltilde \mu^{\tbar}"]
          & \tilde{L} \overline{T} i\tilde{L}\overline{T}F
          \arrow[r, "\tilde{L}\beta^{-1}"]
          & \tilde{L} i\widetilde{T} \tilde{L}\overline{T}F
          \arrow[r, "\epsilon"]
          \arrow[d, "\tilde{L} i\widetilde{T} \tilde{L}\overline{T}\eta^{i\tilde{L}}"]
          & \widetilde{T} \tilde{L}\overline{T}F\arrow[d, "\widetilde{T}\tilde{L}\overline{T}\eta^{i\tilde{L}}"]
          \\
          \tilde{L} \overline{T}F
          \arrow[d, "\ltilde \tbar \eta^{\ltilde}"]
          &
          &\tilde{L} i\widetilde{T} \tilde{L}\overline{T}i\tilde{L}F
          \arrow[r, "\epsilon"]
          \arrow[d, "\ltilde i\ttilde \ltilde \beta^{-1}"]
          &\widetilde{T}\tilde{L}\overline{T}i\tilde{L}F
          \arrow[d, "\ttilde\ltilde\beta^{-1}"]
          \\
          \tilde{L} \overline{T}i \ltilde F
          \arrow[d, "\ltilde \beta^{-1}"]
          &
          & \tilde{L}i\widetilde{T} \tilde{L}i\ttilde\tilde{L}F\arrow[r, "\epsilon"]
          \arrow[d, "\ltilde i \ttilde\epsilon"]
          & \widetilde{T}\tilde{L}i\ttilde\tilde{L}F
          \arrow[d, "\ttilde \epsilon"]
          \\
          \tilde{L}i \widetilde{T} \ltilde F
          \arrow[d, "\epsilon"]
          && \tilde{L}i \ttilde \ttilde  \ltilde F \arrow[r, "\epsilon"]
          \arrow[ll, "\ltilde i \mu^{\ttilde}"]
          & \ttilde \ttilde  \ltilde F
          \arrow[dlll, bend left = 10, "\mu^{\ttilde}"]
          \\
          \widetilde{T} \ltilde F
          &&&
        \end{tikzcd}
      \]
      All of the squares commute by naturality. We just need to show the left diagram commutes.
      It commutes because the following commutes.
      \[
        \begin{tikzcd}
          \tbar \tbar F \arrow[r, "\tbar \eta^{i\ltilde}"]
          \arrow[d, "\mu"]
          \arrow[dr, "\tbar \tbar \eta^{i\ltilde}"]
          & \tbar i\ltilde \tbar F \arrow[rr, "\beta^{-1}"]
          \arrow[dr, "\tbar i\ltilde\tbar\eta^{i\ltilde}"]
          & & i\ttilde \ltilde \tbar F
          \arrow[dd, "i\ltilde \tbar \eta^{i\ltilde}"]
          \\
          \tbar F
          \arrow[d, "\tbar \eta^{i\ltilde}"]
          & \tbar \tbar i\ltilde F\arrow[r, "\tbar \eta^{i\ltilde}"]
          \arrow[dl, "\mu"]
          \arrow[d, "\tbar \beta^{-1}"]
          & \tbar i \ltilde  \tbar i\ltilde F \arrow[dr, "\beta^{-1}"] \arrow[d, "\tbar i \ltilde \beta^{-1}"]&
          \\
          \tbar i \ltilde F
          \arrow[d, "\beta^{-1}"]
          &
          \tbar i \ttilde\ltilde F
          \arrow[r, "\tbar \eta^{i\ltilde}"]
          \arrow[d, "\beta^{-1}"]
          &
          \tbar i\ltilde i \ttilde\ltilde F
          \arrow[dr, "\beta^{-1}"]
          & i\ttilde \ltilde \tbar i\ltilde F
          \arrow[d, "i\ttilde \ltilde\beta^{-1}"]
          \\
          i\ttilde \ltilde F
          &
          i\ttilde \ttilde\ltilde F \arrow[l, "i \mu"]
          &
          & i\ttilde \ltilde i\ttilde \ltilde F
          \arrow[ll, "i\ttilde \epsilon"]
        \end{tikzcd}
      \]
      
      The above diagram commutes, because all the squares commutes
      by naturality. The bottom left corner diagram commutes by (3).
      Note that $i\epsilon \circ \eta^{i\ltilde} = \id$.
    \item If we unfold the definition of $\rho$, we have the following diagram.
      \[
        \begin{tikzcd}
          \tilde{L} i \widetilde{T}F \arrow[r, "\tilde{L}\beta"]
          \arrow[d, "\ltilde \beta"]
          & \tilde{L}\overline{T}iF 
          \arrow[d, "\tilde{L}\overline{T} \eta^{i\tilde{L}}"]
          \\
          \ltilde\tbar i F \arrow[d, "\ltilde \beta^{-1}"]
          & \tilde{L}\overline{T}i\tilde{L}iF \arrow[d, "\tilde{L}\beta^{-1}"] \arrow[l, "\ltilde\tbar i \epsilon"]\\
          \tilde{L} i \widetilde{T}F\arrow[d, "\epsilon"]
          & \tilde{L}i\widetilde{T}\tilde{L}iF \arrow[d, "\epsilon"]
          \arrow[l, "\tilde{L}i\widetilde{T}\epsilon"]
          \\
         \widetilde{T}F & \widetilde{T}\tilde{L}iF \arrow[l, "\widetilde{T}\epsilon"]
        \end{tikzcd}
      \]
      The bottom two squares commute by naturality. The top square commutes because $ i\epsilon \circ \eta^{i\tilde{L}} = \id$.
      
    \item See the following commutative diagram. 
      \[
        \begin{tikzcd}
          \ltilde \tbar (i\ttilde G\otimes_{D}F)
          \arrow[r, "\ltilde \tbar(\beta \otimes_{D} F)"]
          \arrow[dd, "\rho"]
          & \ltilde \tbar (\tbar i G\otimes_{D}F)
          \arrow[r, "\ltilde \tbar \gamma"]
          \arrow[d, "\rho"]
          & \ltilde \tbar (F \otimes_{D}\tbar i G)
          \arrow[dd, "\rho"]
          \\
          & \ttilde \ltilde (\tbar i G\otimes_{D}F) 
          \arrow[dr, "\ttilde \ltilde \gamma"]
          &
          \\
          \ttilde \ltilde (i\ttilde G\otimes_{D}F)
          \arrow[r, "\ttilde\ltilde \gamma"]
          \arrow[ur, "\ttilde \ltilde(\beta\otimes_{D} F)"]
          &
          \ttilde \ltilde (F\otimes_{D}i\ttilde G)
          \arrow[r, "\ttilde \ltilde(F \otimes_{D}\beta)"]
          &
          \ttilde \ltilde (F\otimes_{D}\tbar i G)
        \end{tikzcd}
      \]
    \item
      Consider the following diagram.
      \[
        \begin{tikzcd}
          \ltilde(F \otimes_{D}\tbar G)\arrow[r, "\ltilde \bar{t}"]
          \arrow[d, "\ltilde(F\otimes \eta^{i\ltilde})"]
          \arrow[drr, "\ltilde(F\otimes \tbar \eta^{i\ltilde})"]
          & \ltilde\tbar (F \otimes_{D}G)
          \arrow[r, "\rho"]
          \arrow[drr, "\ltilde \tbar (F\otimes \eta^{i\ltilde})"]
          & \ttilde\ltilde (F \otimes_{D}G)
          \arrow[r, "\ttilde \ltilde(F\otimes \eta^{i\ltilde})"]& \ttilde\ltilde (F \otimes_{D}i\ltilde G)
          \\
          \ltilde(F \otimes_{D}i \ltilde \tbar G)
          \arrow[r, "\ltilde(F \otimes_{D} i\rho)"]
          &
          \ltilde(F \otimes_{D}i\ttilde \ltilde G)
          \arrow[r, "\ltilde(F \otimes_{D} \beta)", swap]
          &
          \ltilde(F \otimes_{D} \tbar i \ltilde G)
          \arrow[r, "\ltilde \bar{t}"]
          &
          \ltilde\tbar(F \otimes_{D}  i \ltilde G)
          \arrow[u, "\rho"]
        \end{tikzcd}
      \]

      Note that the very right diagram and the middle square commute by naturality. The left diagram
      commutes because the following diagram commutes (with $\rho$ unfolded).
      \[
        \begin{tikzcd}
          \ltilde(F\otimes_{D}\tbar G)\arrow[rr, "\ltilde(F\otimes \tbar \eta^{i\ltilde})"]
          \arrow[dr, "\ltilde(F\otimes \tbar \eta^{i\ltilde})"]
          \arrow[d, "\ltilde (F\otimes \eta^{i\ltilde})",swap]& & \ltilde(F\otimes_{D}\tbar i\ltilde G) \\
          \ltilde(F\otimes_{D}i\ltilde \tbar G) \arrow[d, "\ltilde(F\otimes i\ltilde \tbar \eta^{i\ltilde})",swap]
          & \ltilde(F\otimes_{D}\tbar i\ltilde G) \arrow[ur, "\id"]
          \arrow[dl, "\ltilde(F\otimes \eta^{i\ltilde})",swap] \arrow[r, "\ltilde(F\otimes_{D}\beta^{-1})"]
          &  \ltilde(F\otimes_{D}i\ttilde \ltilde G) \arrow[dl,  "\ltilde(F\otimes \eta^{i\ltilde})",swap]\\
          \ltilde(F\otimes_{D}i\ltilde \tbar i \ltilde G) \arrow[r, "\ltilde(F\otimes_{D}i\ltilde \beta^{-1})", swap]
          & \ltilde(F\otimes_{D}i\ltilde i\ttilde \ltilde G) \arrow[r, "\ltilde(F\otimes_{D}i\epsilon)",swap]
          & \ltilde(F\otimes_{D}i\ttilde \ltilde G) \arrow[uu, bend right = 70, "\ltilde(F\otimes_{D}\beta)",swap]
        \end{tikzcd}
      \]
    \end{enumerate}
  \end{proof}

    \begin{theorem}
     The $\V$-functor $\widetilde{T}$ is a commutative strong monad. The strength is given by $\tilde{t}_{F,G} : F \otimes_{\mathrm{Lam}}\widetilde{T} G \to \widetilde{T}(F\otimes_{\mathrm{Lam}} G)$ for any $F, G\in \ctilde$.
  \end{theorem}

  \begin{proof}
    For any $F, G \in \ctilde$, 
    we define $\tilde{t}_{F,G}$ by the following composition.
    \begin{align*}
    F \otimes_{\mathrm{Lam}}\widetilde{T} G &= \tilde{L}(iF \otimes_{\mathrm{Day}}i\widetilde{T}G) \\
      & \stackrel{\beta}{\cong} \tilde{L}(iF \otimes_{\mathrm{Day}}\overline{T}iG) \stackrel{\bar{t}}{\to}\tilde{L}(\overline{T}(iF \otimes_{\mathrm{Day}}iG)) \stackrel{\rho}{\to} \widetilde{T}(\tilde{L}(iF \otimes_{\mathrm{Day}}iG)) = \widetilde{T}(F\otimes_{\mathrm{Lam}} G). 
      \end{align*}

    Now we need to show $\widetilde{T}$ is a commutative strong monad.
    \begin{itemize}
    \item
      \[
        \begin{tikzcd}
          F \otimes_{\mathrm{Lam}} G \arrow[r, "\id_{F} \otimes_{\mathrm{Lam}} \eta" ]
          \arrow[d, "\eta"]
          & F \otimes_{\mathrm{Lam}} \widetilde{T} G \arrow[dl, "\tilde{t}"] \\
          \widetilde{T}(F\otimes_{\mathrm{Lam}} G)
        \end{tikzcd}
      \]    
      The above diagram commutes because the 
      following diagram commutes (by properties of $\bar{t}$ and Theorem~\ref{thm:rho}(1)+(2)). 
      \[
        \begin{tikzcd}
          \tilde{L}(iF \otimes_{\mathrm{Day}} iG)
          \arrow[drr, "\tilde{L}\eta^{\tbar}", bend right = 10, swap, near start]
          \arrow[rr, "\tilde{L}(\id \otimes_{\mathrm{D}} \eta^{\tbar})", bend right = 15]
          \arrow[r, "\tilde{L}(\id \otimes_{\mathrm{D}} i\eta^{\ttilde})" ]
          \arrow[d, "\eta^{\ttilde}"]
          & \tilde{L}(iF \otimes_{\mathrm{Day}} i\widetilde{T}G)
          \arrow[r, "\tilde{L}(\id \otimes_{\mathrm{D}} \beta)" ]
          &
          \tilde{L}(iF \otimes_{\mathrm{Day}} \overline{T} iG)
          \arrow[d, "\tilde{L}\bar{t}"]
           \\
           \widetilde{T}\tilde{L}(iF\otimes_{\mathrm{Day}} iG) & &\tilde{L}\overline{T}(iF \otimes_{\mathrm{Day}}  iG)
           \arrow[ll, "\rho", bend left = 10]
        \end{tikzcd}
      \]          
    \item
      \[
      \begin{tikzcd}
        F \otimes_{\mathrm{Lam}} \widetilde{T} \widetilde{T} G
        \arrow[r, "\tilde{t}"]
        \arrow[d, "F\otimes_{\mathrm{Lam}} \mu^{\ttilde}"]
        & \widetilde{T}(F \otimes_{\mathrm{Lam}} \widetilde{T} G)
        \arrow[r, "\widetilde{T}\tilde{t}"]
        &
        \widetilde{T} \widetilde{T}(F \otimes_{\mathrm{Lam}} G)
        \arrow[d, "\mu^{\ttilde}"]
        \\
        F \otimes_{\mathrm{Lam}} \widetilde{T} G
        \arrow[rr, "\tilde{t}"]
        & & \widetilde{T}(F \otimes_{\mathrm{Lam}} G) 
      \end{tikzcd}
    \]    

    The above diagram commutes because the following diagram commutes (all the inner diagrams commute, by naturality, properties of $\bar{t}$, and Theorem~\ref{thm:rho}(3)+(4)).
    \[
      \begin{tikzcd}
        \ltilde(iF\otimes_{D}i \ttilde \ttilde G)
        \arrow[r, "\ltilde(\id \otimes_{D} \beta)"]
        \arrow[d, "\ltilde(\id \otimes_{D} i \mu^{\ttilde})"]
        &
        \ltilde(iF\otimes_{D}\tbar i\ttilde G)
        \arrow[r, "\ltilde \bar{t}"]
        \arrow[d, "\ltilde(\id\otimes_{D}\tbar \beta)"]
        &
        \ltilde \tbar (iF\otimes_{D} i\ttilde G)
        \arrow[d, "\rho"]
        \arrow[ddl, "\ltilde\tbar (\id\otimes_{D}\beta)", near end]
        \\
        \ltilde(iF\otimes_{D} i \ttilde G)
        \arrow[d, "\ltilde(\id\otimes \beta)"]
        &
        \ltilde(iF\otimes_{D} \tbar \tbar i G)
        \arrow[d, "\ltilde \bar{t}"]
        \arrow[dl, "\ltilde(\id\otimes_{D}\mu^{\tbar})"]
        &
        \ttilde \ltilde(iF\otimes_{D} i\ttilde G)
        \arrow[d, "\ttilde \ltilde(\id \otimes_{D}\beta)"]
        \\
        \ltilde(iF\otimes_{D} \tbar i G)
        \arrow[d, "\ltilde \bar{t}"]
        &
        \ltilde \tbar (iF\otimes_{D}  \tbar i G)
        \arrow[r, "\rho"]
        \arrow[d, "\ltilde \tbar \bar{t}"]
        &
        \ttilde \ltilde(iF\otimes_{D} \tbar i G)
        \arrow[d, "\ttilde \ltilde \bar{t}"]
        \\
        \ltilde\tbar(iF\otimes_{D}iG)
        \arrow[d, "\rho"]
        &
        \ltilde\tbar \tbar(iF\otimes_{D}iG)
        \arrow[r, "\rho"]
        \arrow[l, "\ltilde \mu^{\tbar}"]
        &
        \ttilde \ltilde \tbar(iF\otimes_{D}iG)
        \arrow[d, "\ttilde \rho"]
        \\
        \ttilde\ltilde(iF\otimes_{D}iG)
        &
        &
        \ttilde \ttilde \ltilde(iF\otimes_{D}iG)
        \arrow[ll, "\mu^{\ttilde}"]
      \end{tikzcd}
    \]    
    
  \item
    \[
      \begin{tikzcd}
        (F\otimes_{\mathrm{Lam}} G) \otimes_{\mathrm{Lam}} \widetilde{T}H
        \arrow[rr,"\tilde{t}"]
        \arrow[d, "\alpha"]
        & & \widetilde{T}((F\otimes_{\mathrm{Lam}} G) \otimes_{\mathrm{Lam}} H) \arrow[d, "\widetilde{T}\alpha"]\\
        F\otimes_{\mathrm{Lam}} (G\otimes_{\mathrm{Lam}} \widetilde{T}H)
        \arrow[r, "\id_{F}\otimes_{\mathrm{Lam}} \tilde{t}"]
        &
        F\otimes_{\mathrm{Lam}} \widetilde{T}(G\otimes_{\mathrm{Lam}} H)
        \arrow[r, "\tilde{t}"]
        &
        \widetilde{T}(F\otimes_{\mathrm{Lam}} (G\otimes_{\mathrm{Lam}} H))
      \end{tikzcd}
    \]
    Again, the above diagram commutes because 
    the following diagram commutes (all of the inner diagrams commute by naturality, properties of $\bar{t}$, and Theorem~\ref{thm:rho}(7)). 
           
    \includestandalone[width=0.9\textwidth]{assoc-tilde}

  \item
  \[
    \begin{tikzcd}
      I \otimes_{\mathrm{Lam}} \widetilde{T}F
      \arrow[r, "\tilde{t}_{I, F}"]
      \arrow[dr, "\lambda_{\widetilde{T}F}", swap]
      & \widetilde{T}(I \otimes_{\mathrm{Lam}} F)
      \arrow[d, "\widetilde{T}\lambda_{F}"]
      \\
      & \widetilde{T}F
    \end{tikzcd}
  \]
  The above diagram commutes because the
  following commutes (all the inner diagrams commute, by naturality, property of $\bar{t}$, and Theorem~\ref{thm:rho} (5)). 
  \[
    \begin{tikzcd}
      \ltilde(I \otimes_{D} i \ttilde F) \arrow[r, "\ltilde (\id\otimes_{D}\beta)"]
      \arrow[d, "\ltilde \lambda"]
      & \ltilde(I \otimes_{D} \tbar i F) \arrow[r, "\ltilde \bar{t}"]
      \arrow[d, "\ltilde \lambda"]
      & \ltilde\tbar (I \otimes_{D} i F)
      \arrow[d, "\rho"]
      \arrow[dl, "\ltilde \tbar \lambda", swap]
      \\
      \ltilde i \ttilde F
      \arrow[r, "\ltilde \beta"]
      \arrow[d, "\epsilon"]
      &
      \ltilde \tbar i F
      \arrow[dr, "\rho"]
      &
      \ttilde \ltilde(I\otimes_{D}iF)
      \arrow[d, "\ttilde \ltilde \lambda"]
      \\
      \ttilde F
      &
      &
      \ttilde \ltilde iF
      \arrow[ll, "\ttilde \epsilon"]
    \end{tikzcd}
  \]

  \item
     Lastly, since $\widetilde{\mathbf{C}}$ is symmetric monoidal with $\gamma_{F, G} : F\otimes_{\mathrm{Lam}}G \to G\otimes_{\mathrm{Lam}}F$.
  We define the costrength $\sigma_{F, G} := \widetilde{T} \gamma_{G, F} \circ \tilde{t}_{G, F}\circ \gamma_{\widetilde{T}F, G} : \widetilde{T}F\otimes_{\mathrm{Lam}} G \to \widetilde{T}(F\otimes_{\mathrm{Lam}}G)$.
  We need to show the following diagram commutes.
  \[
    \begin{tikzcd}
      & \widetilde{T} F \otimes_{\mathrm{Lam}} \widetilde{T} G
      \arrow[dl, "\sigma"]
      \arrow[dr, "\tilde{t}"]
      & \\
      \widetilde{T} (F \otimes_{\mathrm{Lam}} \widetilde{T} G)
      \arrow[d, "\widetilde{T}\tilde{t}"]
      & & \widetilde{T} (\widetilde{T}F \otimes_{\mathrm{Lam}}  G)
      \arrow[d, "\widetilde{T}\sigma"]
      \\
      \widetilde{T}\widetilde{T} (F \otimes_{\mathrm{Lam}} G)
      \arrow[dr, "\mu^{\ttilde}"]  
      & & \widetilde{T} \widetilde{T}(F \otimes_{\mathrm{Lam}} G)
      \arrow[dl, "\mu^{\ttilde}"]\\
      & \widetilde{T}(F \otimes_{\mathrm{Lam}} G)&
    \end{tikzcd}
  \]
    Again, the above diagram commutes because 
    the following diagram commutes (all the inner diagrams commute, by naturality, properties of $\bar{t}$, and Theorem~\ref{thm:rho}(4)+(6)). 
    \smallskip

    \includestandalone[width=0.89\textwidth]{commute-tilde}

    \qedhere
    \end{itemize}
  \end{proof}

  \section{Proof of Theorem~\ref{thm:model}}
  \label{app:main}
    \begin{theorem}
  The $\V$-category $\ctilde$ is a model for Proto-Quipper with
  dynamic lifting.
\end{theorem}
\begin{proof}
  We have already shown that $\ctilde$ satisfies conditions~\ref{closed}--\ref{convexity} and \ref{box-unbox}--\ref{dynlift}.
  In the following we will focus on condition~\ref{simple-types}.

    \begin{itemize}
    \item First we need to define a functor $\psi : \m \hookrightarrow V(\ctilde)$. We define it as the following
      composition.
      \[\m \cong V(\C) \stackrel{V\overline{y}}{\hookrightarrow} V(\ctilde)\]
      The functor $\psi$ is strong monoidal, because $V\overline{y}$ is strong monoidal.
      
    \item Next we need to define a functor $\phi : \q \hookrightarrow \Kl_{V\ttilde}(V(\ctilde))$.
      We write
      \[\theta_{F, G} : \widetilde{\mathbf{C}}(F, \tilde{T}G) \cong [\set^{\mathbf{C}^{\mathrm{op}}}]_{\mathrm{prod}}(\overline{U_{0}}'F, \overline{U_{0}}'G)\]
      for any $F, G \in \ctilde$. We also write $\Omega : [\set^{\mathbf{C}^{\mathrm{op}}}]_{\mathrm{prod}} \cong \qtilde$. 
      We have
      \[\Kl_{V\ttilde}(V(\ctilde))(F, G) = \V(1, \widetilde{\mathbf{C}}(F, \tilde{T}G))
      \]
      \[\stackrel{\V(1, \theta_{F,G})}{\cong} \V(1, [\set^{\mathbf{C}^{\mathrm{op}}}]_{\mathrm{prod}}(\overline{U_{0}}'F, \overline{U_{0}}'G)) \stackrel{\Omega_{\overline{U_{0}}'F, \overline{U_{0}}'G}}{\cong} \qtilde(F^{0}, G^{0})\]for any $F, G \in \ctilde$. The category
      $\Kl_{V\ttilde}(V(\ctilde))$ is enriched in convex spaces because $\qtilde$ is enriched in convex spaces.

      Now we define $\phi$. On objects, we define
      \[\phi(S) = \overline{y}S =\mathbf{C}(-, S).\] On morphisms, 
      for any $S, U\in \q$, we define $\phi_{S,U}$ 
      by the following composition of isomorphisms.
      \[\q(S, U) \stackrel{\kappa_{S, U}}{\cong} \qtilde(\kappa S, \kappa U) = \qtilde((\overline{y} S)^{0}, (\overline{y} U)^{0})
        \]
        \[\stackrel{\Omega_{\overline{U_{0}}'\overline{y}S, \overline{U_{0}}'\overline{y}U}^{-1}}{\cong} \V(1, [\set^{\C^{\mathrm{op}}}]_{\mathrm{prod}}(\overline{U_{0}}'\overline{y}S, \overline{U_{0}}'\overline{y}U))
          \stackrel{\V(1, \theta_{\overline{y}S, \overline{y}U}^{-1})}{\cong} \Kl_{V\ttilde}(V(\ctilde))(\overline{y}S, \overline{y}U)\]

      Since the Lambek embedding $\kappa_{S, U}$ preserves convex sum (Theorem~\ref{prop:convex}) and 
      the composition $\V(1, \theta_{\overline{y}S, \overline{y}U}^{-1}) \circ \Omega_{\overline{U_{0}}'\overline{y}S, \overline{U_{0}}'\overline{y}U}^{-1}$
      preserves convex sum, we conclude that $\phi$ preserves convex sum.

      Next we need to show $\phi$ is strong monoidal.
      Since $\kappa$ is strong monoidal, we have
      the natural isomorphisms $I \stackrel{e'}{\cong} \kappa I$ and
      $\kappa S \otimes \kappa U \stackrel{m'_{S, U}}{\cong} \kappa(S \otimes U)$ for any $S, U \in \q$.
      Recall that for any $S\in \q$, $\kappa S = \q(-, S) = \overline{U_{0}}' \C(-, S)$ and $\overline{U_{0}}$
      is strong monoidal.
      Via the isomorphism $\Omega$ (which preserves the monoidal structure) we have the following natural isomorphisms in $[\set^{\C^{\mathrm{op}}}]_{\mathrm{prod}}$.
      \[\Omega^{-1}(e') : \overline{U_{0}}' \C(-, I) \to \overline{U_{0}}' \C(-, I)\]
      \[\Omega^{-1}(m'_{S, U}) : \overline{U_{0}}' (\C(-,S)\otimes \C(-, U)) \to \overline{U_{0}}' \C(-, S\otimes U).\]
      Now let $m_{S, U} = \theta_{\overline{y}S, \overline{y}U}^{-1}(\Omega^{-1}(m'_{S, U})) : \C(-,S)\otimes \C(-, U) \to \ttilde \C(-, S\otimes U)$ and $e = \theta_{\overline{y}I, \overline{y}I}^{-1}(\Omega^{-1}(e')) : \C(-, I) \to \ttilde \C(-, I)$.
      It is obvious that $e$ is an isomorphism in $\Kl_{V\ttilde}(V(\ctilde))$. 
      The inverse of $m_{S, U}$ is defined as $\theta_{\overline{y}S, \overline{y}U}^{-1}(\Omega^{-1}({m'_{S, U}}^{-1}))$, which
      can be verified. 
      We can furthermore show that $m_{S, U}$ is natural and that $e, m_{S, U}$ satisfies the strength diagrams
      for any $S, U\in \q$. For example, showing $m_{S, U}$ is natural
      in $\Kl_{V\ttilde}(V(\ctilde))$, via the adjunction $\overline{U_{0}}' \dashv \overline{\Delta}'$, is
      equivalent to the naturality of $\Omega^{-1}(m'_{S, U})$. 

    \item
      Lastly, we want to show that the following diagram commutes.
      \[
        \begin{tikzcd}
          \m(S, U)
          \arrow[r, "\psi_{S,U}"]
          \arrow[d, "J_{S,U}"]
          &  \V(1,\ctilde(\overline{y}S, \overline{y}U)) = V(\ctilde)(\overline{y}S, \overline{y}U) \arrow[d, "E_{S,U}"]\\
          \q(S, U) & \V(1,\ctilde(\overline{y}S, \widetilde{T} \overline{y}U)) = \Kl_{V\ttilde}(V(\ctilde)(\overline{y}S, \overline{y}U)
          \arrow[l, "\phi_{S,U}^{-1}"]
        \end{tikzcd}
      \]    

      Let $f \in \m(S, U)$. We write $(f, J_{S,U}f)$ for the corresponding map in $\V(1, \mathbf{C}(S, U))$. It corresponds to the following map in $\ctilde$ via the enriched Yoneda embedding.
    \[
      \begin{tikzcd}
        \mathbf{C}(-, S) \arrow[r, "{\overline{y}(f, J_{S,U}f)}"]& \mathbf{C}(-, U)
      \end{tikzcd}
    \]
    Applying $\eta_{U}$ to the above map, we have the following.
    \[
      \begin{tikzcd}
        \mathbf{C}(-, S) \arrow[r, "{\overline{y}(f, J_{S,U}f)}"]& \mathbf{C}(-, U) \arrow[r, "\eta_{U}"]& \ttilde \mathbf{C}(-, U)
      \end{tikzcd}
    \]
    So for any $A\in \mathbf{C}$, we have the following.
    \[
      \begin{tikzcd}
        \m(A, S) \arrow[r, "{\m(A, f)}"]
        \arrow[d, "J_{A,S}"] & \m(A, U) \arrow[r, "J_{A,U}"] \arrow[d, "J_{A,U}"]& \q(A, U)\arrow[d, "\id"] \\
        \q(A, S) \arrow[r, "{\q(A, J_{S,U}f)}"] & \q(A, U) \arrow[r, "\id"]& \q(A, U) 
      \end{tikzcd}
    \]
    Since $\phi_{S,U}^{-1} = \kappa_{S,U}^{-1}\circ \Omega_{\overline{U_{0}}'\overline{y}S, \overline{U_{0}}'\overline{y}U} \circ \V(1, \theta_{\overline{y}S, \overline{y}U})$, we have
    \[\phi_{S,U}^{-1}(\eta_{U}\circ \overline{y}(f, J_{S,U}f)) =\kappa_{S,U}^{-1} \Omega_{\overline{U_{0}}'\overline{y}S, \overline{U_{0}}'\overline{y}U} \V(1, \theta_{\overline{y}S, \overline{y}U})(\eta_{U}\circ \overline{y}(f, J_{S,U}f))\]
    \[= \kappa_{S,U}^{-1} \Omega_{\overline{U_{0}}'\overline{y}S, \overline{U_{0}}'\overline{y}U}(\overline{U_{0}}'\overline{y}(f, J_{S,U}f))
      = \kappa_{S,U}^{-1}(\q(-, J_{S,U}f)) = J_{S,U}f \in \q(S, U).\]
    \end{itemize}

\end{proof}
  
\end{document}

%% file: assoc0.tex

  \begin{tikzcd}
    \int^{B, X, Y}\A(C, (X\otimes Y) \otimes B) \otimes F X \otimes G Y \otimes T H B
    \arrow[rr, "\int t"]
    \arrow[d, "\int \alpha"]
    & &
    \int^{B, X, Y}T(\A(C, (X\otimes Y) \otimes B) \otimes F X \otimes G Y \otimes H B)
    \arrow[dl, "\int T \alpha", swap]
    \arrow[dr, "\xi"] & \\
    \int^{B, X, Y}\A(C, X\otimes (Y \otimes B)) \otimes F X \otimes G Y \otimes T H B
    \arrow[r, "\int t"]
    \arrow[d, "\cong"]
    &\int^{B, X, Y}T(\A(C, X\otimes (Y \otimes B)) \otimes F X \otimes G Y \otimes H B)
    \arrow[dr, "\xi"]
    &
    &
    T\int^{B, X, Y}\A(C, (X\otimes Y) \otimes B) \otimes F X \otimes G Y \otimes H B
    \arrow[dl, "T \int \alpha", swap]
    \\
    \int^{A,X}  \A(C, X\otimes A) \otimes FX \otimes  \int^{Y,B}  \A(A, Y\otimes B) \otimes G Y \otimes THB
    \arrow[d, "\int \id \otimes \int t"]
    & &
    T\int^{B, X, Y}\A(C, X\otimes (Y \otimes B)) \otimes F X \otimes G Y \otimes H B
    \arrow[d, "\cong"]
    & \\
    \int^{A,X}  \A(C, X\otimes A) \otimes FX \otimes  \int^{Y,B} T( \A(A, Y\otimes B) \otimes G Y \otimes HB)
    \arrow[d, "\int \id \otimes \xi"]
    &
    &
    \arrow[d, "\cong"]
    T\int^{X,Y,B} \int^A \A(C, X\otimes A) \otimes \A(A, Y\otimes B) \otimes FX \otimes G Y \otimes HB
    &
    \\
    \int^{A,X}\A(C, X\otimes A) \otimes FX \otimes T \int^{Y,B} \A(A, Y\otimes B) \otimes G Y \otimes HB
    \arrow[r, "\int t"]
    &
    \int^{A,X}T  (\A(C, X\otimes A) \otimes FX \otimes  \int^{Y,B} \A(A, Y\otimes B) \otimes G Y \otimes HB)
    \arrow[r, "\xi"]
    &
   T\int^{A,X}\A(C, X\otimes A) \otimes FX \otimes  \int^{Y,B} \A(A, Y\otimes B) \otimes G Y \otimes HB
    & \\
  \end{tikzcd}


%% file: assoc.tex

  \begin{tikzcd}
    \int^{B, X, Y}\A(C, X\otimes(Y\otimes B)) \otimes FX \otimes GY \otimes THB
    \arrow[rr, "\int t"]
    \arrow[d, "\cong"]
    &
    &
    \int^{B, X, Y}T(\A(C, X\otimes(Y\otimes B)) \otimes FX \otimes GY \otimes HB)
    \arrow[r, "\xi"]
    \arrow[d, "\cong"]
    &
    T\int^{B, X, Y}\A(C, X\otimes(Y\otimes B)) \otimes FX \otimes GY \otimes HB
    \arrow[d, "\cong"]
    \\
    \int^{B, X, Y}(\int^A \A(A, Y\otimes B) \otimes \A(C, X \otimes A)) \otimes FX \otimes GY \otimes THB
    \arrow[rr, "\int t"]
    \arrow[d, "\cong"]
    &
    &
    \int^{B, X, Y}T((\int^A \A(A, Y\otimes B) \otimes \A(C, X \otimes A)) \otimes FX \otimes GY \otimes HB)
    \arrow[r, "\xi"]
    \arrow[d, "\cong"]
    &
    T\int^{B, X, Y}(\int^A \A(A, Y\otimes B) \otimes \A(C, X \otimes A)) \otimes FX \otimes GY \otimes HB
    \arrow[d, "\cong"]
    \\
    \int^{B, X, Y}\int^A (\A(A, Y\otimes B) \otimes \A(C, X \otimes A) \otimes FX \otimes GY \otimes THB)
    \arrow[d, "\cong"]
    \arrow[r, "\int \int t"]
    &
    \int^{B, X, Y}\int^A T(\A(A, Y\otimes B) \otimes \A(C, X \otimes A) \otimes FX \otimes GY \otimes HB)
    \arrow[r, "\int \xi"]
    \arrow[d, "\cong"]
    &
    \int^{B, X, Y}T\int^A (\A(A, Y\otimes B) \otimes \A(C, X \otimes A) \otimes FX \otimes GY \otimes HB)
    \arrow[r, "\xi"]
    &
    T\int^{B, X, Y}\int^A (\A(A, Y\otimes B) \otimes \A(C, X \otimes A) \otimes FX \otimes GY \otimes HB)
    \arrow[d, "\cong"]
\\
    \int^{B, Y}\int^{A,X} (\A(A, Y\otimes B) \otimes \A(C, X \otimes A) \otimes FX \otimes GY \otimes THB)
    \arrow[r, "\int \int t"]
    \arrow[d, "\cong"]
    &
    \int^{B, Y}\int^{A,X} T(\A(A, Y\otimes B) \otimes \A(C, X \otimes A) \otimes FX \otimes GY \otimes HB)
    \arrow[r, "\int \xi"]
    \arrow[d, "\cong"]
    &
    \int^{B, Y}T\int^{A,X}(\A(A, Y\otimes B) \otimes \A(C, X \otimes A) \otimes FX \otimes GY \otimes HB)
    \arrow[r, "\xi"]
    &
    T\int^{B, Y}\int^{A,X}(\A(A, Y\otimes B) \otimes \A(C, X \otimes A) \otimes FX \otimes GY \otimes HB)
    \arrow[d, "\cong"]\\
    \int^{A,X}\int^{B, Y}(\A(C, X \otimes A) \otimes FX \otimes \A(A, Y\otimes B) \otimes GY \otimes THB)
    \arrow[r, "\int \int t"]
    \arrow[d, "\cong"]
    \arrow[dr, "\int \int \id \otimes t"]
    &
    \int^{A,X}\int^{B, Y}T(\A(C, X \otimes A) \otimes FX \otimes \A(A, Y\otimes B) \otimes GY \otimes HB)
    \arrow[r, "\int \xi"]
    &
    \int^{A,X}T\int^{B, Y}(\A(C, X \otimes A) \otimes FX \otimes \A(A, Y\otimes B) \otimes GY \otimes HB)
    \arrow[r, "\xi"]
    \arrow[dd, "\cong"]
    &
    T\int^{A,X}\int^{B, Y}(\A(C, X \otimes A) \otimes FX \otimes \A(A, Y\otimes B) \otimes GY \otimes HB)
    \arrow[dd, "\cong"]
    \\
    \int^{A,X}\A(C, X \otimes A) \otimes FX \otimes \int^{B, Y}(\A(A, Y\otimes B) \otimes GY \otimes THB)
    \arrow[d, "\int \id \otimes \int t"]
    &
    \int^{A,X}\int^{B, Y} (\A(C, X \otimes A) \otimes FX \otimes T(\A(A, Y\otimes B) \otimes GY \otimes HB))
    \arrow[dl, "\cong"]
    \arrow[u, "\int \int t"]
    &
    &
    \\
    \int^{A,X}\A(C, X \otimes A) \otimes FX \otimes \int^{B, Y}T(\A(A, Y\otimes B) \otimes GY \otimes HB)
    \arrow[r, "\int \id \otimes \xi"]
    &
    \int^{A,X}\A(C, X \otimes A) \otimes FX \otimes T\int^{B, Y}(\A(A, Y\otimes B) \otimes GY \otimes HB)
    \arrow[r, "\int t"]
    &
    \int^{A,X}T(\A(C, X \otimes A) \otimes FX \otimes \int^{B, Y}(\A(A, Y\otimes B) \otimes GY \otimes HB))
    \arrow[r, "\xi"]
    &
    T\int^{A,X}(\A(C, X \otimes A) \otimes FX \otimes \int^{B, Y}(\A(A, Y\otimes B) \otimes GY \otimes HB))
  \end{tikzcd}



%% file: commute.tex

    \begin{tikzcd}
      &
      &
      &
      &
      \int^{A, B}\A(C, A\otimes B)\otimes TFA \otimes TGB
      \arrow[ddrr, "\int t"]
      \arrow[ddll, "\int \gamma", swap]
      &
      &
      &
      \\
      &&&&&&&
      \\
      &
      &
      \int^{A, B}\A(C, B\otimes A)\otimes TGB \otimes TFA
      \arrow[dl, "\int t"]
      &
      &
      &
      &
      \int^{A, B}T(\A(C, A\otimes B)\otimes TFA \otimes GB)
      \arrow[dr, "\xi"]
      \arrow[dl, "\int T \gamma"]
      &
      \\
      &
      \int^{A, B}T(\A(C, B\otimes A)\otimes TGB \otimes FA)
      \arrow[dl, "\xi"]
      \arrow[dr, "\int T \gamma"]
      &&&&
      \int^{A, B}T(\A(C, B\otimes A)\otimes GB\otimes TFA)
      \arrow[dl, "\int T t"]
      \arrow[dr, "\xi"]
      &
      &
      T\int^{A, B}(\A(C, A\otimes B)\otimes TFA \otimes GB)
      \arrow[dl, "T\int \gamma"]
      \\
      T\int^{A, B}(\A(C, B\otimes A)\otimes TGB \otimes FA)
      \arrow[dr, "T\int \gamma"]
      &&
      \int^{A, B}T(\A(C, A\otimes B)\otimes FA \otimes TGB)
      \arrow[dr, "\int T t"]
      \arrow[dl, "\xi"]
      &&
      \int^{A, B}TT(\A(C, B\otimes A)\otimes GB\otimes FA)
      \arrow[dl, "\int T T \gamma"]
      \arrow[dr, "\xi"]
      &&
      T\int^{A, B}(\A(C, B\otimes A)\otimes GB\otimes TFA)
      \arrow[dl, "T\int t"]
      &
      \\
      &
      T\int^{A, B}(\A(C, A\otimes B)\otimes FA\otimes TGB)
      \arrow[dr, "T\int t"]
      &&
      \int^{A, B}TT(\A(C, A\otimes B)\otimes FA\otimes GB)
      \arrow[dl, "\xi"]
      \arrow[ddr, bend left = 5, "\int \mu"]
      &&
      T\int^{A, B}T(\A(C, B\otimes A)\otimes GB\otimes FA)
      \arrow[dl, "T\xi"]
      &&
      \\
      &&
      T\int^{A, B}T(\A(C, A\otimes B)\otimes FA \otimes GB)
      \arrow[dr, "T\xi"]
      &
      &
      TT\int^{A, B}(\A(C, B\otimes A)\otimes GB\otimes FA)
      \arrow[dl, "TT\int \gamma"]
      &&&
      \\
      &&&
      TT\int^{A, B}(\A(C, A\otimes B)\otimes FA \otimes GB)
      \arrow[d, "\mu"]
      &
      \int^{A, B}T(\A(C, A\otimes B)\otimes FA\otimes GB)
      \arrow[dl, "\xi"]
      &&&
      \\
      &&&
      T\int^{A, B}(\A(C, A\otimes B)\otimes FA \otimes GB)
      &&&
    \end{tikzcd}


%% file: assoc-tilde.tex

      \begin{tikzcd}
        \ltilde(i\ltilde (iF\otimes_{D}iG)\otimes_{D}i\ttilde H) \arrow[r, "\ltilde (\id\otimes_{D} \beta)"]
        \arrow[d, "(\ltilde(\eta^{i\ltilde} \otimes_D \id))^{-1}"]
        & \ltilde(i\ltilde (iF\otimes_{D}iG)\otimes_{D}\tbar i H)
        \arrow[rr, "\ltilde \bar{t}"]
        \arrow[d, "(\ltilde(\eta^{i\ltilde} \otimes_D \id))^{-1}"]
        & &\ltilde\tbar(i\ltilde (iF\otimes_{D}iG)\otimes_{D} i H)
        \arrow[d, "\rho"]
        \\
        \ltilde((iF\otimes_{D}iG)\otimes_{D}i\ttilde H)
        \arrow[r, "\ltilde (\id\otimes_{D} \beta)"]
        \arrow[d, "\ltilde \alpha"]
        & \ltilde((iF\otimes_{D}iG)\otimes_{D}\tbar i H)
        \arrow[r, "\ltilde \bar{t}"]
        \arrow[d, "\ltilde \alpha"]
        &
        \ltilde\tbar((iF\otimes_{D}iG)\otimes_{D} i H)
        \arrow[dr, "\rho"]
        \arrow[dddl, "\ltilde \tbar \alpha", bend left = 10]
        & \ttilde\ltilde(i\ltilde (iF\otimes_{D}iG)\otimes_{D} i H)
        \arrow[d, "\ttilde (\ltilde(\eta^{i\ltilde} \otimes_D \id))^{-1}"]
        \\
        \ltilde(iF\otimes_{D}(iG\otimes_{D}i\ttilde H))
        \arrow[r, "\ltilde (\id\otimes_{D} (\id\otimes \beta))"]
        \arrow[d, "\ltilde(\id\otimes_D \eta^{i\ltilde})"]
        &
        \ltilde(iF\otimes_{D}(iG\otimes_{D}\tbar i H))
        \arrow[d, "\ltilde(\id\otimes_{D} \bar{t})"]
        \arrow[ddl, "\ltilde(\id\otimes_D \eta^{i\ltilde})", near start, swap]
        &
        &
        \ttilde\ltilde((iF\otimes_{D}iG)\otimes_{D} i H)
        \arrow[d, "\ttilde \ltilde \alpha"]
        \\
        \ltilde(iF\otimes_{D} i\ltilde (iG\otimes_{D}i\ttilde H))
        \arrow[d, "\ltilde(\id \otimes_{D} i \ltilde (\id \otimes_{D} \beta))", swap]
        &
        \ltilde(iF\otimes_{D}\tbar(iG\otimes_{D} i H))
        \arrow[d, "\ltilde \bar{t}"]
        \arrow[ddl, "\ltilde(\id\otimes_D \eta^{i\ltilde})", near end]
        &
        &
        \ttilde\ltilde(iF\otimes_{D} (iG\otimes_{D} i H))
        \arrow[d, "\ttilde(\ltilde(\id \otimes_D \eta^{i\ltilde}))"]
        \\
        \ltilde(iF\otimes_{D} i\ltilde (iG\otimes_{D}\tbar i H))
        \arrow[d, "\ltilde (\id \otimes_{D} i \ltilde \bar{t})", swap]
        &
        \ltilde\tbar(iF\otimes_{D}(iG\otimes_{D} i H))
        \arrow[urr, "\rho"]
        &
        &
        \ttilde\ltilde(iF\otimes_{D} i\ltilde(iG\otimes_{D} i H))
        \\
        \ltilde(iF\otimes_{D} i\ltilde \tbar (iG\otimes_{D} i H))
        \arrow[r, "\ltilde(\id \otimes_{D} i \rho)"]
        &
        \ltilde(iF\otimes_{D} i \ttilde \ltilde (iG\otimes_{D} i H))
        \arrow[r, "\ltilde(\id \otimes_{D} \beta)"]
        &
        \ltilde(iF\otimes_{D} \tbar i \ltilde (iG\otimes_{D} i H))
        \arrow[r, "\ltilde \bar{t}"]
        &
        \ltilde\tbar(iF\otimes_{D}  i \ltilde (iG\otimes_{D} i H))
        \arrow[u, "\rho"]
      \end{tikzcd}



%% file: commute-tilde.tex

\begin{tikzcd}
  & & & & \ltilde(i \ttilde G \otimes_{D} i\ttilde F)
  \arrow[ddl, "\ltilde(\id\otimes_D \beta)", swap]
  &\ltilde(i\ttilde F \otimes_{D} i \ttilde G)
  \arrow[dr, "\ltilde(\id \otimes_D \beta)"]
  \arrow[dl, "\ltilde(\beta \otimes_D \id)", swap]
  \arrow[l, "\ltilde \gamma", swap]
  & & & & 
  \\
  &&& 
  &\ltilde(\tbar i F \otimes_{D} i \ttilde G)
  \arrow[dr, "\ltilde(\id \otimes_D \beta)"]
  \arrow[dl, "\ltilde\gamma"]
  &
  &\ltilde(i\ttilde F \otimes_D \tbar i G)
  \arrow[dl, "\ltilde(\beta \otimes_D \id)", swap]
  \arrow[dr, "\ltilde \bar{t}"]
  &&&
  \\
  &&&\ltilde(i\ttilde G\otimes_D \tbar i F)
  \arrow[dr, "\ltilde(\beta \otimes_D \id)"]
  \arrow[dl, "\ltilde \bar{t}", swap]
  && \ltilde (\tbar i F \otimes_D \tbar i G)
  \arrow[dr, "\ltilde \bar{t}"]
  \arrow[dl, "\ltilde \gamma"]
  &
  &\ltilde\tbar (i\ttilde F \otimes_D  i G) \arrow[dl, "\ltilde\tbar(\beta \otimes_D \id)"]
  \arrow[dr, "\rho"]
  &&
  \\
  &&\ltilde\tbar(i\ttilde G\otimes_D  i F) \arrow[dr, "\ltilde \tbar (\beta \otimes_D \id)"]
  \arrow[dl, "\rho"]
  && \ltilde(\tbar i G\otimes_D \tbar i F)
  \arrow[dl, "\ltilde \bar{t}"]
  & & \ltilde\tbar (\tbar i F \otimes_D  i G) \arrow[dr, "\rho"]
  \arrow[dl, "\ltilde \tbar \gamma"]&
  &\ttilde\ltilde(i\ttilde F\otimes_D iG) \arrow[dl, "\ttilde\ltilde(\beta\otimes_D \id)"]
  \arrow[dr, "\ttilde \ltilde \lambda"]
  &
  \\
  &\ttilde \ltilde(i\ttilde G \otimes_D iF)
  \arrow[dl, "\ttilde \ltilde \gamma"]&& \ltilde\tbar( \tbar i G \otimes_D i F) \arrow[dl, "\ltilde \tbar \gamma"]
  && \ltilde\tbar(iG\otimes_D \tbar i F)\arrow[dl, "\ltilde \tbar \bar{t}"]\arrow[drrr, "\rho"]
  &&\ttilde \ltilde(\tbar i F\otimes_D iG)\arrow[dr, "\ttilde\ltilde \lambda"] &
  &\ttilde \ltilde( iG\otimes_D i\ttilde F)\arrow[dl, "\ttilde\ltilde(\id \otimes_D \beta)"]
  \\
  \ttilde \ltilde(iF\otimes_D i\ttilde G) \arrow[dr, "\ttilde \ltilde(\id \otimes_D \beta)"] &
  & \ltilde\tbar(i F \otimes_D \tbar i G)
  \arrow[dl, "\rho"]\arrow[dr, "\ltilde \tbar \bar{t}"] && \ltilde\tbar \tbar(iG\otimes_D  i F)\arrow[dl, "\ltilde\tbar \tbar \gamma"]
  \arrow[drrr, "\rho"]&&&& \ttilde \ltilde( iG\otimes_D \tbar i F) \arrow[dl, "\ttilde\ltilde \bar{t}"]&
  \\
  &\ttilde \ltilde(iF\otimes_D \tbar i G) \arrow[dr, "\ttilde \ltilde \bar{t}"] &
  &\ltilde\tbar \tbar(i F \otimes_D  i G) \arrow[dl, "\rho"] \arrow[dd, "\ltilde \mu"]&&&& \ttilde \ltilde\tbar ( iG\otimes_D  i F)\arrow[dl, "\ttilde \rho"]&&
  \\
  &&\ttilde \ltilde \tbar (iF\otimes_D i G) \arrow[drrr, "\ttilde \rho"]&&&&\ttilde \ttilde\ltilde ( iG\otimes_D  i F) \arrow[dl, "\ttilde \ttilde \ltilde \gamma"]&&&
  \\
  &&&\ltilde \tbar(iF \otimes_D iG)\arrow[dr, "\rho"] & &\ttilde \ttilde\ltilde (iF\otimes_D  i G) \arrow[dl, "\mu"]&&&&
  \\
  &&&&\ttilde\ltilde (iF\otimes_D  i G)  &&&&&
\end{tikzcd}
